\documentclass[11pt,a4,epsf,fleqn]{article}
\usepackage{verbatim}
\usepackage{eurosym,rotating,float}
\usepackage{wrapfig, framed, caption}
\usepackage{caption}
\usepackage{graphicx}
\usepackage{multirow}
\usepackage{booktabs}
\usepackage{array}
\usepackage{soul}
\usepackage{todonotes}
\usepackage{graphicx}
\usepackage{mathcomp}
\usepackage{afterpage}
\usepackage{bigints}
\usepackage{textcomp}
\usepackage[para,online,flushleft]{threeparttable}
\usepackage{pdflscape}
\usepackage{booktabs}
\usepackage{caption}
\usepackage{relsize} 
\usepackage{array}
\usepackage{longtable}
\usepackage{placeins}
\usepackage{bm}
\usepackage{array}
\newcolumntype{K}[1]{>{\centering\arraybackslash}m{#1}}
\usepackage[sort&compress]{natbib}
\usepackage[final]{pdfpages}
\usepackage{booktabs}
\usepackage{arydshln}
\usepackage[bookmarksopen,bookmarksnumbered,colorlinks,linkcolor={blue},citecolor={blue}]{hyperref}
\usepackage{cleveref}[2012/02/15]
\crefformat{footnote}{#2\footnotemark[#1]#3}
\usepackage{graphicx}
\usepackage{color}
\usepackage[displaymath,mathlines,pagewise]{lineno}
\newcolumntype{P}[1]{>{\centering\arraybackslash}p{#1}}
\newcolumntype{L}{>{\centering\arraybackslash}m{3cm}}
\newcolumntype{M}[1]{>{\centering\arraybackslash}m{#1}}
\usepackage{multirow}
\usepackage{lscape}
\usepackage{pdflscape}
\usepackage[T1]{fontenc}
\usepackage{pdflscape}
\usepackage{adjustbox}
\usepackage{mathtools}
\usepackage{enumitem}
\setlist[description]{leftmargin=\parindent,labelindent=\parindent-0.2cm}
\setlist{noitemsep}
\DeclareMathSymbol{\mathdblquotechar}{\mathalpha}{letters}{`"}

\newcommand{\mathdblquote}{\mathtt{\mathdblquotechar}}
\begingroup\lccode`~=`"\lowercase{\endgroup
  \let~\mathdblquote
}
\AtBeginDocument{\mathcode`"="8000 }
\usepackage{amsmath}
\usepackage{amssymb}
\newtheorem{result}{Identity}
\newtheorem{proposition}{Proposition}

 \usepackage{listings}
\begin{document}
\title{Model fit vs prediction performance: contrasting fixed-coefficient models with mixture models}
\author{Stephane Hess\thanks{University of Leeds \& Delft University of Technology, s.hess@leeds.ac.uk} \and Sander van Cranenburgh\thanks{Delft University of Technology, S.vanCranenburgh@tudelft.nl}}

\maketitle
\begin{abstract}
Mixture models such as mixed logit and latent class routinely achieve large gains in fit on panel data, yet several studies have found that these gains do not carry over to out-of-sample prediction. This paper explains why, building on the distinction between population-level and person-level prediction familiar from statistics. The panel log-likelihood of a mixture is exactly the cumulative score of forecasts that condition each choice on the person's earlier choices. Models are instead usually validated and applied with unconditional forecasts, which use only the estimated distribution of tastes and are scored by the cross-sectional log-likelihood. The difference between the two log-likelihoods, which we call choice consistency, comes from reproducing the dependence between a person's choices, and unconditional forecasts cannot use it. Decomposing the fit gain into choice consistency and heterogeneity, we show that when the assumed mixing distribution cannot reproduce this dependence exactly, pursuing choice consistency distorts the estimated heterogeneity, so that a mixture may predict single choices worse than a model without mixing. In two case studies, on revealed preference and stated choice data, choice consistency accounts for more than $90\%$ of the fit gain for each of the fourteen panel mixtures. In our data, the part of the fit gain that helps to forecast single choices shrinks as people make more choices, and with many choices per person it becomes negative, leaving the mixture worse than a model without mixing. Mixtures forecast single choices clearly better only when the forecast can condition on the person's own earlier choices, which is often not possible in practice. For forecasting, the score of a panel mixture on single choices should therefore be reported alongside its fit, which requires only one further evaluation of the likelihood.\\

\textbf{Keywords}: choice modelling; mixed logit; latent class; logit mixed logit; prediction; forecasting; panel data
\end{abstract}

\section{Introduction}

Forecasting future choices is a key aim of choice modelling \citep{Chapter26Forecastingchoice}, yet far less effort has gone into what makes a good forecasting model than into behavioural insights such as willingness to pay. One reason is that forecasts often look so far ahead that the outcomes are unknown when models are developed and tested. A small body of work has examined temporal transferability, that is, how well a model predicts choices in a later period, and the accuracy of demand predictions from different model structures \citep[see][for a review]{wrro43607}. \citet{FoxDalyHessMiller2014} showed that a better specification can improve not only a model's estimation fit but also its transferability. Much less is however known about whether more complex model structures, as opposed to better utility specifications, predict better. This is an important knowledge gap.

Applications increasingly rely on models with random heterogeneity, often chosen precisely for their superior fit in estimation, which is one of the most reliable findings in applied choice modelling. On panel data, with repeated choices by the same people, letting coefficients vary randomly across people, continuously as in mixed logit or discretely as in latent class, typically improves the log-likelihood by a wide margin. A growing number of studies has however found that this gain does not carry over to prediction. \citet{CherchiCirillo2010} contrasted multinomial logit (MNL), cross-sectional and panel mixed logit models on simulated data and a six-week travel diary, and found that the panel model fitted much better without predicting better, concluding that ``a model with a better fit might provide a worse validation and forecast''. \citet{Zhao2020} found a mixed logit less accurate than MNL out of sample despite a much better fit, and attributed this to overfitting. \citet{Krueger2021} noted that mixed logit models ``do not offer substantial improvements in unconditional predictive accuracy over standard logit''. In a benchmark of over a hundred models, \citet{Wang2024} found mixed logit the weakest of the three discrete choice models they included, the other two having no random heterogeneity, and suggested that this was because prediction ``implies averaging again over the random components''. Studies of personalised prediction \citep{Danaf2019,Xie2020} report gains for forecasts conditioned on each person's earlier choices, but no clear advantage of mixtures over models without mixing when forecasts do not use that information.

The empirical finding is therefore well established, but it lacks an explanation. The two reasons offered so far are informal. Overfitting is not needed to explain the gap, which is present even at the true parameters of a correctly specified model, and the averaging of random components points in the right direction but does not say what is lost, how much, or when it matters.

The distinction behind the gap between fit and prediction is well known in statistics. For random-effects models of longitudinal data, which include panel mixture models, population-averaged and subject-specific predictions, roughly our unconditional and conditional forecasts, are different objects that answer different questions \citep{Zeger1988,HeagertyZeger2000}. The consequences of misspecifying the random-effects distribution for population-level quantities have been studied in detail and depend on which part of the specification is wrong and which quantity is of interest \citep{Neuhaus1992,HeagertyKurland2001,McCullochNeuhaus2011}. The cross-sectional likelihood used below, which treats each of a person's choices as if it were independent of the others, is the independence composite likelihood of that literature \citep{Lindsay1988,VarinReidFirth2011}, a simpler relative of the pairwise composite marginal likelihood familiar in choice modelling \citep{Bhat2011}. In econometrics, separating persistent heterogeneity from state dependence, the influence of past choices on current ones, is a long-standing problem \citep{769,DubeHitschRossi2010}. In choice modelling, however, these ideas have not been connected to the observed gap between fit and prediction, nor used to measure how much of the fit gain of a mixture comes from reproducing the dependence between a person's choices.

A simple example shows the mechanism. Suppose there are two equally common types of people, one choosing the first of two options with probability $0.9$ and the other with probability $0.1$. For one choice by a person of unknown type, the mixture predicts the first option with probability $0.5$ ($0.5\times0.9+0.5\times0.1$), exactly as a model without heterogeneity would. Its advantage appears only across choices. A person makes the same choice twice with probability $0.82$ under the mixture ($2\times0.5\times(0.9^2+0.1^2)$), against $0.5$ for two independent predictions ($2\times0.5^2$), because the mixture explains why a person's choices resemble one another. That advantage is invisible to a forecast of a single choice made without knowledge of the person's other choices, which is how mixtures are usually validated and used.

We formalise this observation, showing that it follows from how the likelihood of a mixture model is constructed, not from any particular specification. First, the fit gain of a mixture model splits exactly into two parts. The first, \emph{choice consistency}, is the difference between the panel and cross-sectional log-likelihoods at the same estimates. It measures how far the mixture explains the correlation between a person's choices, and is worth nothing to a forecast of a single choice made without knowledge of the others. We use the term in its everyday sense of a person's choices resembling one another, not in the sense of a consistent estimator, and from here on call it consistency for short. The second, which we call \emph{realised heterogeneity}, is the gain in predicting single choices one at a time over a model without mixing with the same covariates, and is the part that unconditional forecasts can use. It mainly reflects heterogeneity across people, together with the more flexible substitution patterns that mixing allows, and it can be negative when the model is misspecified, for reasons given in the third step. We split it further into what the mixture could achieve if it were estimated to fit single choices and what panel estimation gives away in pursuing consistency.

Second, we show that the panel log-likelihood is the cumulative score of one-step-ahead forecasts, each conditioned on the person's earlier choices, so that consistency is exactly a sum of log ratios of conditional to unconditional forecast probabilities. Its expectation under the model is the total correlation of each person's choice sequence, a measure of how much a person's choices reveal about one another, for any number of choices and any mixing distribution. Panel maximum likelihood is thus the right criterion for a different task, forecasting each choice from the person's earlier ones, so the apparent tension between fit and prediction is a mismatch between tasks rather than a misleading criterion.

Third, we show that a mixture can predict single choices worse than a model without mixing because, when the assumed structure of heterogeneity is wrong, panel estimation distorts the estimated heterogeneity across people. Under correct specification this cannot happen in a large enough sample, provided heterogeneity is identified from single choices, since the panel and cross-sectional estimators then target the same mixing distribution. In practice a person's choices are rarely exactly as consistent as stable tastes of the assumed form would make them, and the panel estimator can match the consistency it observes only through the dispersion of the mixing distribution. We show this locally, for changes in the spread of the mixing distribution. The dispersion estimated by a panel mixture, often interpreted substantively, therefore need not measure how much tastes differ across people, and can overstate it.

Fourth, scoring an observed sequence of choices, which the likelihood does, is not the same as forecasting one, which needs a probability for every possible sequence. Single-choice forecasts do not determine these probabilities, so a forecast that is to reflect within-person dependence, as in a microsimulation, has to be built from whole sequences. These can be simulated by drawing each person's tastes once, but the number of possible sequences grows rapidly with their length.

Using two case studies and five-fold validation, we estimate seven mixture specifications per dataset, including logit mixed logit with a flexible mixing distribution \citep{Train2016}, each on both likelihoods, giving fourteen panel/cross-sectional pairs. Consistency accounts for more than $90\%$ of the fit gain in every pair. On the revealed preference data, with a median of $23$ choices per person in estimation, the usable part of the gain is negative for all seven specifications in the mean over folds. The mixed logit, for example, fits the estimation data $986$ log-likelihood points better than MNL, yet assigns the chosen alternatives of held-out individuals a probability about four percentage points lower ($0.739$ against $0.778$) and a worse log score. In shortened panels the usable part is positive, turning negative at between ten and fifteen choices per person for the mixed logit and somewhat later for latent class. On the stated choice data, with five choices per person in estimation, it is positive but only $2\%$ to $8\%$ of the fit gain, and the mixed nested logit, $1{,}475$ points better than nested logit in fit, predicts held-out respondents no better on the mean probability and only marginally better on the log score.

Mixture models do pay off once a person's own choices can be conditioned on. The ranking of models then reverses, and the gain grows with each observed choice, although for new people this requires some of their choices to have been observed. The finding is thus not that mixture models are poor forecasting tools, but that their headline fit largely measures something that unconditional forecasts do not use.

The remainder of this paper is organised as follows. Section \ref{sec:theory} sets out the framework and the theoretical results. Section \ref{sec:case_studies} describes the data, models and validation design of the case studies, with the results presented in Section \ref{sec:results}. Section \ref{sec:conclusions} concludes.

\section{Theory}\label{sec:theory}

This section defines the panel and cross-sectional likelihoods and the forecasting targets (Sections \ref{sec:notation} and \ref{sec:targets}), shows that the panel likelihood is a score of conditional forecasts (Section \ref{sec:prequential}), decomposes the fit gain of a mixture (Section \ref{sec:consistency}), and explains when the two estimators disagree (Section \ref{sec:overdispersion}).

\subsection{Setting and notation}\label{sec:notation}

We have an estimation sample of $N$ individuals, with person $n$ making $T_n$ choices. Let $Y_{nt}$ denote the alternative chosen by person $n$ in choice situation $t$, with attributes $x_{nt}$, let $z_n$ denote the person's covariates, such as socio-demographic characteristics, and let $\beta$ denote the vector of taste coefficients. We write
\begin{equation}
P_{nt}(\beta)=P\left(Y_{nt}\mid x_{nt},z_n,\beta\right)
\end{equation}
for the probability of the choice actually observed in situation $t$, conditional on $\beta$. Heterogeneity is represented by a mixing distribution $G(\beta\mid\Omega)$ with parameters $\Omega$. In a continuous mixture, such as mixed logit, $G$ has a density $f(\beta\mid\Omega)$, while in a discrete mixture, such as latent class, it places probability $\pi_{nq}$ on the coefficients $\beta_q$ of class $q=1,\dots,Q$, where the class allocation probabilities may depend on the covariates, $\pi_{nq}=\pi_q(z_n\mid\Omega)$. The conditional probability $P_{nt}(\beta)$, the kernel of the mixture, need not be a logit, and nothing in this section depends on its form, which is why we speak of continuous and discrete mixtures rather than of mixed logit and latent class.

On panel data, the standard specification \citep{44} assumes that the same coefficients govern all of a person's choices. Writing $\mathbb{E}_G[\cdot]$ for the expectation over $\beta\sim G(\beta\mid\Omega)$, an integral for a continuous mixture and a sum over classes for a discrete one, this gives the likelihood contribution
\begin{subequations}\label{eq:LP}
\begin{alignat}{2}
&\text{general:}      &\quad L^{P}_n(\Omega) &= \mathbb{E}_G\!\left[\prod_{t=1}^{T_n}P_{nt}(\beta)\right] \label{eq:LPa}\\
&\text{continuous mixture:}  &\quad L^{P}_n(\Omega) &= \int_\beta\prod_{t=1}^{T_n}P_{nt}(\beta)\,f(\beta\mid\Omega)\,\mathrm{d}\beta \label{eq:LPb}\\
&\text{discrete mixture:} &\quad L^{P}_n(\Omega) &= \sum_{q=1}^{Q}\pi_{nq}\prod_{t=1}^{T_n}P_{nt}(\beta_q). \label{eq:LPc}
\end{alignat}
\end{subequations}
The alternative is to treat each choice in isolation, which gives
\begin{subequations}\label{eq:LCS}
\begin{alignat}{2}
&\text{general:}      &\quad L^{CS}_n(\Omega) &= \prod_{t=1}^{T_n}\mathbb{E}_G\!\left[P_{nt}(\beta)\right] \label{eq:LCSa}\\
&\text{continuous mixture:}  &\quad L^{CS}_n(\Omega) &= \prod_{t=1}^{T_n}\int_\beta P_{nt}(\beta)\,f(\beta\mid\Omega)\,\mathrm{d}\beta \label{eq:LCSb}\\
&\text{discrete mixture:} &\quad L^{CS}_n(\Omega) &= \prod_{t=1}^{T_n}\sum_{q=1}^{Q}\pi_{nq}\,P_{nt}(\beta_q). \label{eq:LCSc}
\end{alignat}
\end{subequations}
We refer to these as the \emph{panel} (P) and \emph{cross-sectional} (CS) likelihoods. The only difference between them is the position of the product relative to the expectation. The discrete form also covers nonparametric mixing distributions, which approximate a continuous distribution by a large fixed set of support points with estimated probabilities, as in logit mixed logit \citep{Train2016}, and discrete mixtures without the joint class structure of a latent class model \citep{880}. 

Mixtures in which tastes also vary within people \citep[see, e.g.,][]{849,Krueger2021} are rarely used, because the extra expectation inside the product makes them hard to estimate \citep{HESS2011973}. We concentrate on the two forms above because the panel form is what is estimated in practice \citep{44,Train2009} and, as Section \ref{sec:targets} shows, the cross-sectional form is what forecasts are scored on. 

Estimation maximises the log-likelihood over the sample, which for the two specifications is
\begin{subequations}\label{eq:LL}
\begin{alignat}{3}
LL^{P}(\Omega)&=\sum_{n=1}^{N}LL^{P}_n(\Omega), &\quad\text{with}\quad LL^{P}_n(\Omega)&=\log L^{P}_n(\Omega)&&=\log\mathbb{E}_G\!\left[\prod_{t=1}^{T_n}P_{nt}(\beta)\right], \label{eq:LLP}\\
LL^{CS}(\Omega)&=\sum_{n=1}^{N}LL^{CS}_n(\Omega), &\quad\text{with}\quad LL^{CS}_n(\Omega)&=\log L^{CS}_n(\Omega)&&=\sum_{t=1}^{T_n}\log\mathbb{E}_G\!\left[P_{nt}(\beta)\right], \label{eq:LLCS}
\end{alignat}
\end{subequations}
where $LL^{P}_n$ and $LL^{CS}_n$ are the contributions of person $n$, and $\widehat\Omega_{P}$ and $\widehat\Omega_{CS}$ are the maximum likelihood estimates of \eqref{eq:LLP} and \eqref{eq:LLCS} respectively.

For a model without mixing, $G$ places all its mass on a single value of $\beta$, so that it makes no difference whether the expectation is taken inside or outside the product, and \eqref{eq:LP} and \eqref{eq:LCS} both reduce to the likelihood of a fixed-coefficient (FC) model,
\begin{equation}\label{eq:nomix}
L^{FC}_n(\beta)=\prod_{t=1}^{T_n}P_{nt}(\beta).
\end{equation}
For such a model the panel and cross-sectional likelihoods coincide exactly, and its log-likelihood $LL^{FC}(\widehat\beta_{FC})=\sum_n\log L^{FC}_n(\widehat\beta_{FC})$, at its own maximum likelihood estimates $\widehat\beta_{FC}$, can serve as a benchmark for either.

\subsection{Prediction targets}\label{sec:targets}

After estimation, a model can be used to predict choices in three situations: for the choices used in estimation, for individuals not included in estimation, and for new choices by individuals who were included. What the prediction can use differs between them. Forecasts are probabilistic, and scoring a forecast means evaluating the probability it assigned to the alternative actually chosen, which is what the likelihood does. We therefore speak of the \emph{fit} of a model only on the likelihood on which it was estimated, and of its \emph{score} when its estimates are evaluated on another, as when the panel estimates are evaluated on single choices.

Forecasts are made with the estimated model. We write $\widehat G=G(\beta\mid\widehat\Omega)$ for the estimated mixing distribution, where $\widehat\Omega$ is whichever estimate is in use, in practice $\widehat\Omega_P$.

Three cases arise, as discussed next.

\subsubsection{Unconditional forecasts of a single choice} 

Without information on the person's other choices, the predicted probability of alternative $j$ in situation $t$ is
\begin{subequations}\label{eq:uncond}
\begin{alignat}{2}
&\text{general:}      &\quad P\left(j\mid x_{nt},z_n\right) &= \mathbb{E}_{\widehat G}\!\left[P\left(j\mid x_{nt},z_n,\beta\right)\right] \label{eq:unconda}\\
&\text{continuous mixture:}  &\quad P\left(j\mid x_{nt},z_n\right) &= \int_\beta P\left(j\mid x_{nt},z_n,\beta\right)f(\beta\mid\widehat\Omega)\,\mathrm{d}\beta \label{eq:uncondb}\\
&\text{discrete mixture:} &\quad P\left(j\mid x_{nt},z_n\right) &= \sum_{q=1}^{Q}\widehat\pi_{nq}\,P\left(j\mid x_{nt},z_n,\widehat\beta_q\right), \label{eq:uncondc}
\end{alignat}
\end{subequations}
which uses the population distribution of tastes, and the person's covariates where these enter the model, but none of the person's observed choices. It is the only forecast available for a person whose previous choices have not been observed, and in practice also the one usually made for people in the estimation sample. Scoring such forecasts one choice at a time by their log-likelihood gives $\sum_n\sum_t\log\mathbb{E}_{\widehat G}[P_{nt}(\beta)]$, which is exactly the cross-sectional log-likelihood \eqref{eq:LLCS} at the estimates $\widehat\Omega$ in use, computed on the choices being forecast, whether or not they were used in estimation. In standard practice, a model is therefore estimated on \eqref{eq:LP} but scored on \eqref{eq:LCS}, giving $LL^{CS}(\widehat\Omega_P)$ rather than either $LL^{P}(\widehat\Omega_P)$, the objective it was estimated on, or $LL^{CS}(\widehat\Omega_{CS})$, the maximum of the objective it is scored on.

\subsubsection{Conditional forecasts} 

Let $\mathcal{O}_n$ denote the set of choice situations of person $n$ in which the choice has been observed. These observed choices can be used to update the distribution of tastes. By Bayes' theorem, the posterior (or conditional) distribution is
\begin{subequations}\label{eq:posterior}
\begin{alignat}{2}
&\text{general:}      &\quad \mathrm{d}\widehat G(\beta\mid \mathcal{O}_n) &= \frac{\left[\prod_{t\in \mathcal{O}_n}P_{nt}(\beta)\right]\mathrm{d}G(\beta\mid\widehat\Omega)}{\mathbb{E}_{\beta'\sim\widehat G}\left[\prod_{t\in \mathcal{O}_n}P_{nt}(\beta')\right]} \label{eq:posteriora}\\
&\text{continuous mixture:}  &\quad \widetilde f(\beta\mid \mathcal{O}_n) &= \frac{\left[\prod_{t\in \mathcal{O}_n}P_{nt}(\beta)\right]f(\beta\mid\widehat\Omega)}{\int_{\beta'}\left[\prod_{t\in \mathcal{O}_n}P_{nt}(\beta')\right]f(\beta'\mid\widehat\Omega)\,\mathrm{d}\beta'} \label{eq:posteriorb}\\
&\text{discrete mixture:} &\quad \widetilde\pi_{nq} &= \frac{\widehat\pi_{nq}\prod_{t\in \mathcal{O}_n}P_{nt}(\widehat\beta_q)}{\sum_{q'=1}^{Q}\widehat\pi_{nq'}\prod_{t\in \mathcal{O}_n}P_{nt}(\widehat\beta_{q'})}, \label{eq:posteriorc}
\end{alignat}
\end{subequations}
where $\mathrm{d}G(\beta\mid\Omega)$ is a density for a continuous mixing distribution and a probability mass for a discrete one, $\beta'$ is the variable of integration or summation, $\widetilde f$ is the posterior density of the coefficients, $\widetilde\pi_{nq}$ the posterior class allocation probabilities, and the denominator is the panel likelihood of the observed choices. 

The predicted probability of alternative $j$ in a new choice situation $t^*$ is then
\begin{subequations}\label{eq:cond}
\begin{alignat}{2}
&\text{general:}      &\quad P\left(j\mid x_{nt^*},z_n,\mathcal{O}_n\right) &= \mathbb{E}_{\widehat G\mid \mathcal{O}_n}\!\left[P\left(j\mid x_{nt^*},z_n,\beta\right)\right] \label{eq:conda}\\
&\text{continuous mixture:}  &\quad P\left(j\mid x_{nt^*},z_n,\mathcal{O}_n\right) &= \int_\beta P\left(j\mid x_{nt^*},z_n,\beta\right)\widetilde f(\beta\mid \mathcal{O}_n)\,\mathrm{d}\beta \label{eq:condb}\\
&\text{discrete mixture:} &\quad P\left(j\mid x_{nt^*},z_n,\mathcal{O}_n\right) &= \sum_{q=1}^{Q}\widetilde\pi_{nq}\,P\left(j\mid x_{nt^*},z_n,\widehat\beta_q\right). \label{eq:condc}
\end{alignat}
\end{subequations}
For a continuous mixture the integrals are approximated by weighting draws from $f(\beta\mid\widehat\Omega)$ by the likelihood of the person's observed choices \citep[][ch.~11]{Train2009}. Conditional forecasts are available for anyone whose earlier choices are observed at the time of forecasting, including people outside the estimation sample, for whom the posterior \eqref{eq:posterior} can be computed without re-estimating the model.

\subsubsection{Scoring and forecasting a sequence}

Finally, the model can also be used to calculate a probability for any sequence of alternatives $(j_1,\dots,j_{T_n})$ that person $n$ might choose in situations $1,\dots,T_n$,
\begin{equation}\label{eq:path}
P\left(j_1,\dots,j_{T_n}\mid x_{n1},\dots,x_{nT_n},z_n\right)=\mathbb{E}_{\widehat G}\!\left[\prod_{t=1}^{T_n}P\left(j_t\mid x_{nt},z_n,\beta\right)\right],
\end{equation}
with the continuous and discrete forms following \eqref{eq:LP}. Evaluated at the alternatives actually chosen, \eqref{eq:path} is the panel likelihood \eqref{eq:LP} at the estimates. 

For a single choice, forecasting and scoring use the same $J$ probabilities, the forecast being the full vector and the score the entry for the alternative chosen. For a sequence they part company. Scoring an observed sequence needs only \eqref{eq:path} at the observed path, which is cheap for any $T_n$. Forecasting a sequence, for example in a microsimulation of switching between alternatives, needs \eqref{eq:path} for every possible path, or $J^{T_n}$ probabilities. These are not determined by the single-choice forecasts, since the joint probability differs from their product by exactly the dependence that Section \ref{sec:consistency} quantifies as consistency, so sampling each choice independently from its own forecast would discard the dependence the panel model was estimated to capture. Enumerating all paths is feasible for short sequences, such as six tasks with five options ($5^6=15{,}625$ paths) in the stated choice data of Section \ref{sec:case_studies}, but not for long ones, such as $27$ trips among up to six modes ($6^{27}\approx10^{21}$) for the median respondent in the revealed preference data. Sequences can still be simulated, by drawing each person's tastes once and then their choices given those tastes. The ``sequence'' results in this paper are scores of realised sequences, not forecasts of them.

\subsection{The panel likelihood as a sequence of conditional forecasts}\label{sec:prequential}

The panel and conditional objects are closely linked. Let $Y_{n,1:t-1}$ denote the first $t-1$ observed choices of person $n$, and write $\mathbb{E}_{G\mid Y_{n,1:t-1}}$ for an expectation under the posterior given those choices, formed as in \eqref{eq:posterior} but with $G$ in place of $\widehat G$ and the first $t-1$ choices as $\mathcal{O}_n$, with the prior $G$ at $t=1$.

\begin{result}\label{res:prequential}
For any mixing distribution $G$ and any $T_n$,
\begin{equation}\label{eq:prequential}
LL^{P}_n=\sum_{t=1}^{T_n}\log\mathbb{E}_{G\mid Y_{n,1:t-1}}\left[P_{nt}(\beta)\right].
\end{equation}
\end{result}

\noindent\emph{Proof.} Let $A_{nt}=\mathbb{E}_G[\prod_{s\le t}P_{ns}(\beta)]$, with $A_{n0}=1$. By \eqref{eq:posterior}, $\mathbb{E}_{G\mid Y_{n,1:t-1}}[P_{nt}(\beta)]=A_{nt}/A_{n,t-1}$. In the product of these ratios over $t=1,\dots,T_n$, each intermediate $A_{nt}$ appears once in a numerator and once in a denominator and cancels, leaving $A_{nT_n}/A_{n0}=L^{P}_n$. $\square$

\medskip
Each term in \eqref{eq:prequential} is the log of the probability that a forecast of choice $t$, conditioned on everything the person chose before, assigns to the choice actually made. For any given $\Omega$, the panel log-likelihood is therefore exactly the cumulative score of one-step-ahead conditional forecasts, which is the prequential form of the likelihood \citep{Dawid1984}. Maximising the panel likelihood thus chooses the mixing distribution that makes these one-step-ahead forecasts as good as possible. By the same token, maximising the cross-sectional likelihood makes unconditional forecasts \eqref{eq:uncond} as good as possible. The two estimators are each the right criterion for a forecasting problem, but for different problems.

\subsection{Choice consistency and the decomposition of the fit gain}\label{sec:consistency}

Subtracting the cross-sectional log-likelihood of the same person, $LL^{CS}_n$ in \eqref{eq:LLCS}, from \eqref{eq:prequential} gives the exact difference between the two objects.

\begin{result}\label{res:telescope}
For any mixing distribution $G$ and any $T_n$,
\begin{equation}\label{eq:telescope}
LL^{P}_n-LL^{CS}_n=\sum_{t=2}^{T_n}\log\frac{\mathbb{E}_{G\mid Y_{n,1:t-1}}\left[P_{nt}(\beta)\right]}{\mathbb{E}_G\left[P_{nt}(\beta)\right]}.
\end{equation}
\end{result}

\noindent The term for $t=1$ vanishes because the first forecast has no history to condition on. We call this sum \emph{choice consistency}, or consistency for short, as it measures how much better the person's choices are predicted once their other choices are known. Nothing in the model is sequential. The panel likelihood \eqref{eq:LP} does not depend on the order of a person's choices, so \eqref{eq:telescope} holds for any ordering, and the sum measures how much a person's choices resemble one another, not an influence of earlier choices on later ones. Summed over the sample, it is
\begin{equation}\label{eq:C}
C(\Omega)=\sum_{n=1}^{N}\sum_{t=2}^{T_n}\log\frac{\mathbb{E}_{G\mid Y_{n,1:t-1}}\left[P_{nt}(\beta)\right]}{\mathbb{E}_G\left[P_{nt}(\beta)\right]}=LL^{P}(\Omega)-LL^{CS}(\Omega),
\end{equation}
the difference between the two log-likelihoods of \eqref{eq:LL}. Consistency compares the two likelihoods of one model at the same parameter values, not each at its own maximum, and, like Identities \ref{res:prequential} and \ref{res:telescope}, \eqref{eq:C} holds at any such value. The quantity of practical interest, and the one we report in Section \ref{sec:results}, is consistency at the panel estimates, $C(\widehat\Omega_P)$, since these are the estimates a mixture model is most commonly used with.

The difference between the likelihoods themselves is easiest to interpret for two choice situations, where it reduces to
\begin{equation}\label{eq:cov}
L^{P}_n-L^{CS}_n=\mathbb{E}_G\left[P_{n1}(\beta)P_{n2}(\beta)\right]-\mathbb{E}_G\left[P_{n1}(\beta)\right]\mathbb{E}_G\left[P_{n2}(\beta)\right]=\operatorname{Cov}_G\left(P_{n1}(\beta),P_{n2}(\beta)\right),
\end{equation}
so the panel likelihood exceeds the cross-sectional one when the same tastes make both observed choices likely, that is, when the covariance is positive. The corresponding term of \eqref{eq:telescope} is exactly $\log\left(1+\operatorname{Cov}_G\left(P_{n1}(\beta),P_{n2}(\beta)\right)/\left(\mathbb{E}_G\left[P_{n1}(\beta)\right]\mathbb{E}_G\left[P_{n2}(\beta)\right]\right)\right)$, which is close to the covariance divided by the product of the two expected probabilities when the dependence is weak. For more than two choices the difference between the likelihoods involves co-moments of all orders and has no simple form, whereas \eqref{eq:telescope} has one term per choice, which is why we work with log-likelihoods.

The expectation of \eqref{eq:telescope} has a standard interpretation in terms of two quantities from information theory \citep{Shannon1948,CoverThomas2006}. The entropy of a random variable, $\mathcal{H}(Y)=-\mathbb{E}[\log P(Y)]$, measures how unpredictable it is, and is largest, namely $\log J$, for a choice among $J$ alternatives, when all outcomes are equally likely. The mutual information between two random variables, $I(Y;Z)=\mathcal{H}(Y)-\mathcal{H}(Y\mid Z)$, is the reduction in the entropy of $Y$ from knowing $Z$, which equals the expected value of $\log P(Y\mid Z)-\log P(Y)$. Each term of \eqref{eq:telescope} is such a log ratio, so if the choices are generated by the model, the expected value of the term for choice $t$ is the mutual information between choice $t$ and the choices before it, $I(Y_{nt};Y_{n,1:t-1})$, and the expected value of the sum is
\begin{equation}\label{eq:totalcorr}
\sum_{t=2}^{T_n}I\left(Y_{nt};Y_{n,1:t-1}\right)=\sum_{t=1}^{T_n}\mathcal{H}\left(Y_{nt}\right)-\mathcal{H}\left(Y_{n,1:T_n}\right),
\end{equation}
where the right-hand side follows by expanding each mutual information as a difference of entropies and applying the chain rule for entropy, $\mathcal{H}(Y_{n,1:T_n})=\sum_t\mathcal{H}(Y_{nt}\mid Y_{n,1:t-1})$. It is the \emph{total correlation} of the choice sequence implied by the model \citep{Watanabe1960}, the Kullback--Leibler divergence between the joint distribution of a person's choices and the product of its marginals. All these quantities are conditional on the attributes and covariates, so the choices need not be identically distributed. If the model is correctly specified, consistency as measured in a sample is therefore an estimate of the total correlation of individuals' choice sequences.

If the model is misspecified, the expected value of consistency under the true distribution of choices is the true total correlation, less the extent to which the model misrepresents the joint distribution of a person's choices, plus the extent to which it misrepresents each choice on its own, and it can be larger or smaller than the true total correlation, or even negative. Under the model, non-negativity holds only in expectation, not term by term, since a person whose history makes their observed choice less likely contributes a negative term. Since the mutual information cannot exceed the entropy of $Y_{nt}$, which is at most $\log J_{nt}$ with $J_{nt}$ available alternatives, consistency can contribute at most $\log J_{nt}$ for choice $t$ in expectation under the model.

Three properties of Identities \ref{res:prequential} and \ref{res:telescope} matter for what follows. First, they hold for any mixing distribution, parametric or not. A more flexible mixing distribution can improve the fit to single choices, but it cannot make consistency available to an unconditional forecast. Second, the difference is not overfitting, the explanation offered by \citet{Zhao2020}. It is present at the true parameters of a correctly specified model, and in expectation it is a sum of $T_n-1$ non-negative terms, so its expected total does not fall as the number of choices per person rises, although its average per choice can, and it does not depend on the number of parameters. Identity \ref{res:telescope} also makes precise the suggestion of \citet{Wang2024} that prediction ``implies averaging again over the random components'', since what is lost in that averaging is exactly consistency. Third, the difference is directly measurable. Consistency at the panel estimates, $C(\widehat\Omega_P)=LL^{P}(\widehat\Omega_P)-LL^{CS}(\widehat\Omega_P)$, requires only one evaluation of the model beyond estimation, of the cross-sectional log-likelihood at those estimates, which amounts to predicting each estimation choice unconditionally.

\label{sec:decomposition}%

Let $LL^{FC}(\widehat\beta_{FC})$ be the log-likelihood at convergence of a fixed-coefficient (FC) model with the same deterministic part of utility as the mixture, including any socio-demographic interactions, estimated by maximum likelihood. By \eqref{eq:nomix}, this benchmark is the same on either likelihood. The fit gain of a panel mixture then splits exactly as
\begin{equation}\label{eq:decomp}
\underbrace{LL^{P}(\widehat\Omega_P)-LL^{FC}(\widehat\beta_{FC})}_{\text{fit gain, }\Delta}=\underbrace{LL^{P}(\widehat\Omega_P)-LL^{CS}(\widehat\Omega_P)}_{\text{consistency, }C}+\underbrace{LL^{CS}(\widehat\Omega_P)-LL^{FC}(\widehat\beta_{FC})}_{\text{realised heterogeneity, }R}.
\end{equation}
The second term, \emph{realised heterogeneity}, is the gain in the score on single choices, the part that unconditional prediction can use. It mainly reflects heterogeneity across people, but also any gain from the more flexible substitution patterns that mixing allows, which one choice per person cannot separate from it.

Realised heterogeneity need not be positive, because $\widehat\Omega_P$ was not chosen to fit single choices (Section \ref{sec:overdispersion}). To see how far the panel estimates are from the best possible fit to single choices, we use the cross-sectional estimates $\widehat\Omega_{CS}$ of the same mixture model. By construction, $LL^{CS}(\widehat\Omega_{CS})$ is the best fit to single choices that the mixture family can achieve. Adding and subtracting it splits realised heterogeneity into two parts,
\begin{equation}\label{eq:ladder}
R=\underbrace{\left[LL^{CS}(\widehat\Omega_{CS})-LL^{FC}(\widehat\beta_{FC})\right]}_{H^*}-\underbrace{\left[LL^{CS}(\widehat\Omega_{CS})-LL^{CS}(\widehat\Omega_P)\right]}_{E},
\end{equation}
where $H^*$ is how much better the mixture family can fit single choices than the fixed-coefficient model, when it is estimated for that purpose, and so the most that heterogeneity can contribute to that fit, which we refer to as the \emph{ceiling}. $E$ is how much of this ceiling the panel estimates give away. $E\geq0$ because $\widehat\Omega_{CS}$ maximises $LL^{CS}$, provided the maximum found is the global one and the cross-sectional likelihood is simulated with the same draws at both sets of estimates. $H^*$ is non-negative whenever the fixed-coefficient model is a restricted case of the mixture family (for example, with all standard deviations at zero, or a support of zero width for the logit mixed logit), provided the cross-sectional maximum found is the global one. 

If $E<H^*$, realised heterogeneity is positive. Panel estimation gives away part of what heterogeneity could add, but the mixture still scores better on single choices than the model without mixing. If $E>H^*$, realised heterogeneity is negative. What reproducing the dependence between a person's choices costs on single choices is then more than heterogeneity could add, and the mixture scores worse on single choices than the model without mixing, although the mixture family could fit them better. It still fits the panel data better, because consistency exceeds the net loss, $E-H^*$. The cross-sectional estimates are used here only as a reference point for measuring the panel estimates, not as a recommended estimator (Section \ref{sec:overdispersion}).

Combining \eqref{eq:decomp} and \eqref{eq:ladder}, the fit gain of a panel mixture is made up of three terms, two gains and one loss, giving
\begin{equation}\label{eq:three}
\Delta=C(\widehat\Omega_P)+H^*-E.
\end{equation}

The split of the fit gain into consistency and the score on single choices is not unique, because consistency depends on the parameter values at which the two likelihoods are compared. Comparing them at the cross-sectional estimates $\widehat\Omega_{CS}$ instead would give a different value of consistency, $C(\widehat\Omega_{CS})$, and a correspondingly different remainder. We use the panel estimates for the reason given after \eqref{eq:C}, so that \eqref{eq:three} describes the model the analyst actually has.

The terms of \eqref{eq:three} can also be grouped differently. Combining $C$ and $-E$, rather than $H^*$ and $-E$, gives
\begin{equation}\label{eq:regroup}
\Delta=\underbrace{\left[LL^{CS}(\widehat\Omega_{CS})-LL^{FC}(\widehat\beta_{FC})\right]}_{H^*}+\underbrace{\left[LL^{P}(\widehat\Omega_P)-LL^{CS}(\widehat\Omega_{CS})\right]}_{C-E},
\end{equation}
which compares the two estimators of the same mixture, each at its own maximum. Here, $H^*$ is what the cross-sectional estimator gains over the fixed-coefficient model, and $C-E$ is how much higher the maximised panel log-likelihood is than the maximised cross-sectional one. The difference between the two maxima, which is what an analyst sees when estimating a model both ways, is thus consistency less what panel estimation gives away on single choices, not consistency alone. This grouping is used only for that comparison.

For an analyst who wants to know how much of the fit gain of a panel mixture will carry over to unconditional prediction, the quantity of interest is realised heterogeneity, which we summarise as a share of the fit gain, the \emph{usable share},
\begin{equation}\label{eq:usable}
S=\frac{R}{\Delta}=1-\frac{C(\widehat\Omega_P)}{\Delta}=\frac{H^*-E}{C(\widehat\Omega_P)+H^*-E}.
\end{equation}
The fit gain in the denominator is positive whenever the fixed-coefficient model is a restricted case of the mixture and the panel maximum is the global one. The second form follows from \eqref{eq:decomp} and shows that $S$ is one minus the share of the fit gain accounted for by consistency. The third follows from \eqref{eq:ladder} and \eqref{eq:three}. Expressing the result as a share rather than in log-likelihood points makes it scale-free, so that it can be compared across datasets that differ in the number of people and the number of alternatives, and across models whose fit gains differ in size, although by construction it depends on the number of choices per person. A usable share of one, $S=1$, would mean that the whole gain comes from a better score on single choices, as must be the case with one choice per person, where $C(\widehat\Omega_P)=0$. A usable share of zero, $S=0$, means that the whole gain comes from consistency, so that the unconditional forecasts of the mixture fit the estimation choices no better than those of the model without mixing, even though its panel fit is better. A negative usable share, $S<0$, means that they score worse on them. Consistency then exceeds the whole fit gain, so that the share it accounts for is above one. The third form requires a second, cross-sectional estimation, while the first two do not. Since all three are computed on the estimation data, the usable share describes how well unconditional forecasts fit the choices used in estimation. Whether the same holds for new people is an out-of-sample question, which we examine in Section \ref{sec:results}.

\subsection{The role of misspecification}\label{sec:overdispersion}

Whether realised heterogeneity is positive depends on how far the two estimators disagree. Suppose first that the panel mixture is correctly specified, with stable tastes for each person drawn from the assumed mixing distribution. The panel estimator is then the maximum likelihood estimator. The cross-sectional estimator maximises a sum of single-choice log-likelihoods, each of which is correctly specified, so provided the parameters are identified from a single choice per person, it also converges to the true parameters \citep[Lemma 15.2]{Ruud2000}, as \citet[Section 2.2]{HESS2011973} show. It discards the information in the dependence between a person's choices and is therefore less efficient. Identification is the weak point in practice. With one choice per person, heterogeneity is identified only through the shape of the single-choice probabilities, which is why estimating it from cross-sectional data requires large samples \citep{HESS2011973}. Both estimators then converge to the same $G$. As a result, $E$ reflects only in-sample optimism, the extra fit $\widehat\Omega_{CS}$ gains from being estimated on the same data on which it is evaluated, and realised heterogeneity is positive in large samples, provided the heterogeneity changes the probabilities of single choices and not only the relation between a person's choices. $H^*$ then grows with the sample size, while $E$ does not. Out of sample, unconditional forecasts from the panel estimates are if anything better than those from the cross-sectional estimates, since both target the same distribution and the panel estimator estimates it more precisely.

In practice, the assumption that separates the two estimators, that a single stable vector of tastes governs all of a person's choices, is rarely exactly true. The cross-sectional estimator ignores the relation between a person's choices altogether. The panel estimator treats all of it as stable heterogeneity, and has only one instrument with which to reproduce it, the dispersion of $G$. This dispersion determines two things at once, how single choices vary across people and how strongly each person's choices are related to one another. Under correct specification the dispersion that matches the first also matches the second. Under misspecification it need not. If a person's choices are more strongly related than the model implies at the dispersion that best fits single choices, the panel estimator can reproduce the relation only by increasing the dispersion of $G$. This happens when part of the relation does not come from stable tastes, as with habit or state dependence, or when the true distribution of tastes has a shape the assumed family cannot represent, for instance more people with strong and persistent preferences for one alternative than it can accommodate while also fitting single choices. If the relation is weaker than the model implies, for instance because tastes also vary from one choice to the next, the panel estimator reduces the dispersion instead. Since the panel likelihood weights the relation between a person's choices by their number, the distortion, and with it $E$, can be expected to grow with $T_n$. The panel estimate of the dispersion, often read as a measure of how much tastes differ across people, is then larger than the dispersion that best describes single choices in the first case and smaller in the second. Neither estimate should be read directly as a measure of heterogeneity across people. Under habit or state dependence, the panel estimate overstates it. When tastes also vary from one choice to the next, it is the cross-sectional estimate that overstates it, since it mixes variation across and within people, and the panel estimate, although smaller, may be the closer of the two.

This is the mechanism behind a negative realised heterogeneity, which arises when the loss $E$ caused by the distortion exceeds the ceiling $H^*$. A more flexible mixing distribution can remove the part of the distortion that comes from the shape of $G$, but not the part that comes from dependence that stable tastes cannot produce, such as habit or tastes that vary from one choice to the next. Specifications that allow tastes to vary within as well as across people address the second of these \citep{HESS2011973,Krueger2021}, but they are rarely used in practice, and \citet{Krueger2021} find that they do not improve prediction over standard mixed logit. In a large enough sample, comparing the two estimators gives an indication of which of the two cases above applies.

The case in which the panel likelihood favours more dispersion can be stated more precisely by asking whether, starting from the mixing distribution that best fits single choices, the panel likelihood would increase if that distribution were made slightly more spread out. Let $G_\lambda$ be a path of mixing distributions within the family, indexed by a scalar $\lambda$, that passes through $\widehat G_{CS}$ at $\lambda=1$. The natural example is a dilation, which stretches $\widehat G_{CS}$ about its mean $\mu$ by a factor $\lambda$, leaving the mean and the shape of the distribution unchanged, so that $G_\lambda$ is the distribution of $\mu+\lambda(\beta-\mu)$ for $\beta\sim\widehat G_{CS}$. Here, $\lambda=1$ returns $\widehat G_{CS}$, $\lambda=0$ is a model without mixing with coefficients $\mu$, which need not be the estimated fixed-coefficient model, and for normal mixing distributions the dilation simply scales every standard deviation by $\lambda$. Write $\psi(\lambda)=LL^{CS}(G_\lambda)$, $\phi(\lambda)=LL^{P}(G_\lambda)$ and $D(\lambda)=\phi(\lambda)-\psi(\lambda)$ for consistency \eqref{eq:C} along the path.

\begin{proposition}\label{prop:local}
Let $\widehat G_{CS}$ maximise the cross-sectional likelihood within a family of mixing distributions, let $G_\lambda$, for $\lambda$ in an open interval containing $1$, be a path within the family with $G_1=\widehat G_{CS}$, and suppose $\psi$ and $D$ are differentiable at $\lambda=1$. Then $\phi'(1)=D'(1)$. Hence if consistency increases along the path at $\widehat G_{CS}$, so does the panel likelihood, and $\widehat G_{CS}$ is not a stationary point of the panel objective.
\end{proposition}

\noindent\emph{Proof.} Since the path stays within the family, $\psi$ has a maximum at $\lambda=1$, so $\psi'(1)=0$, and $\phi=\psi+D$. $\square$

\medskip
In words, $\widehat G_{CS}$ is the best fit to single choices, so it sits at the peak of the cross-sectional likelihood. At a peak the likelihood is flat, so stretching or shrinking $\widehat G_{CS}$ slightly hardly changes the fit to single choices, which is what $\psi'(1)=0$ says. Since the panel likelihood is the cross-sectional likelihood plus consistency, near $\widehat G_{CS}$ it rises or falls exactly as consistency does. The only requirement on the path is that it stays within the family. A dilation does so for any location-scale family, such as normal mixing distributions, and so does stretching only some of the coefficients, as in Section \ref{sec:res_dispersion}. The sign of $D'(1)$ links this to the cases above. Under correct specification, both likelihoods are maximised at the same $G$ in large samples, so $D'(1)=0$ in the limit and the two estimators differ only through sampling variation. In a finite sample $D'(1)$ varies even under correct specification, and this variation can be large, so its sign indicates misspecification only when the slope lies well outside the values that a correctly specified model produces. If $D'(1)>0$ along a path that increases dispersion, the choices in the sample are more strongly related than the model implies at $\widehat G_{CS}$, and the panel likelihood rises with more dispersion. If $D'(1)<0$, it rises with less. The result requires no concavity, uniqueness or single-peakedness of either likelihood.

The result has three limits. It is local, saying in which direction the panel likelihood rises away from $\widehat G_{CS}$, not how far the panel estimate lies from it. It concerns one direction of change at a time, and so says nothing about the shape the panel estimate takes. And it does not say that the panel estimate lies on this path, since in a parametric family the panel and cross-sectional estimates generally differ in their means as well. A stronger, global result, in which the panel estimate is always the more dispersed of the two, does not hold in general. In a parametric family the means and dispersion parameters are estimated jointly, so an individual dispersion parameter can move in either direction even when consistency increases with dispersion overall, and in the nonparametric class many distributions of different dispersion fit single choices equally well. How large the difference is, and in which direction it goes, therefore has to be measured in each dataset, which we do in Section \ref{sec:res_dispersion}.

Because Identities \ref{res:prequential} and \ref{res:telescope} hold whether or not the mixing distribution is correctly specified, they say nothing about which of two models is closer to the true data-generating process. Where the two estimators disagree materially, neither recovers the true mixing distribution, and our results describe what the panel estimates, which are the ones used in practice, deliver on each forecasting object.

\section{Case study setup}\label{sec:case_studies}

This section describes the two datasets and how they are split for estimation and validation (Section \ref{sec:data}), the models (Section \ref{sec:models}), their specification (Section \ref{sec:specification}) and their estimation (Section \ref{sec:estimation}).

\subsection{Data}\label{sec:data}

We use two datasets, one revealed preference (RP) and one stated choice (SC). 

\subsubsection{DECISIONS data}

The first dataset comes from a large-scale RP survey conducted as part of the DECISIONS project carried out by the Choice Modelling Centre at the University of Leeds \citep{calastri2020we}. We use the observed mode choices from the GPS tracking component of this survey, collected with the rMove smartphone app. Data cleaning and enrichment, including the derivation of travel times and costs for every trip, were carried out by \citet{tsoleridis2022deriving}. For each trip, individuals travelled by one of six modes: car (47.5\%), bus (14.2\%), rail (5\%), taxi (3.4\%), cycling (3.4\%) or walking (26.5\%). The attributes of the alternatives used in the models are in-vehicle travel time (IVT), out-of-vehicle travel time (OVT), and travel cost. Not every mode is available for every trip, with car available for $53\%$ of trips, rail for $28\%$, cycling for $39\%$, walking for $42\%$, and bus and taxi for about $90\%$. For the present analysis, we retain the $373$ individuals with at least $14$ recorded trips, so that four trips per person can be held out for validation while leaving at least ten for estimation, and for individuals with more than $50$ trips we keep the first $50$ in chronological order. This gives $11{,}022$ trips, with between $14$ and $50$ per individual and a median of $27$. Successive trips by the same person are not linked only by stable tastes, however. The outbound and return trips of a tour, for example, usually share a mode, which is one source of the dependence discussed in Section \ref{sec:overdispersion} that stable tastes cannot produce.

\subsubsection{COVID-19 data}\label{sec:data_covid}

\begin{figure}[h]
\centering
 \includegraphics[width=0.4\textwidth]{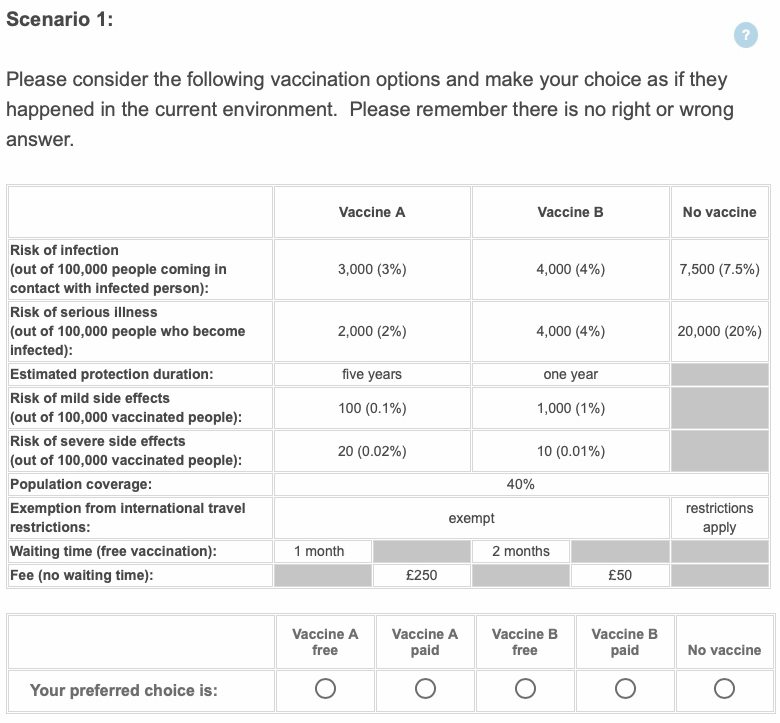}
    \caption{Example of SC choice task from COVID-19 study}
    \label{fig:covid}
\end{figure}

The second dataset comes from a global SC survey looking at the potential uptake of COVID-19 vaccines early on in the pandemic \citep{HESS2022114800}. Respondents were presented with six choice tasks, each giving a choice between two vaccines, with a paid and free version of each, or no vaccination (cf. Figure \ref{fig:covid}). This gives a choice between five options, where the combined market share for the free vaccines was 63.4\%, with 29.8\% for the paid vaccines, and 6.8\% for no vaccination. The alternatives were described by nine attributes, defined in Section \ref{sec:utilities}. We use the UK sample from the overall study, with $12{,}882$ observations from $2{,}147$ respondents. The sixth task differs from the others in a way that matters for what follows. Rather than being drawn from the experimental design, it is a single fixed scenario, identical for all respondents and always presented last, in which the free vaccines carry long waiting times and the paid vaccines high fees \citep[cf.][]{HESS2022114800}. This places it at the edge of the design space on most attributes, and the observed share of the free vaccines is $0.72$ in this task, against $0.62$ in the first five. We use it as the in-sample holdout (Section \ref{sec:splitting}).

\subsubsection{Data splitting and validation design}\label{sec:splitting}

Each dataset is split by individual, at random, into five folds. In each fold, four fifths of the individuals form the estimation sample and the remaining fifth are held out entirely, so that every individual is out of sample exactly once. Within the estimation sample, the last choices of each person are also held out, namely the final four trips, in chronological order, in the DECISIONS data, and the fixed sixth task in the COVID-19 data. All models are estimated on the remaining choices of the estimation individuals, which leaves a median of $23$ trips per person (between $10$ and $46$) in the DECISIONS data and five tasks per respondent in the COVID-19 data, and every result reported below is a mean over the five folds, with the standard deviation across folds given where it is informative.

This gives three sets of choices on which a model can be scored, corresponding to the three situations of Section \ref{sec:targets}. The estimation rows are the choices the model was fitted to, on which Section \ref{sec:decomposition} decomposes the fit. The in-sample holdout consists of later choices by individuals in the estimation sample, for whom a history exists, so that unconditional and conditional forecasts can be compared, and, in the DECISIONS data, sequences of up to four choices can be scored. The out-of-sample set consists of all choices by held-out individuals, for whom only unconditional forecasts are available. In the DECISIONS data, where the trips of a held-out individual form a sequence in time, we also condition each trip on the earlier ones, to show how quickly conditioning pays off for a new person. Table \ref{tab:data_splitting} gives the sizes.

Both holdouts are systematic rather than random, but in different ways. The DECISIONS holdout is a forecast forward in time, one to four trips beyond the estimation window, in the same region of attribute space. The COVID-19 holdout is the fixed sixth task (Section \ref{sec:data_covid}), a combination of attribute levels not seen in estimation. Because every respondent faces the same task, unconditional forecasts differ between respondents only through the covariates, so this holdout is mainly informative about aggregate share recovery and the value of conditioning, not about how people differ in their response to the attributes.

\begin{table}[htbp]
  \centering\small
  \caption{Data split per fold (ranges over the five folds)}
    \begin{adjustbox}{max width=0.8\textwidth}
\begin{tabular}{lcccc}
\toprule          & \multicolumn{2}{c}{\textbf{DECISIONS}} & \multicolumn{2}{c}{\textbf{COVID-19}} \\
          & Individuals & Choices & Individuals & Choices \\
\midrule
    Estimation rows & 298--299 & 7,516--7,766 & 1,717--1,718 & 8,585--8,590 \\
    In-sample holdout & 298--299 & 1,192--1,196 & 1,717--1,718 & 1,717--1,718 \\
    Out of sample & 74--75 & 2,064--2,314 & 429--430 & 2,574--2,580 \\
\midrule
    Total & 373 & 11,022 & 2,147 & 12,882 \\
\bottomrule
    \end{tabular}
\end{adjustbox}%
  \label{tab:data_splitting}%
\end{table}%

\subsection{Models}\label{sec:models}

The base specification is a Multinomial Logit (MNL) model in the DECISIONS case study and a Nested Logit (NL) model in the COVID-19 case study, nesting together the four vaccine alternatives. Starting from each base model, we estimate the following specifications on every fold:

\begin{description}
	\item[Models without random heterogeneity:] the base model without covariates, and the base model with deterministic heterogeneity through socio-demographic covariates. These are the fixed-coefficient benchmarks $LL^{FC}$ of Section \ref{sec:decomposition}, and each mixture model is compared with the benchmark that uses the same covariates.
	\item[Models with continuous random heterogeneity:] mixed versions of the base models, Mixed MNL (MMNL) for DECISIONS and Mixed NL for COVID-19, with and without covariates.
	\item[Models with discrete random heterogeneity:] latent class (LC) versions of the base models, with three classes, chosen on the basis of BIC in preliminary estimation on the full sample with the panel likelihood, which also supplied the starting values, so that the held-out individuals informed these choices, although they did not enter the likelihood maximised on any fold. They are estimated in three variants: without covariates, with covariates in the utilities, and with covariates in the class allocation model. In the tables and figures, ``with socios'' denotes a specification with the socio-demographic covariates, and ``socios in utility'' and ``socios in class allocation'' the two latent class variants that use them.
	\item[Models with nonparametric random heterogeneity:] logit mixed logit (LML) versions of the base models \citep{Train2016}, with and without covariates, described below. With the nested logit kernel, the COVID-19 version is strictly a logit-mixed nested logit, but for simplicity we use the name LML for both datasets.
\end{description}

\noindent Each of the seven mixture specifications per dataset is estimated twice, on the panel likelihood \eqref{eq:LP}, which is the standard practice, and on the cross-sectional likelihood \eqref{eq:LCS}. We refer to the latter as the cross-sectional (CS) version of the model. Each dataset therefore has seven pairs of a panel and a cross-sectional version of the same mixture model, fourteen pairs in total, and with the two fixed-coefficient benchmarks, sixteen models are estimated on each fold of each dataset.

\subsection{Model specification}\label{sec:specification}

\subsubsection{Utility functions}\label{sec:utilities}

All attributes enter the utilities linearly, apart from an indicator for unknown duration of protection in the COVID-19 study, and the constants of car and of no vaccination are normalised to zero. The three binary covariates, female, aged over 60, and annual household income over $\pounds 60$K, are interacted with the constants. Dropping the indices for the person and the choice situation, the utilities of the models without random heterogeneity in the DECISIONS case study are
\begin{equation}\label{eq:util_dec}
\begin{aligned}
V_{\text{car}}&=\beta_{\text{IVT},\text{car}}\,\text{IVT}_{\text{car}}+\beta_{\text{cost}}\,\text{cost}_{\text{car}},\\
V_{m}&=\delta_{m}+\sum_{w}\kappa_{m,w}z_{n,w}+\beta_{\text{IVT},m}\,\text{IVT}_{m}+\beta_{\text{OVT},m}\,\text{OVT}_{m}+\beta_{\text{cost}}\,\text{cost}_{m}, &\quad& m\in\{\text{bus},\text{rail}\},\\
V_{\text{taxi}}&=\delta_{\text{taxi}}+\sum_{w}\kappa_{\text{taxi},w}z_{n,w}+\beta_{\text{IVT},\text{taxi}}\,\text{IVT}_{\text{taxi}}+\beta_{\text{cost}}\,\text{cost}_{\text{taxi}},\\
V_{m}&=\delta_{m}+\sum_{w}\kappa_{m,w}z_{n,w}+\beta_{\text{IVT},m}\,\text{IVT}_{m}, && m\in\{\text{cycle},\text{walk}\},
\end{aligned}
\end{equation}
where IVT and OVT are in-vehicle and out-of-vehicle travel time in minutes and cost is travel cost in \pounds, $z_{n,w}$, $w\in\{\text{fem},\text{age},\text{inc}\}$, are the three binary covariates of person $n$, and $\kappa_{m,w}$ are their effects on the constant of mode $m$, which are set to zero in the specifications without covariates in the utilities. In the COVID-19 case study, for the two vaccines $v\in\{A,B\}$, with $\text{pos}_{A}=1$ and $\text{pos}_{B}=-1$,
\begin{equation}\label{eq:util_cov}
\begin{aligned}
V_{\text{free},v}&=\delta_{\text{free}}+\delta_{\text{pos}}\,\text{pos}_{v}+\sum_{w}\kappa_{\text{free},w}z_{n,w}+W_{v}+\beta_{\text{wait}}\,\text{wait}_{v},\\
V_{\text{paid},v}&=\delta_{\text{paid}}+\delta_{\text{pos}}\,\text{pos}_{v}+\sum_{w}\kappa_{\text{paid},w}z_{n,w}+W_{v}+\beta_{\text{fee}}\,\text{fee}_{v},\\
W_{v}&=\beta_{\text{inf}}\,\text{inf}_{v}+\beta_{\text{ill}}\,\text{ill}_{v}+\beta_{\text{unk}}\,\mathrm{I}[\text{prot}_{v}=0]+\beta_{\text{prot}}\,\text{prot}_{v}\\
V_{\text{none}}&=\beta_{\text{inf}}\,\text{inf}_{0}+\beta_{\text{ill}}\,\text{ill}_{0},\\
&\quad+\beta_{\text{mild}}\,\text{mild}_{v}+\beta_{\text{sev}}\,\text{sev}_{v}+\beta_{\text{cov}}\,\text{cov}+\beta_{\text{ex}}\,\text{ex},
\end{aligned}
\end{equation}
where inf and ill are the remaining risks of infection and of serious illness, in percent (with subscript $0$ for the no vaccination option), prot is the duration of protection in months, with $\text{prot}_v=0$ denoting an unknown duration and $\mathrm{I}[\cdot]$ the indicator function, mild and sev are the risks of mild and severe side effects, in percent, cov is the existing population coverage in percent, ex indicates exemption from travel restrictions, and wait and fee are the waiting time in months and the fee in \pounds. The four vaccine alternatives share a nest with nesting parameter $\theta$.

\subsubsection{Random heterogeneity}\label{sec:randhet}

For the DECISIONS case study, the continuous mixture models use normal distributions for the constants, $\delta_m\sim N(\mu_{\delta,m}+\sum_{w}\kappa_{m,w}z_{n,w},\,\sigma_{\delta,m})$, where the second argument is the standard deviation, and negative log-uniform distributions for the coefficients of time and cost. A negative log-uniform distribution is the negative exponential of a uniform distribution, $\beta_k=-e^{a_k+b_k\xi_k}$ with $\xi_k\sim U\left[0,1\right]$ \citep[cf.][]{Hess_et_al_VTT2017}, so that the logarithm of $|\beta_k|$ is uniform between $a_k$ and $a_k+b_k$, with $b_k$ of either sign. We use mode and attribute specific location parameters but a spread parameter, and a draw $\xi_k$, that are generic across modes within an attribute, so that $\beta_{\text{IVT},m}=-e^{a_{\text{IVT},m}+b_{\text{IVT}}\xi_{\text{IVT}}}$, $\beta_{\text{OVT},m}=-e^{a_{\text{OVT},m}+b_{\text{OVT}}\xi_{\text{OVT}}}$ and $\beta_{\text{cost}}=-e^{a_{\text{cost}}+b_{\text{cost}}\xi_{\text{cost}}}$. A person who is more sensitive to in-vehicle time with one mode is therefore equally so, in relative terms, with every mode. The latent class models have three classes, with class-specific values $\delta_m^{(q)}$ and $\beta^{(q)}$ of all constants and coefficients, and the covariate effects $\kappa_{m,w}$, where included in the utilities, generic across classes. The class allocation probabilities are $\pi_{nq}=\exp(\zeta_q+\sum_{w}\omega_{q,w}z_{n,w})/\sum_{q'}\exp(\zeta_{q'}+\sum_{w}\omega_{q',w}z_{n,w})$, with $\zeta_1=0$ and $\omega_{1,w}=0$, and with $\omega_{q,w}=0$ for all classes except in the specification with covariates in the class allocation model.

For the COVID-19 case study, the continuous mixture models use normal distributions for the constants of the free and paid vaccines and for the coefficients of population coverage and exemption, and log-uniform distributions for the other coefficients, negative ones for the risks of infection and of illness, the side effects, waiting time and fee, and a positive one for the duration of protection. The coefficient for unknown duration is non-random. The latent class models again have three classes, with all parameters, including the nesting parameter $\theta^{(q)}$, class-specific except for the position constant $\delta_{\text{pos}}$ and, where included in the utilities, the covariate effects $\kappa$, which are generic across classes as for the DECISIONS data. The coefficient for exemption is fixed to zero in the first two classes and that for unknown duration in the third, as their estimates tended to zero. With these restrictions the fixed-coefficient model is not strictly a special case of the COVID-19 latent class models, so $H^*$ and the fit gain are not guaranteed to be non-negative for them, although both are positive on every fold. The class allocation model is the same as for the DECISIONS data.

The logit mixed logit models let the same coefficients vary as the continuous mixtures, but replace the parametric distributions by a discrete distribution over a large fixed set of points mapped to estimated bounds, with the probability of each point given by a logit whose utility is a polynomial in the coefficient values\footnote{Our logit mixed logit follows \citet{Train2016} in representing the mixing distribution as a logit over a discrete support, with Legendre polynomials of degree six in each random dimension, the order used in his application. It differs in three respects. The support is a single set of $5{,}000$ MLHS points shared by all respondents, rather than a random subset of a fine grid drawn for each respondent, so the mixing distribution is exactly discrete on that support rather than simulated. The bounds of the support are estimated rather than fixed by the analyst. And there are no cross-product terms, so that, like the mixed logit it is compared with, the specification allows no correlation between coefficients.}. This adds $48$ parameters in the DECISIONS models and $66$ in the COVID-19 models relative to the corresponding continuous mixture. The constants and the normally distributed coefficients of the continuous mixtures are distributed over supports $[l,u]$, for example $\delta_m\in[l_{\delta,m},u_{\delta,m}]$. For the coefficients that enter through an exponential, $\beta_k=\pm\exp(a_k+\eta_d)$ with $\eta_d$ distributed over a support $[l_d,u_d]$ shared by the coefficients of the same attribute $d$ (for example $\beta_{\text{IVT},m}=-\exp(a_{\text{IVT},m}+\eta_{\text{IVT}})$ for all modes $m$), only $a_k+l_d$ and $u_d-l_d$ are identified, so we normalise the lower bound of each such support, and the parameter counts reported here are those of the identified model.

\subsection{Estimation}\label{sec:estimation}

The core models were coded and estimated using \emph{Apollo} v0.3.9 \citep{hess_palma_apollo}. The continuous mixtures use $5{,}000$ MLHS draws \citep{266} per individual in estimation and application, and the cross-sectional likelihood is simulated with $5{,}000$ MLHS draws per choice, the same draws being used at the panel and at the cross-sectional estimates. The logit mixed logit models were estimated with purpose-written code in R, which uses a single set of $5{,}000$ MLHS support points shared by all individuals, as in \citet{Train2016}, and analytic gradients, making estimation on every fold and both likelihoods feasible. The same support is used for both likelihoods. Estimation proceeded first by maximising over the blocks of parameters in turn and then by a joint quasi-Newton optimisation over all parameters from that solution, stopped when the largest absolute gradient fell below $0.01$. This criterion was met for every panel model on every fold except the logit mixed logit without covariates on fold 5 of the COVID-19 data, which was still improving slowly when the evaluation budget was reached (largest absolute gradient $4.9$). A further gain of a few log-likelihood points on that fold would change the five-fold means for this model by about one point. 

Cross-sectional estimation of a mixture is numerically harder than panel estimation, because heterogeneity is only weakly identified from single choices (Section \ref{sec:overdispersion}). Where the default optimiser failed, we used BFGS instead, and for the logit mixed logit, whose cross-sectional likelihood has several local maxima, we use the best of several runs from different starting values. Any shortfall in a cross-sectional maximum would lower $H^*$ and $E$ by the same amount and leave realised heterogeneity, which depends only on the panel estimates, unchanged, so $H^*$ and $E$ are best read as lower bounds. Convergence to an inferior local optimum, which cannot be excluded for the latent class and logit mixed logit models, would instead lower the fit gain and change consistency and realised heterogeneity.

\section{Results}\label{sec:results}

This section focuses on prediction performance and on testing the ideas of Section \ref{sec:theory}, rather than on the estimates themselves. The estimates of all panel models on the first fold, with robust $t$-ratios, and the implied distributions of the coefficients in the logit mixed logit models are given in the supplementary material.

\subsection{Model fit}

Table \ref{tab:model_fit} reports the three log-likelihoods used in the decomposition of Section \ref{sec:decomposition}, $LL^{P}(\widehat\Omega_P)$, on which the models were estimated, $LL^{CS}(\widehat\Omega_P)$, which scores their unconditional forecasts of the estimation choices, and $LL^{CS}(\widehat\Omega_{CS})$. For the models without random heterogeneity the three coincide, by \eqref{eq:nomix}. The table also gives the number of estimated parameters and the Bayesian Information Criterion (BIC) for the panel and for the cross-sectional estimates, computed from the log-likelihood that was maximised and the number of estimation rows. Every quantity is computed on each fold separately and then averaged over the five folds, with the standard deviation across folds in brackets. These standard deviations overstate the uncertainty in comparisons between models, since the folds differ in which individuals they contain, which moves the log-likelihood of every model in the same direction.

\begin{table}[t!]
\small\centering
\caption{Model fit on the estimation rows, mean (standard deviation) over five folds, best value in each column in bold}
\label{tab:model_fit}
\resizebox{0.8\textwidth}{!}{\begin{tabular}{lrrrrrr}
\toprule
\multicolumn{7}{c}{\textbf{DECISIONS}}\\
 & & \multicolumn{3}{c}{panel estimates} & \multicolumn{2}{c}{cross-sectional estimates}\\ \cmidrule(lr){3-5}\cmidrule(lr){6-7}
 & parameters & $LL^{P}(\widehat\Omega_P)$ & BIC & $LL^{CS}(\widehat\Omega_P)$ & $LL^{CS}(\widehat\Omega_{CS})$ & BIC \\
\midrule
MNL & 14 & $-3{,}236.2$ (100.0) & 6,597.6 & $-3{,}236.2$ (100.0) & $-3{,}236.2$ (100.0) & 6,597.6 \\
MNL with socios & 29 & $-3{,}147.2$ (92.2) & 6,553.5 & $\mathbf{-3{,}147.2}$ (92.2) & $-3{,}147.2$ (92.2) & 6,553.5 \\
MMNL & 22 & $-2{,}250.6$ (61.7) & \textbf{4,697.8} & $-3{,}364.5$ (94.5) & $-3{,}121.2$ (95.0) & 6,439.1 \\
MMNL with socios & 37 & $-2{,}232.8$ (62.8) & 4,796.3 & $-3{,}289.9$ (99.0) & $-3{,}034.0$ (88.5) & 6,398.6 \\
LC & 44 & $-2{,}551.8$ (80.7) & 5,496.9 & $-3{,}268.4$ (122.1) & $-2{,}979.0$ (103.4) & 6,351.3 \\
LC, socios in utility & 59 & $-2{,}514.3$ (77.2) & 5,556.0 & $-3{,}227.0$ (121.6) & $\mathbf{-2{,}893.6}$ (101.7) & \textbf{6,314.6} \\
LC, socios in class allocation & 50 & $-2{,}546.1$ (79.6) & 5,539.1 & $-3{,}255.5$ (128.3) & $-2{,}936.5$ (88.0) & 6,320.0 \\
LML & 70 & $-2{,}179.1$ (57.6) & 4,984.0 & $-3{,}249.9$ (87.1) & $-3{,}006.5$ (94.6) & 6,638.7 \\
LML with socios & 85 & $\mathbf{-2{,}146.4}$ (61.6) & 5,052.5 & $-3{,}187.6$ (76.4) & $-2{,}918.5$ (88.1) & 6,596.8 \\
\midrule
\multicolumn{7}{c}{\textbf{COVID-19}}\\
 & & \multicolumn{3}{c}{panel estimates} & \multicolumn{2}{c}{cross-sectional estimates}\\ \cmidrule(lr){3-5}\cmidrule(lr){6-7}
 & parameters & $LL^{P}(\widehat\Omega_P)$ & BIC & $LL^{CS}(\widehat\Omega_P)$ & $LL^{CS}(\widehat\Omega_{CS})$ & BIC \\
\midrule
NL & 14 & $-11{,}059.1$ (38.9) & 22,245.0 & $-11{,}059.1$ (38.9) & $-11{,}059.1$ (38.9) & 22,245.0 \\
NL with socios & 20 & $-10{,}994.5$ (45.2) & 22,170.2 & $-10{,}994.5$ (45.2) & $-10{,}994.5$ (45.2) & 22,170.2 \\
Mixed NL & 25 & $-9{,}584.1$ (22.6) & \textbf{19,394.6} & $-10{,}959.3$ (47.1) & $-10{,}892.1$ (48.2) & 22,010.7 \\
Mixed NL with socios & 31 & $-9{,}565.4$ (24.9) & 19,411.5 & $-10{,}909.0$ (53.3) & $-10{,}832.9$ (47.0) & \textbf{21,946.5} \\
LC NL & 39 & $-9{,}779.6$ (20.6) & 19,912.5 & $-11{,}004.1$ (40.0) & $-10{,}835.5$ (33.8) & 22,024.3 \\
LC NL, socios in utility & 45 & $-9{,}764.6$ (19.7) & 19,936.8 & $-10{,}965.1$ (46.3) & $-10{,}771.8$ (49.4) & 21,951.3 \\
LC NL, socios in class allocation & 45 & $-9{,}755.0$ (23.4) & 19,917.5 & $-10{,}936.0$ (46.2) & $-10{,}813.8$ (51.1) & 22,035.2 \\
LML NL & 91 & $-9{,}419.6$ (28.6) & 19,663.5 & $-10{,}920.1$ (43.3) & $-10{,}776.3$ (37.9) & 22,376.8 \\
LML NL with socios & 97 & $\mathbf{-9{,}397.8}$ (26.6) & 19,674.2 & $\mathbf{-10{,}871.6}$ (50.9) & $\mathbf{-10{,}724.6}$ (43.1) & 22,327.8 \\
\bottomrule
\end{tabular}}
\end{table}

In both case studies every mixture model fits far better on the panel likelihood than the models without random heterogeneity, and the ordering within the mixtures is the same. The logit mixed logit fits best, ahead of the continuous mixture, ahead of latent class. The gain of the continuous mixture over the fixed-coefficient model that uses the same covariates is $986$ log-likelihood points on the DECISIONS data and $1{,}475$ on the COVID-19 data. The logit mixed logit improves on the continuous mixture by a further $71$ points on the DECISIONS data and $165$ on the COVID-19 data, positive on every fold in both, and by $86$ and $168$ points when covariates are included, so the flexibility of the mixing distribution is worth no less once observed heterogeneity is also in the model. BIC nevertheless gives preference to the continuous mixture in both datasets, by about $286$ and $269$ points, because the LML gain costs $48$ and $66$ additional parameters. 

The other columns tell a different story. On the cross-sectional likelihood ($LL^{CS}(\widehat\Omega_{CS})$) every mixture fits worse than on the panel likelihood, since the cross-sectional likelihood does not reward the dependence between a person's choices, and the preferred model changes. When estimated on that likelihood, the latent class model with socios in utility fits best on the DECISIONS data, and is also preferred there on BIC, while on the COVID-19 data the logit mixed logit with socios fits best and BIC prefers Mixed NL with socios. The BIC of the cross-sectional estimates is given for comparison only, since the usual penalty does not strictly apply to a likelihood that treats a person's choices as independent \citep{GaoSong2010}. 

For every mixture, the panel estimates score worse on the cross-sectional likelihood than the cross-sectional estimates of the same model, $LL^{CS}(\widehat\Omega_P)<LL^{CS}(\widehat\Omega_{CS})$, as they must since the latter maximise it. At the panel estimates no mixture scores better on single choices on the DECISIONS data than MNL with socios, and on the COVID-19 data the mixtures score better than the fixed-coefficient model with the same covariates by at most $139$ points, against up to $1{,}640$ on the panel likelihood. In short, judged on the panel likelihood the mixtures are far better than the models without random heterogeneity, and the more flexible the mixing distribution, the better they fit. Judged one choice at a time, at the same estimates, this advantage shrinks to less than a tenth on the COVID-19 data and disappears altogether on the DECISIONS data. Panel fit alone therefore does not show how much of a mixture's advantage will carry over to forecasting the choices of people whose earlier choices are unknown. Section \ref{sec:res_decomp} separates the two sources of fit that produce this difference.

\subsection{Decomposing the fit gain}\label{sec:res_decomp}

Table \ref{tab:decomp} evaluates the decomposition \eqref{eq:decomp}--\eqref{eq:ladder} for all fourteen panel/cross-sectional pairs. Each panel mixture is compared with the fixed-coefficient model that uses the same covariates (MNL or NL for the specifications without covariates, MNL or NL with socios otherwise), so that the fit gain reflects mixing rather than the covariates.

\begin{table}[t!]
\centering
\caption{Decomposition of the fit gain of each panel mixture over its fixed-coefficient benchmark, mean (standard deviation) over five folds}
\label{tab:decomp}
\resizebox{0.8\textwidth}{!}{%
\begin{tabular}{llrrrrrr}
\toprule
 & & & \multicolumn{3}{c}{heterogeneity} & & \\
\cmidrule(lr){4-6}
Dataset & Specification & fit gain & $H^{*}$ & $-E$ & realised & $C$ & usable share \\
\midrule
DECISIONS & MMNL & 985.6 (73.0) & 115.0 (17.5) & $-$243.3 (10.8) & $-128.3$ (19.8) & 1{,}113.9 (66.0) & $-$13\% (2.7) \\
DECISIONS & MMNL with socios & 914.4 (56.9) & 113.2 (12.9) & $-$256.0 (19.9) & $-142.8$ (23.9) & 1{,}057.1 (58.3) & $-$16\% (3.1) \\
DECISIONS & LC & 684.4 (36.7) & 257.2 (23.7) & $-$289.5 (20.8) & $-32.2$ (27.9) & 716.7 (52.9) & $-$5\% (4.0) \\
DECISIONS & LC, socios in utility & 632.9 (36.3) & 253.5 (22.6) & $-$333.4 (21.1) & $-79.9$ (36.7) & 712.8 (61.1) & $-$13\% (5.6) \\
DECISIONS & LC, socios in class allocation & 601.1 (27.4) & 210.6 (15.0) & $-$318.9 (44.7) & $-108.3$ (41.9) & 709.4 (56.1) & $-$18\% (6.8) \\
DECISIONS & LML & 1{,}057.1 (74.6) & 229.7 (34.1) & $-$243.4 (13.3) & $-13.7$ (39.9) & 1{,}070.8 (68.8) & $-$1\% (3.8) \\
DECISIONS & LML with socios & 1{,}000.8 (67.4) & 228.7 (24.8) & $-$269.1 (14.8) & $-40.5$ (30.8) & 1{,}041.3 (59.4) & $-$4\% (3.1) \\
\midrule
COVID-19 & Mixed NL & 1{,}475.0 (28.4) & 167.0 (12.0) & $-$67.2 (8.3) & $99.7$ (8.5) & 1{,}375.2 (36.1) & 7\% (0.7) \\
COVID-19 & Mixed NL with socios & 1{,}429.2 (29.8) & 161.7 (13.9) & $-$76.1 (14.8) & $85.5$ (9.0) & 1{,}343.6 (38.7) & 6\% (0.7) \\
COVID-19 & LC NL & 1{,}279.5 (43.9) & 223.6 (15.1) & $-$168.6 (18.0) & $55.0$ (8.4) & 1{,}224.5 (47.1) & 4\% (0.7) \\
COVID-19 & LC NL, socios in utility & 1{,}229.9 (44.3) & 222.9 (10.3) & $-$193.4 (18.3) & $29.4$ (9.6) & 1{,}200.5 (49.1) & 2\% (0.8) \\
COVID-19 & LC NL, socios in class allocation & 1{,}239.6 (46.0) & 180.9 (21.6) & $-$122.3 (28.8) & $58.6$ (9.0) & 1{,}181.0 (48.6) & 5\% (0.8) \\
COVID-19 & LML NL & 1{,}639.5 (27.3) & 282.8 (15.6) & $-$143.9 (17.1) & $139.0$ (5.0) & 1{,}500.5 (29.9) & 8\% (0.4) \\
COVID-19 & LML NL with socios & 1{,}596.8 (29.8) & 270.0 (12.1) & $-$147.0 (14.2) & $123.0$ (7.6) & 1{,}473.8 (35.3) & 8\% (0.6) \\
\bottomrule
\end{tabular}}
\end{table}

Three findings are common to every pair in both datasets. First, consistency accounts for more than $90\%$ of the fit gain in all fourteen pairs, and for more than $100\%$ in the seven DECISIONS pairs, and the share of the fit gain that unconditional forecasts can use lies between $-18\%$ and $8\%$. What a mixture model gains in fit on panel data over a model without mixing thus comes mainly from its ability to reproduce the dependence between a person's choices. On the DECISIONS data this part is even larger than the whole gain. Taken one choice at a time, the panel mixtures there fit worse than the model without mixing, and they fit the panel data better only because they capture how alike each person's choices are. Second, the ceiling $H^*$ is positive in every pair, including the COVID-19 latent class models, for which this is not guaranteed (Section \ref{sec:randhet}), so heterogeneity across people is never worthless for single choices. A mixture estimated to fit single choices always fits them better than the model without mixing, so the problem is not a lack of heterogeneity for unconditional forecasts to use. Third, panel estimation always gives some of this away, because the panel estimates were chosen to fit sequences rather than single choices. The panel estimates are therefore not the best version of the mixture for unconditional forecasts, and Section \ref{sec:res_prediction} shows that the cross-sectional estimates of the same model also forecast single choices better out of sample. Per estimation choice, which makes the two datasets comparable, $H^*$ is $0.015$ to $0.034$ log-likelihood points on the DECISIONS data and $0.019$ to $0.033$ on the COVID-19 data, and $E$ is $0.032$ to $0.044$ and $0.008$ to $0.023$.

The datasets differ in the sign of what remains. On the DECISIONS data realised heterogeneity is negative in all seven pairs, since $E$ exceeds $H^*$, and every panel mixture scores worse than the fixed-coefficient model on the single choices in its own estimation sample, by $14$ to $143$ log-likelihood points. On the COVID-19 data realised heterogeneity is positive in all seven pairs, by $29$ to $139$ points, which is $2\%$ to $8\%$ of the fit gain. In practical terms, for a person whose earlier choices are unknown, the panel mixtures forecast worse than a model without mixing on the DECISIONS data, and only slightly better on the COVID-19 data. As Section \ref{sec:overdispersion} anticipates, the opposite signs in the two datasets are consistent with the difference in the length of the panel, which Section \ref{sec:res_panel_length} supports. The DECISIONS individuals make a median of $23$ choices in the estimation rows, the COVID-19 respondents five, and $E$ per estimation choice is correspondingly larger, at $0.037$ log-likelihood points on average across the seven DECISIONS pairs against $0.015$ across the seven COVID-19 pairs. The longer each person's sequence of choices, the more the panel estimates are bent towards reproducing it, at the expense of predicting single choices.

Measured per choice, excluding each person's first, consistency is similar in the two datasets, at $0.152$ and $0.200$ log-likelihood points for the continuous mixtures. It cannot exceed $\log J$, the uncertainty of a choice among $J$ equally likely alternatives (Section \ref{sec:consistency}), with $J=6$ modes on the DECISIONS data and $J=5$ alternatives on the COVID-19 data, and these values are only about a tenth of that maximum ($8.5\%$ and $12.4\%$). Knowing a person's earlier choices thus removes only a modest part of the uncertainty about their next one, but summed over many choices this adds up to most of the fit gain, and the two datasets differ mainly in how many choices each person makes.

Within each dataset, the shape of the mixing distribution matters for the two heterogeneity terms but not for the conclusion. The nonparametric mixing distribution raises the ceiling substantially, from $115$ to $230$ points on the DECISIONS data and from $167$ to $283$ on the COVID-19 data. On the DECISIONS data it leaves $E$ essentially unchanged ($243$ points for both), so the whole of the ceiling's increase carries through, and realised heterogeneity narrows from $-128$ to $-14$. The LML value is within one fold-to-fold standard deviation of zero and positive on two of the five folds, so its sign is not settled, although its usable share is close to zero, at $-1\%$. On the COVID-19 data the LML raises $E$, from $67$ to $144$ points, but raises the ceiling by more, so that realised heterogeneity still increases, from $100$ to $139$. In both datasets realised heterogeneity is higher for the LML than for the continuous mixture on every fold.

On the DECISIONS data the latent class models, the coarsest representation, give the most away, with $E$ of $290$ to $333$ points against $243$ to $269$ for the continuous and nonparametric mixtures, so $E$ is not a constant but depends on both the panel length and how much freedom the mixing distribution has to serve the panel likelihood. Since $E$ is a lower bound (Section \ref{sec:estimation}), differences in $E$ of a few tens of points between families should be read with caution. In plain terms, a more flexible mixing distribution improves a mixture's score on single choices as well as its fit to sequences, but it does not remove the cost of panel estimation, which is largest for the coarsest representation, latent class. No choice of mixing distribution makes the panel estimates the best ones for unconditional forecasts.

The standard deviations across folds in Table \ref{tab:decomp} show which of these findings depend on the particular split. Consistency is stable across folds, with a standard deviation of $2\%$ to $9\%$ of its mean, and the usable share of the fit gain has a standard deviation across folds of at most $7$ percentage points. Realised heterogeneity is the least stable relative to its size, since it is a small difference between two larger terms. The findings above nonetheless hold fold by fold. $H^*$ and $E$ are positive in all $70$ combinations of pair and fold, and the usable share lies between $-26\%$ and $9\%$ in every one of them. Realised heterogeneity is negative on all five folds for six of the seven DECISIONS pairs and positive on all five folds for all seven COVID-19 pairs. In short, none of these conclusions depends on how the data were split, except whether the LML without covariates scores slightly better or slightly worse on single choices than the model without mixing. Because the folds share individuals in their estimation samples, these standard deviations describe the spread across splits rather than a standard error, and we do not test on them.

\subsection{Impact of panel length}\label{sec:res_panel_length}

As the two datasets differ in more than panel length, we also vary panel length within the DECISIONS data. For each person we keep only the last $k$ estimation trips before the holdout, for $k=2$, $5$, $10$ and $15$, so that the people, the holdout and the specification are unchanged and only the number of choices per person differs, and re-estimate MNL and the panel and cross-sectional versions of the mixed logit and latent class models on all five folds. The logit mixed logit is not included, since re-estimating it at each panel length would take several days per fold, and with $70$ parameters, against $22$ for the mixed logit, it would be even less well identified at short panel lengths.\footnote{The cross-sectional likelihood of two latent class models at $k=2$ cannot be evaluated at their panel estimates, so that point averages three folds. Restricting every $k$ to the same three folds leaves the pattern unchanged. At $k=15$, $13\%$ to $17\%$ of people have fewer than $15$ estimation trips and keep all of theirs. Consistency per choice in the figure counts each person's first choice too, so it is slightly lower than the per-choice values in Section \ref{sec:res_decomp}.} 

\begin{figure}[t!]\centering
\includegraphics[width=0.7\textwidth]{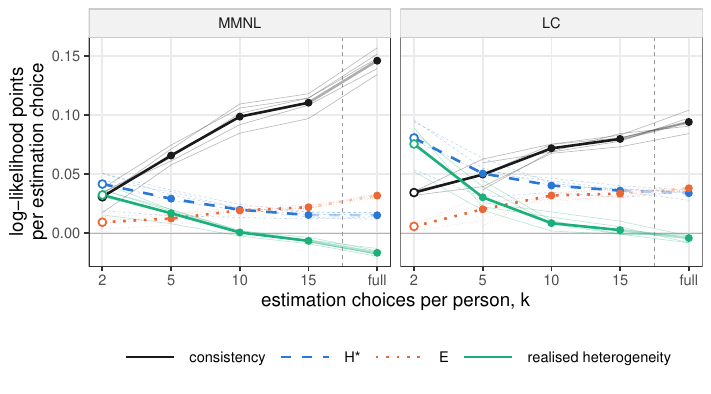}
\caption{Choice consistency and the other terms of the fit gain per estimation choice, against the number of estimation choices per person, DECISIONS data, fold means (thick) and individual folds (thin).}
\label{fig:panel_length}
\end{figure}

Figure \ref{fig:panel_length} shows the result. In the figure the last point is each person's whole panel, and $k=2$, the least reliable point, is marked with open symbols. Per estimation choice, consistency rises steeply with $k$, the ceiling $H^*$ falls, and $E$ rises, from $0.009$ at $k=2$ to $0.032$ on the full panel for the mixed logit and from $0.006$ to $0.038$ for latent class. Realised heterogeneity therefore falls with $k$ on every fold of both families and changes sign, between $k=10$ and $k=15$ for the mixed logit and between $k=15$ and the full panel for latent class. Out of sample the same reversal appears. The panel mixed logit assigns the chosen alternative a higher mean probability than MNL at $k=2$ (by $0.008$) and a lower one from $k=5$ onwards, by up to $0.039$ on the full panel.

A reader may wonder whether $E$ could grow with panel length even if the model were correct. Some growth is expected in any case, because the cross-sectional estimates are fitted to the same choices on which $E$ is measured, and this in-sample advantage increases with the correlation between a person's choices. To measure it, we generated data from a model that is correct by construction. Taking the fold-1 panel estimates of the mixed logit as the true model, once with ten choices per person and once with the full panel, we simulated twenty datasets from each with the observed attributes and re-estimated the panel and cross-sectional mixtures on every one. The full-panel simulation is used again in Section \ref{sec:res_dispersion}.

In the simulated data, $E$ per choice is $0.0035$ with ten choices per person and $0.0046$ with the full panel. In the real data it is $0.020$ and $0.032$. A correct model thus produces a small $E$ that grows only slightly with panel length, and this accounts for about a tenth of the absolute growth seen in the real data ($0.0011$ \emph{vs} $0.012$). It is the absolute growth that matters, since realised heterogeneity is the difference $H^*-E$. The rest is the effect of misspecification described in Section \ref{sec:overdispersion}. The comparison also shows that with ten choices per person, where realised heterogeneity is still positive, $E$ is already almost six times its value under a correct model ($0.020$ \emph{vs} $0.0035$). A mixture that still scores better on single choices than the model without mixing can therefore already be misspecified.

\subsection{Differences in the estimated mixing distribution}\label{sec:res_dispersion}

Table \ref{tab:dispersion} compares the spread of the random coefficients under the two objectives, for the continuous mixture without covariates in each dataset, to show the direction of the distortion described in Section \ref{sec:overdispersion}. For the normally distributed coefficients the spread is the standard deviation. For the log-uniform coefficients, $\beta=\mp e^{a+b\xi}$ with $\xi$ uniform on $[0,1]$, it is the width $|b|$ of the interval over which $\log|\beta|$ is uniform.

\begin{table}[htbp]
\centering\small
\caption{Spread of the random coefficients under panel and cross-sectional estimation, mean of absolute values over five folds}
\label{tab:dispersion}
\begin{adjustbox}{max width=0.8\textwidth}
\begin{tabular}{llrr}
\toprule
Dataset & Coefficient & Panel & CS\\
\midrule
\multicolumn{4}{l}{\emph{Normally distributed coefficients, standard deviation}}\\
DECISIONS & constant, cycling & 4.509 & 2.582\\
DECISIONS & constant, bus & 2.675 & 0.024\\
DECISIONS & constant, taxi & 2.669 & 1.278\\
DECISIONS & constant, rail & 2.538 & 0.065\\
DECISIONS & constant, walking & 1.647 & 0.011\\
COVID-19 & coefficient, exemption from travel restrictions & 3.295 & 3.338\\
COVID-19 & constant, paid vaccines & 0.787 & 0.177\\
COVID-19 & coefficient, population coverage & 0.156 & 0.121\\
COVID-19 & constant, free vaccines & 0.025 & 0.066\\
\midrule
\multicolumn{4}{l}{\emph{Log-uniform coefficients, width $|b|$ of the interval of $\log|\beta|$}}\\
DECISIONS & cost & 4.884 & 6.537\\
DECISIONS & out-of-vehicle time & 2.665 & 1.738\\
DECISIONS & in-vehicle time & 0.957 & 0.334\\
COVID-19 & duration of protection & 7.216 & 11.704\\
COVID-19 & severe side effects & 6.252 & 6.888\\
COVID-19 & risk of illness & 5.455 & 5.712\\
COVID-19 & fee & 5.306 & 4.926\\
COVID-19 & waiting time & 5.074 & 4.880\\
COVID-19 & risk of infection & 4.605 & 6.792\\
COVID-19 & mild side effects & 3.656 & 4.002\\
\bottomrule
\end{tabular}
\end{adjustbox}
\end{table}

On the DECISIONS data every standard deviation of the constants of the modes is larger under panel estimation, on all five folds, and the differences are large. The cross-sectional estimator drives three of the five to almost zero and halves the other two, so that the mixing distribution that best fits single choices has nearly fixed constants, while the panel estimator needs large dispersion in the constants to reproduce the fact that people keep choosing the same mode. This is the first of the two cases of Section \ref{sec:overdispersion}, persistence in mode choice that stable tastes of the assumed form cannot reproduce. It may come from tours whose outbound and return trips share a mode (Section \ref{sec:data}), from the constraints of a person's schedule and location, from errors in the derived travel times and costs of the alternatives not chosen \citep{tsoleridis2022deriving}, or from the specification itself, such as uncorrelated constants or a shape of taste distribution that the assumed mixing cannot represent. 

On the COVID-19 data, with five choices per person in the estimation rows, the two estimators want nearly the same mixing distribution. Apart from the constant for the paid vaccines, whose standard deviation is over four times larger under panel estimation ($0.787$ against $0.177$), the standard deviations are close under the two estimators, or negligible under both (the constant for the free vaccines). This is consistent with the small $E$ and positive realised heterogeneity in Table \ref{tab:decomp}.

The log-uniform coefficients show that the panel estimator does not widen every distribution. On the DECISIONS data it widens those for in-vehicle and out-of-vehicle time, which are specific to each mode, by factors of almost three and one and a half, but narrows the one for cost, which is common to all modes, on every fold in both cases, consistent with the persistence in mode choice discussed above. On the COVID-19 data the cross-sectional estimator gives the wider distribution for five of the seven log-uniform coefficients, most clearly for the risk of infection and the duration of protection. Since the means and spreads are estimated jointly, individual spreads can move in either direction (Section \ref{sec:overdispersion}).

Whether the gap between the two estimators goes beyond sampling variation can be checked by simulation. Taking the fold-1 panel estimates of the DECISIONS mixed logit as the true model, we generated twenty datasets with the observed attributes and availability, and re-estimated MNL and the panel and cross-sectional mixtures on each.\footnote{Each simulated person has one draw of the coefficients, used for all of their choices. Seeds are fixed, and the specifications, draws and starting values are those of the estimated models.} In these simulations $E$ is $34$ on average, with a standard deviation of $9$ and a largest value of about $46$, against $242$ observed on the same fold, and realised heterogeneity is positive in $18$ of the $20$ replications, with a mean of $+10$ against $-112$ observed. Out of sample, the simulated panel mixture is level with MNL and with its own cross-sectional version, while on the real data it is $3.3$ percentage points below MNL and $4.0$ below the cross-sectional version on that fold (Section \ref{sec:res_prediction}), and no replication comes close to either gap. On this fold, what panel estimation gives away on single choices is therefore far beyond what a correctly specified mixture produces. The comparison of dispersion is less informative, since the cross-sectional estimates of the standard deviations are noisy even when the mixture is correct, and against that spread only the collapse to zero of the constants for bus, rail and walking is unusual.

\begin{figure}[t]\centering
\includegraphics[width=0.7\textwidth]{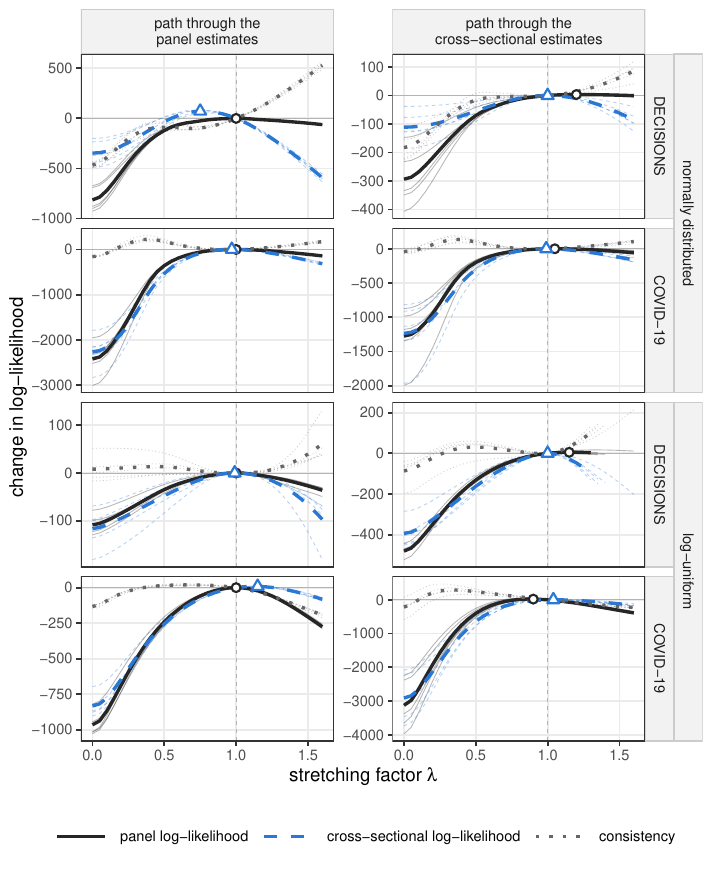}
\caption{Change relative to $\lambda=1$ in the two log-likelihoods and in choice consistency, stretching the normally distributed (top) or the log-uniform coefficients (bottom), fold means (thick) and individual folds (thin).}
\label{fig:dilation}
\end{figure}

Figure \ref{fig:dilation} evaluates both likelihoods along stretching paths of the kind described in Section \ref{sec:overdispersion}, with everything not stretched held at its estimate. In the upper half the standard deviations of all normally distributed coefficients are scaled by a common factor $\lambda$. In the lower half the log-uniform coefficients are stretched instead, by widening the interval of $\log|\beta|$ for each about its centre, so that the median of $\beta$ is unchanged.\footnote{The normally distributed coefficients are the constants of the five modes on the DECISIONS data, and the constants of the free and paid vaccines and the coefficients of population coverage and exemption from travel restrictions on the COVID-19 data. Since the coefficients not being stretched keep their estimated distributions, the mixture still generates much of the dependence between a person's choices at $\lambda=0$, for instance $58\%$ to $66\%$ of consistency at $\lambda=1$ on the DECISIONS data and $88\%$ to $96\%$ on the COVID-19 data on the normal path. Both likelihoods are evaluated with the same $5{,}000$ draws per person used in the panel estimation, and at $\lambda=1$ on the path through the panel estimates $D$ reproduces the consistency in Table \ref{tab:decomp}. The cross-sectional estimation used different draws for each choice of a person, so the cross-sectional likelihood computed here is slightly below its maximum at the cross-sectional estimates, and its slope there is not exactly zero. The vertical scales of the panels differ, so heights are not comparable across them.} In each half, one path passes through the panel estimates and the other through the cross-sectional estimates, the setting of Proposition \ref{prop:local}. Circles mark the maximum of the panel likelihood and triangles that of the cross-sectional likelihood. The dotted line is consistency along the path, $D(\lambda)$, the difference between the two. Since the panel likelihood is the cross-sectional likelihood plus consistency, the panel likelihood peaks to the right of the cross-sectional one where consistency rises with $\lambda$, and to the left where it falls, which is what Proposition \ref{prop:local} states at the cross-sectional maximum. At $\lambda=1$ each path returns the estimates it starts from, so on the path through the panel estimates the panel likelihood peaks there, as it must.

For the normally distributed coefficients, the cross-sectional likelihood on the path through the panel estimates peaks below $\lambda=1$ on every fold, at $0.75$ to $0.80$ on the DECISIONS data and at $0.97$ on the COVID-19 data. Single choices are thus fitted best with less dispersion than the panel estimates have, although by much less than Table \ref{tab:dispersion} suggests, since the path scales all standard deviations together and keeps the means at their panel values. On the path through the cross-sectional estimates the panel likelihood correspondingly favours more dispersion. On the DECISIONS data the slope of consistency at $\lambda=1$ is positive on every fold (mean $D'(1)=42$), and the panel likelihood peaks at $\lambda=1.22$ on average ($1.15$ to $1.30$ across folds). On the COVID-19 data the slope of consistency is also positive on every fold ($D'(1)=50$), but the cross-sectional likelihood falls by about as much on three folds, so the panel likelihood rises with more dispersion on only two folds and peaks at about $\lambda=1$ on the other three.

For the log-uniform coefficients the picture follows Table \ref{tab:dispersion} rather than the normal path. On the path through the panel estimates, the cross-sectional likelihood peaks close to $\lambda=1$ on the DECISIONS data ($0.92$ to $1.04$ across folds) and above it on the COVID-19 data, at $1.12$ on average, so that there single choices are fitted best with more spread in these coefficients than the panel estimates have. On the path through the cross-sectional estimates, the panel likelihood rises with more spread on the DECISIONS data, peaking at $\lambda=1.17$ on average, and falls on the COVID-19 data, peaking at $\lambda=0.88$, with $D'(1)$ positive on every DECISIONS fold and negative on every COVID-19 fold. Whether the panel likelihood favours more dispersion thus depends on which coefficients are stretched.

Three caveats apply to both halves. First, the slope $D'(1)$ is not by itself evidence of misspecification. In the simulation above, where the mixture is correct by construction, $D'(1)$ on the normal path at the cross-sectional estimates ranges from $-70$ to $366$ across replications, and the $48$ observed on the same fold lies well within that range. Second, at the DECISIONS cross-sectional estimates the standard deviations of the constants for bus, rail and walking are close to zero (Table \ref{tab:dispersion}), so the normal path there mainly stretches the constants for taxi and cycling. Third, $D(\lambda)$ is not monotone, rising, falling and rising again in $17$ of the $20$ curves for the normal coefficients and $18$ of $20$ for the log-uniform ones (two datasets, two paths and five folds each). More dispersion in the mixing distribution therefore does not always mean a stronger relation between a person's choices, and the slope of $D$ at one point on a path does not tell its sign elsewhere.

A better specification, for instance one allowing for habit or for tastes that vary from one choice to the next, could reduce $E$ and turn realised heterogeneity positive. Our point is not that the models could be improved, which is almost always true, but that a better fit on panel data is not evidence of better predictions of single choices. Even a correctly specified mixture cannot make consistency available to an unconditional forecast.

\subsection{Predicting single choices}\label{sec:res_prediction}

\begin{figure}[htbp]\centering
\includegraphics[width=0.7\textwidth,height=0.82\textheight,keepaspectratio]{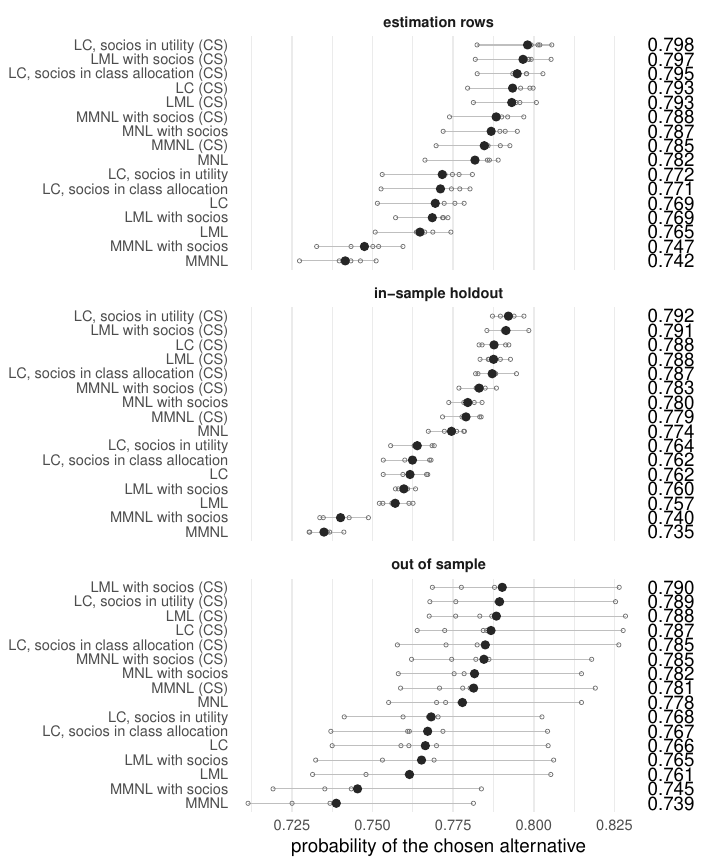}
\caption{DECISIONS: probability of the chosen alternative under unconditional forecasts, fold means (filled) and individual folds (open).}
\label{fig:prediction}
\end{figure}

\begin{figure}[htbp]\centering
\includegraphics[width=0.7\textwidth,height=0.82\textheight,keepaspectratio]{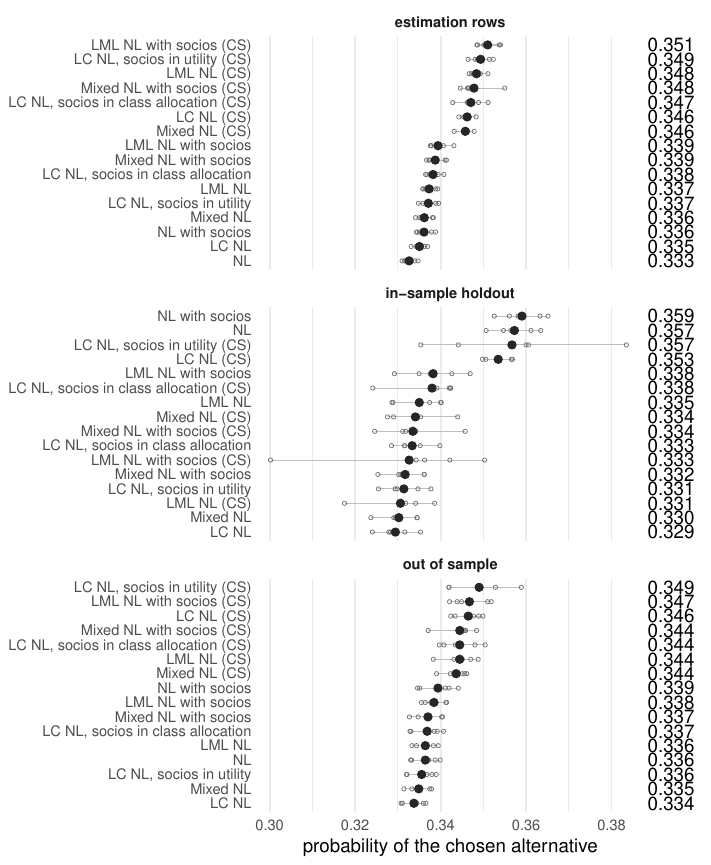}
\caption{COVID-19: probability of the chosen alternative under unconditional forecasts, same construction as Figure \ref{fig:prediction}.}
\label{fig:prediction_covid}
\end{figure}

Tables \ref{tab:pred_decisions} and \ref{tab:pred_covid} report the unconditional forecasts \eqref{eq:uncond} of single choices, on which mixture models are usually validated and used. For every specification they give the mean probability assigned to the chosen alternative and the mean log-probability (the log score) on the in-sample holdout and on the out-of-sample individuals, together with the root mean squared error (RMSE) between the predicted and the observed market shares, where the predicted share of an alternative is the average of its forecast probabilities over all the choices being forecast. Figures \ref{fig:prediction} and \ref{fig:prediction_covid} show the probabilities with the individual folds. We report the mean probability because it is easily interpreted and widely used, and the log score because it is a proper scoring rule \citep{Gneiting2007} and, per choice, is the out-of-sample counterpart of the cross-sectional log-likelihood. Rows marked (CS) give the same specification estimated on the cross-sectional likelihood. The RMSE of the market shares is computed over the six modes on the DECISIONS data, and over free vaccines, paid vaccines and no vaccine on the COVID-19 data.

\begin{table}[t!]
\centering
\caption{Prediction of single choices, DECISIONS data, probability of the chosen alternative (prob.), log score and share RMSE, means over five folds}
\label{tab:pred_decisions}
\resizebox{0.8\textwidth}{!}{\begin{tabular}{lrrrrrr}
\toprule
 & \multicolumn{3}{c}{In-sample holdout} & \multicolumn{3}{c}{Out of sample} \\
\cmidrule(lr){2-4}\cmidrule(lr){5-7}
Specification & prob. & log score & share RMSE & prob. & log score & share RMSE \\
\midrule
MNL & 0.7745 & $-$0.431 & 0.0065 & 0.7779 & $-$0.440 & 0.0137 \\
MNL with socios & 0.7795 & $-$0.421 & 0.0064 & 0.7816 & $-$0.589 & 0.0132 \\
\midrule
MMNL & 0.7349 & $-$0.451 & 0.0132 & 0.7388 & $-$0.450 & 0.0190 \\
MMNL (CS) & 0.7790 & $-$0.420 & 0.0069 & 0.7813 & $-$0.424 & 0.0139 \\
MMNL with socios & 0.7401 & $-$0.442 & 0.0130 & 0.7453 & $-$0.903 & 0.0178 \\
MMNL with socios (CS) & 0.7831 & $-$0.409 & 0.0068 & 0.7845 & $-$0.879 & 0.0136 \\
LC & 0.7616 & $-$0.438 & 0.0131 & 0.7664 & $-$0.443 & 0.0166 \\
LC (CS) & 0.7877 & $-$0.400 & 0.0072 & 0.7867 & $-$0.421 & 0.0134 \\
LC, socios in utility & 0.7639 & $-$0.432 & 0.0126 & 0.7682 & $-$0.480 & 0.0164 \\
LC, socios in utility (CS) & 0.7921 & $-$0.394 & 0.0067 & 0.7894 & $-$0.468 & 0.0131 \\
LC, socios in class allocation & 0.7624 & $-$0.436 & 0.0126 & 0.7671 & $-$0.447 & 0.0161 \\
LC, socios in class allocation (CS) & 0.7871 & $-$0.402 & 0.0072 & 0.7850 & $-$0.446 & 0.0128 \\
LML & 0.7570 & $-$0.439 & 0.0112 & 0.7615 & $-$0.437 & 0.0168 \\
LML (CS) & 0.7875 & $-$0.408 & 0.0071 & 0.7884 & $-$0.417 & 0.0135 \\
LML with socios & 0.7597 & $-$0.433 & 0.0114 & 0.7652 & $-$0.488 & 0.0161 \\
LML with socios (CS) & 0.7914 & $-$0.400 & 0.0075 & 0.7902 & $-$0.498 & 0.0133 \\
\bottomrule
\end{tabular}}
\end{table}

\begin{table}[t!]
\centering
\caption{Prediction of single choices, COVID-19 data, probability of the chosen alternative (prob.), log score and share RMSE, means over five folds}
\label{tab:pred_covid}
\resizebox{0.8\textwidth}{!}{\begin{tabular}{lrrrrrr}
\toprule
 & \multicolumn{3}{c}{In-sample holdout (fixed task)} & \multicolumn{3}{c}{Out of sample} \\
\cmidrule(lr){2-4}\cmidrule(lr){5-7}
Specification & prob. & log score & share RMSE & prob. & log score & share RMSE \\
\midrule
NL & 0.3573 & $-$1.355 & 0.0858 & 0.3364 & $-$1.301 & 0.0205 \\
NL with socios & 0.3590 & $-$1.343 & 0.0853 & 0.3394 & $-$1.294 & 0.0197 \\
\midrule
Mixed NL & 0.3302 & $-$1.315 & 0.0092 & 0.3349 & $-$1.284 & 0.0163 \\
Mixed NL (CS) & 0.3341 & $-$1.326 & 0.0269 & 0.3436 & $-$1.280 & 0.0192 \\
Mixed NL with socios & 0.3317 & $-$1.310 & 0.0088 & 0.3370 & $-$1.279 & 0.0154 \\
Mixed NL with socios (CS) & 0.3335 & $-$1.316 & 0.0241 & 0.3445 & $-$1.275 & 0.0173 \\
LC NL & 0.3294 & $-$1.318 & 0.0135 & 0.3338 & $-$1.290 & 0.0168 \\
LC NL (CS) & 0.3535 & $-$1.409 & 0.0189 & 0.3465 & $-$1.292 & 0.0159 \\
LC NL, socios in utility & 0.3314 & $-$1.312 & 0.0124 & 0.3356 & $-$1.286 & 0.0160 \\
LC NL, socios in utility (CS) & 0.3567 & $-$1.361 & 0.0572 & 0.3490 & $-$1.280 & 0.0167 \\
LC NL, socios in class allocation & 0.3334 & $-$1.305 & 0.0138 & 0.3368 & $-$1.282 & 0.0154 \\
LC NL, socios in class allocation (CS) & 0.3380 & $-$1.347 & 0.0472 & 0.3445 & $-$1.280 & 0.0167 \\
LML NL & 0.3350 & $-$1.315 & 0.0098 & 0.3364 & $-$1.282 & 0.0164 \\
LML NL (CS) & 0.3306 & $-$1.348 & 0.0422 & 0.3445 & $-$1.276 & 0.0153 \\
LML NL with socios & 0.3383 & $-$1.311 & 0.0113 & 0.3385 & $-$1.277 & 0.0159 \\
LML NL with socios (CS) & 0.3326 & $-$1.324 & 0.0371 & 0.3467 & $-$1.268 & 0.0178 \\
\bottomrule
\end{tabular}}
\end{table}

\FloatBarrier

Figures \ref{fig:prediction} and \ref{fig:prediction_covid} add the estimation rows and the spread across folds. On the DECISIONS data the picture is the same on all three sets of choices. Every cross-sectional mixture is above MNL and every panel mixture below it. Even on the data the models were estimated on, the advantage of the panel mixtures thus disappears once their choices are predicted one at a time. Out of sample the spread across folds is wide, because the folds differ in which individuals are held out. On the COVID-19 data the cross-sectional versions are again above the panel versions on the estimation rows and out of sample. On the fixed task the two models without mixing come first, and several cross-sectional mixtures vary widely across folds (see further discussion in Section \ref{sec:res_shares}).

\subsubsection{Panel \emph{vs} cross-sectional estimates}

The comparison that Section \ref{sec:decomposition} makes on the estimation rows carries over to data the models have not seen. On the out-of-sample individuals, the cross-sectional version of every specification assigns a higher mean probability to the chosen alternative than its own panel version, in all fourteen pairs, by $1.8$ to $4.3$ percentage points on the DECISIONS data and by $0.8$ to $1.3$ points on the COVID-19 data. On the log score for the same individuals the cross-sectional version is better in twelve of the fourteen pairs. The two exceptions, the DECISIONS LML with socios and the COVID-19 latent class model without covariates, are within about $0.01$. 

On the in-sample holdout the cross-sectional version is again better in every pair on both measures on the DECISIONS data, while on the fixed COVID-19 task, which lies at the edge of the design space, the comparison is mixed (Section \ref{sec:res_shares}). In the terms of Section \ref{sec:decomposition}, this gap is the out-of-sample counterpart of $E$, free of the in-sample optimism of the cross-sectional fit. For the mixed logit, latent class and LML without covariates on the DECISIONS data it is $0.026$, $0.022$ and $0.020$ per choice on the log score, positive on every fold and about three fifths to four fifths of $E$ per estimation choice. The same specification, with the same mixing distribution, forecasts single choices better when it is estimated on the object those forecasts are scored on. This is the finding of \citet{CherchiCirillo2010}, who compared panel and cross-sectional mixed logit on one RP dataset, reproduced across two datasets, three families of mixing distribution and five folds. It is not what a correctly specified model would produce, since the panel estimator is then the more precise of the two (Section \ref{sec:overdispersion}). The consistent advantage of the cross-sectional estimates is therefore evidence that the assumed structure of heterogeneity does not hold, clearly on the DECISIONS data and more mildly on the COVID-19 data, where it rests on the consistency of the sign rather than the size of the gaps.

\subsubsection{Mixtures \emph{vs} fixed-coefficient models}

On the DECISIONS data the ordering of the panel models on unconditional single-choice prediction is almost the reverse of their ordering on fit. On fit, the LML is best, then the mixed logit, then latent class, then MNL. In prediction, MNL is best, then latent class, then LML, then the mixed logit, on both holdouts on the mean probability and on the in-sample log score. The mixed logit, $986$ points better than MNL on the panel likelihood, predicts the chosen alternative with a mean probability of $0.739$ against $0.778$ out of sample, and its log score is worse, $-0.450$ against $-0.440$. Since the mixture nests MNL, a worse proper score out of sample, on four of the five folds and on all five for the in-sample holdout, is a second sign that the assumed structure of heterogeneity does not hold. The LML is marginally ahead of MNL on the log score out of sample ($-0.437$ against $-0.440$) and behind on the in-sample holdout ($-0.439$ against $-0.431$), as its realised heterogeneity close to zero in Table \ref{tab:decomp} suggests. On the COVID-19 data the picture is the one the positive realised heterogeneity implies. The panel mixtures are level with the fixed-coefficient models on the mean probability out of sample and slightly better on the log score, a difference of the order of $0.01$ per choice against a fit advantage of $1{,}230$ to $1{,}640$ points. In neither dataset does the fit advantage of the mixtures translate into more than a marginal advantage in predicting single choices, and in the dataset with long panels it becomes a disadvantage.

\subsubsection{Covariates and the two measures}

Including the socio-demographic covariates improves the mean probability of the chosen alternative for every panel specification on both holdouts in both datasets, by small amounts. The out-of-sample log scores of the DECISIONS models with covariates in the utilities are nevertheless much worse ($-0.589$ against $-0.440$ for MNL, $-0.903$ against $-0.450$ for the mixed logit), because on one fold a handful of held-out individuals with a covariate profile rare in the estimation sample receive probabilities below $10^{-6}$ for their chosen alternative. On that fold the mean log score falls by $0.75$ for MNL and $2.27$ for the mixed logit when covariates are added, while on the other four folds it changes by about $0.01$. The log score, while proper, is dominated out of sample by its tail, which is why we report both measures. None of the comparisons above depend on it, since they are made within a covariate structure.

\subsubsection{Market shares and the fixed COVID-19 task}\label{sec:res_shares}

On the DECISIONS out-of-sample individuals the cross-sectional models recover the mode shares better than their panel counterparts in all seven pairs (RMSE $0.013$ to $0.014$ against $0.016$ to $0.019$), and MNL is as good as any of them. On the COVID-19 out-of-sample individuals all models are within about $0.005$ of one another. On the fixed sixth task, however, the panel mixtures recover the shares far better than any other model, in all seven pairs, with an RMSE of $0.009$ to $0.014$ against $0.085$ to $0.086$ for NL and $0.019$ to $0.057$ for the cross-sectional mixtures. NL predicts a share of $0.83$ for the free vaccines against an observed $0.72$, and the panel Mixed NL $0.72$.

We report this as an observation on a single task, not as a general result. On that task the panel mixtures are better than NL on the log score ($-1.315$ against $-1.355$) but worse on the mean probability ($0.330$ against $0.357$), since NL concentrates probability on the alternative most respondents chose. The theory of Section \ref{sec:theory} says nothing about forecasts at the edge of the design space, and the forecast forward in time on the DECISIONS data shows no corresponding effect. The fixed task has no design variation with which to separate heterogeneity from calibration, and since it was always presented last, learning and fatigue cannot be separated from the change in design. Whether a mixture's account of heterogeneity helps aggregate forecasts under a shift in the design needs testing on designs built for it.

\subsection{Scoring sequences}\label{sec:res_sequences}

\begin{figure}[t!]\centering
\includegraphics[width=0.7\textwidth]{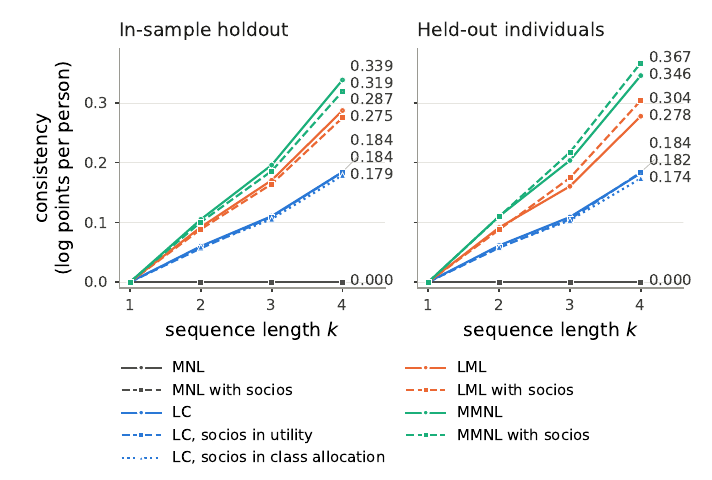}
\caption{Choice consistency on held-out choices against sequence length $k$, DECISIONS data, per person, means over individuals and folds.}
\label{fig:covgap}
\end{figure}

The DECISIONS data hold out the last four trips of every estimation individual, and all trips of the held-out individuals, so the difference between the panel and cross-sectional objects can be measured in prediction rather than in estimation, namely, for the first $k$ held-out trips of a person, the log of the probability of the sequence \eqref{eq:path} minus the log of the product of the single-choice forecasts \eqref{eq:uncond}. By Identity \ref{res:telescope} this is consistency evaluated on new choices, zero at $k=1$ by construction, and identically zero for a model without mixing, which serves as a check on the computation. Figure \ref{fig:covgap} reports it for both holdouts. This scores the paths people actually took rather than forecasting them (Section \ref{sec:targets}). The test is out of sample, but a panel mixture that does well on it is being rewarded for what it maximised, so it is not independent support for the panel model. Models are compared on this difference rather than on the probability of the sequence itself, which falls with $k$ simply because it is a product of more probabilities.

Consistency grows with the length of the sequence, as the sum in Identity \ref{res:telescope} says it must, and its size is almost the same for individuals in and outside the estimation sample, at $0.339$ and $0.346$ per person at $k=4$ for the mixed logit, since it depends mainly on the estimated mixing distribution. The ordering across families is the ordering of Table \ref{tab:decomp}. The mixed logit reproduces the most within-person dependence, then LML, then latent class, and MNL none. This is exactly the reverse of their ordering on unconditional single-choice prediction in Table \ref{tab:pred_decisions}, on the mean probability, both for the specifications without covariates and for those with them. The more dependence a specification reproduces, the worse it forecasts single choices. The comparison only works among specifications with the same covariates. Adding the covariates changes consistency at $k=4$ by only about $0.02$, but improves out-of-sample accuracy for every family, because the covariates describe differences between people that a single-choice forecast can use, whereas the dependence between a person's choices cannot be used without knowing their earlier choices.

\subsection{Conditional forecasts}\label{sec:res_conditioning}

\begin{figure}[t!]\centering
\includegraphics[width=0.7\textwidth,height=0.82\textheight,keepaspectratio]{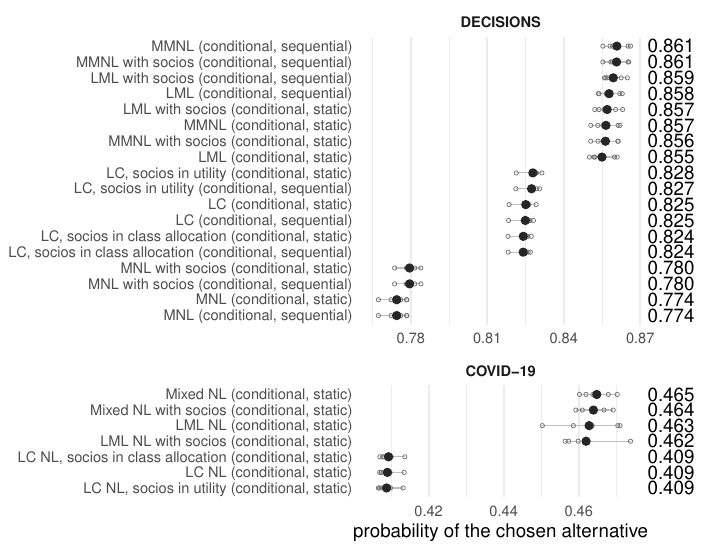}
\caption{Probability of the chosen alternative under conditional forecasts, in-sample holdout, fold means (filled) and individual folds (open).}
\label{fig:prediction_cond}
\end{figure}

\begin{figure}[t!]\centering
\includegraphics[width=0.7\textwidth]{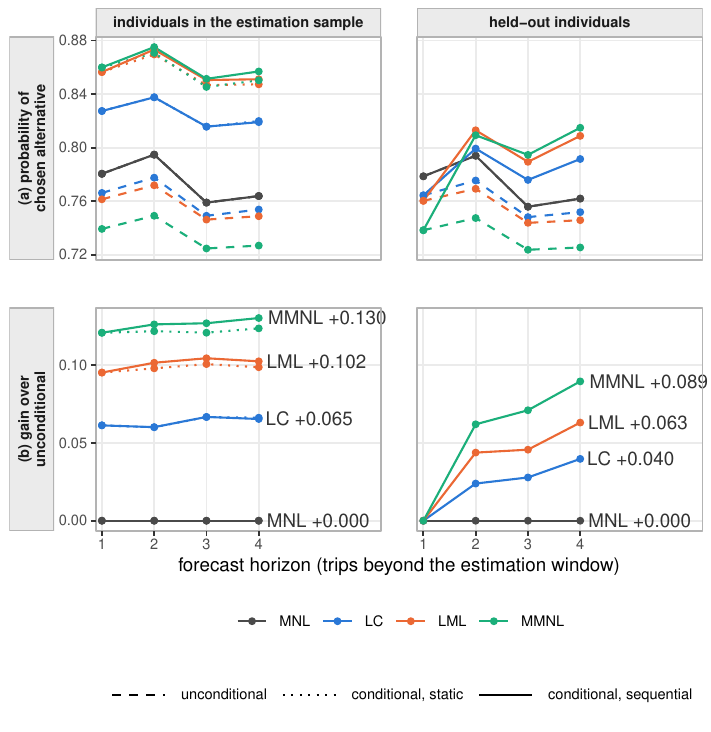}
\caption{Unconditional and conditional forecasts against forecast horizon, DECISIONS data, one specification per family without covariates, means over five folds.}
\label{fig:conditioning}
\end{figure}

Conditional forecasts are where the consistency captured by a panel mixture becomes usable, and where the three situations of Section \ref{sec:targets} separate. Table \ref{tab:horizon} reports, for the DECISIONS data, the mean probability of the chosen alternative on the held-out trips under three forecasts: unconditional, static (conditioning through \eqref{eq:cond} on the individual's estimation-sample trips) and sequential (conditioning also on the held-out trips already observed). For held-out individuals there is no estimation history, so only sequential updating can help, and the table reports their fourth held-out trip, forecast after three have been observed. Figure \ref{fig:prediction_cond} shows the conditional forecasts for every specification in both datasets. On the DECISIONS data the MNL forecasts equal their unconditional values, since a model without mixing has nothing to update. On the COVID-19 data, with a single held-out task per person, only the static forecast is available, and the models without mixing are not shown. Figure \ref{fig:conditioning} shows the same by forecast horizon, one to four trips beyond the estimation window, for one specification per family. Its top row gives the probability of the chosen alternative under the unconditional and the two conditional forecasts, and its bottom row the gain of each conditional forecast over the model's own unconditional one, for individuals in the estimation sample (left) and for held-out individuals (right). The unconditional forecasts show no systematic decay with the horizon. They rise and fall together from one trip to the next, and the ordering of the models (MNL, latent class, LML, mixed logit) is the same at every horizon, so forecasting forward in time within the same design region does not change which model predicts single choices best.

\begin{table}[htbp]
\centering
\caption{Conditional forecasts, DECISIONS data, probability of the chosen alternative, mean over five folds, families ordered by consistency}
\label{tab:horizon}
\resizebox{0.8\textwidth}{!}{\begin{tabular}{lrrrrrr}
\toprule
 & \multicolumn{3}{c}{In-sample holdout (mean of trips 1--4)} & \multicolumn{3}{c}{Out of sample (trip 4)}\\
\cmidrule(lr){2-4}\cmidrule(lr){5-7}
 & & \multicolumn{2}{c}{Conditional} & & Conditional & \\
\cmidrule(lr){3-4}\cmidrule(lr){6-6}
Specification & unconditional & static & sequential & unconditional & sequential & gain from conditioning \\
\midrule
MNL & 0.774 & 0.774 & 0.774 & 0.762 & 0.762 & +0.000 \\
MNL with socios & 0.780 & 0.780 & 0.780 & 0.766 & 0.766 & +0.000 \\
LC & 0.762 & 0.825 & 0.825 & 0.752 & 0.791 & +0.040 \\
LC, socios in utility & 0.764 & 0.828 & 0.827 & 0.754 & 0.794 & +0.040 \\
LC, socios in class allocation & 0.762 & 0.824 & 0.824 & 0.753 & 0.792 & +0.039 \\
LML & 0.757 & 0.855 & 0.858 & 0.746 & 0.809 & +0.063 \\
LML with socios & 0.760 & 0.857 & 0.859 & 0.749 & 0.815 & +0.066 \\
MMNL & 0.735 & 0.857 & 0.861 & 0.725 & 0.815 & +0.089 \\
MMNL with socios & 0.740 & 0.856 & 0.861 & 0.731 & 0.816 & +0.085 \\
\bottomrule
\end{tabular}}
\end{table}

\begin{table}[htbp]
\centering\small
\caption{Conditional forecasts of the fixed task, COVID-19 data, mean over five folds}
\label{tab:cond_covid}
\begin{adjustbox}{max width=0.8\textwidth}
\begin{tabular}{lrrrr}
\toprule
 & \multicolumn{2}{c}{Probability of chosen alternative} & \multicolumn{2}{c}{Log score}\\
\cmidrule(lr){2-3}\cmidrule(lr){4-5}
Specification & unconditional & conditional & unconditional & conditional\\
\midrule
NL & 0.357 & 0.357 & $-$1.355 & $-$1.355\\
NL with socios & 0.359 & 0.359 & $-$1.343 & $-$1.343\\
LC NL & 0.329 & 0.409 & $-$1.318 & $-$1.169\\
LC NL, socios in utility & 0.331 & 0.409 & $-$1.312 & $-$1.174\\
LC NL, socios in class allocation & 0.333 & 0.409 & $-$1.305 & $-$1.166\\
LML NL & 0.335 & 0.463 & $-$1.315 & $-$1.170\\
LML NL with socios & 0.338 & 0.462 & $-$1.311 & $-$1.164\\
Mixed NL & 0.330 & 0.465 & $-$1.315 & $-$1.080\\
Mixed NL with socios & 0.332 & 0.464 & $-$1.310 & $-$1.081\\
\bottomrule
\end{tabular}
\end{adjustbox}
\end{table}

For an individual in the estimation sample, conditioning on their history lifts the mean probability from $0.735$ to $0.857$ for the mixed logit (Figure \ref{fig:prediction_cond}), a gain of $12$ percentage points, against a spread of $4$ to $5$ percentage points between the best and worst unconditional forecasts. Sequential updating on the held-out trips then adds almost nothing ($+0.004$ and $+0.005$ for the two mixed logit specifications, less for the others), since with a median of $23$ estimation trips the posterior is already concentrated, so one to three further trips barely move it. This is Identity \ref{res:prequential} seen from the other side, since the later terms of the panel score are small once a person has made many choices. For a held-out individual only sequential updating is available, and on their fourth trip it rises from $0.725$ unconditionally to $0.815$. For the mixed logit the gain is $+0.062$ after one observed trip and $+0.089$ after three, so the returns diminish, and three trips already give two thirds to three quarters of the gain from a full history. Together, the two panels of Figure \ref{fig:conditioning} trace a single curve, rising steeply for a new person and levelling off once around twenty choices have been observed. A single observed choice from a new person is worth more than the whole spread between the unconditional forecasts of the different models.

The ranking of the models reverses under conditioning. Unconditionally the order is MNL, latent class, LML, mixed logit, while conditioned it is mixed logit and LML, then latent class, then MNL, and the gains for held-out individuals after three observed trips follow the same order. The specifications that reproduce the most within-person consistency lose the most when it cannot be used and gain the most when it can, and the ordering of Figure \ref{fig:covgap} appears a third time, now in conditional prediction. After conditioning, the residual dependence is nearly exhausted. The conditional probability of a sequence and the product of the conditional single-choice forecasts agree to three decimals at every $k$, so the consistency that Figure \ref{fig:covgap} measures is almost entirely accounted for by the posterior.

The COVID-19 data show the same in a short panel. Conditioning on each respondent's five estimation tasks raises the mean probability on the fixed sixth task from $0.33$ to $0.46$ for the continuous mixture and the LML, and to $0.41$ for latent class (Table \ref{tab:cond_covid}), a far larger change than any difference between mixing specifications or between panel and cross-sectional estimates. On the log score the Mixed NL improves from $-1.315$ to $-1.080$, while the LML gains less, to $-1.170$, despite the same gain in probability, so its conditional forecasts more often give the chosen alternative a very low probability. With five observed choices per person the COVID-19 respondents sit on the steep part of the curve, which is why conditioning is worth $+0.13$ there against $+0.09$ after three trips on the DECISIONS data. Two further observations hold in both datasets. Once a person's choices are conditioned on, the choice of mixing distribution matters little. The mixed logit and LML, $71$ and $165$ log-likelihood points apart in fit, end within $0.01$ of each other on the DECISIONS data and $0.002$ on the COVID-19 data, and the three latent class variants are indistinguishable from one another. And the socio-demographic covariates add essentially nothing to a conditioned forecast, at most $0.006$ in either dataset. The conditional forecasts of a latent class model are weaker than those of the continuous and nonparametric mixtures in both datasets, consistent with three classes carrying less recoverable information about an individual than a continuous distribution.

In short, an unconditional forecast can draw only on realised heterogeneity, the second term of \eqref{eq:decomp}, whereas a conditional forecast can also draw on consistency, the first and much larger term. This is why the specifications that forecast single choices worst when a person's earlier choices are unknown forecast best once they are known.

\section{Conclusions}\label{sec:conclusions}

This paper set out to explain why mixture models generally fit panel data far better than models without random heterogeneity, yet may forecast individual choices no better, and sometimes worse. The panel log-likelihood is exactly the cumulative score of forecasts conditioned on each person's earlier choices, whereas forecasts in practice are unconditional and scored on the cross-sectional likelihood. The difference, which we term \emph{choice consistency}, comes from reproducing the dependence between a person's choices, and unconditional forecasts cannot use it. The remaining gain, which we term \emph{realised heterogeneity}, need not be positive, because when the assumed structure of heterogeneity cannot reproduce that dependence exactly, the panel estimator distorts the heterogeneity across people in order to do so, by more the longer the panel in the revealed preference data.

We examined this on two datasets, revealed preference data on mode choice, with a median of $23$ trips per person in estimation, and stated choice data on COVID-19 vaccines, with five tasks per respondent in estimation. On each, seven mixture specifications with continuous, discrete and nonparametric mixing distributions were estimated on both the panel and the cross-sectional likelihood. In all fourteen pairs, consistency accounts for more than $90\%$ of the fit gain, and the part usable by unconditional forecasts lies between $-18\%$ and $8\%$. It is negative in the mean over folds on the revealed preference data, with long panels, and small but positive on the stated choice data, with short ones, and shortening the revealed preference panels turns it positive, for the two families we checked. In every pair, in both datasets, the same specification forecasts the single choices of new people better when estimated on the cross-sectional likelihood than when estimated on the panel likelihood, as measured by the mean probability of the chosen alternative. A correctly specified mixture would not do this in a large sample, since the panel estimator would then be the more precise of the two. On the revealed preference data, what panel estimation gives away on single choices is also far beyond what a simulation from the fitted mixture produces, on the fold we simulated. A more flexible mixing distribution raises what heterogeneity can contribute, but no mixing distribution can make consistency available to an unconditional forecast.

Where a person's own choices can be conditioned on, the picture reverses. For the continuous mixture, conditional forecasts gain twelve percentage points on the revealed preference data and thirteen on the stated choice data, the ranking of models turns around so that the specifications capturing the most within-person dependence forecast best, and the gain accumulates with every observed choice, quickly for a new person and slowly once many choices are known. This places the results of \citet{Krueger2021}, \citet{Danaf2019} and \citet{Xie2020} within the same account. The argument extends to other model classes. Hierarchical Bayes models used with each person's own posterior, or hybrid choice models with attitudinal indicators observed at the time of forecasting, produce conditional forecasts, which can use consistency. Applied to new individuals, for whom no such information is available, the same models produce unconditional forecasts, which cannot.

Three implications follow for practice. Model selection by panel fit can favour specifications that forecast single choices worse, and the two-term decomposition, which requires only that the estimated model be evaluated on the cross-sectional likelihood, shows how much of a fit gain is of the usable kind. Wherever a mixture is estimated with forecasting in mind, we therefore recommend reporting, alongside its fit, its score on single choices, which is the cross-sectional log-likelihood at the panel estimates, and, where feasible, also re-estimating it on the cross-sectional likelihood. The value of a mixture model for forecasting depends on whether the forecast can condition on an individual's choices. For individuals with no observed choices it lies entirely in the heterogeneity term, which may be small or negative, whereas for individuals with observed choices it is large and grows with their number. We also note that a large gap between the panel and cross-sectional estimates of a mixture, or a panel mixture that predicts single choices worse than a model without mixing on a proper score, is, in a large sample, evidence that the assumed structure of heterogeneity is misspecified. It also means that the dispersion estimated by a panel mixture should not be read directly as heterogeneity across people, since the panel estimator adjusts it to reproduce how alike a person's choices are, and can overstate it when habit or state dependence makes them more alike than stable tastes of the assumed form allow. This distortion is not a uniform widening. On the revealed preference data the panel estimator widens the distributions of the mode constants and the travel time coefficients but narrows that of cost, and on the stated choice data it narrows most of the attribute distributions. This matters wherever the estimated spread is itself a result of interest. Studies of the value of travel time, for example, often report the distribution of valuations from panel mixtures estimated on stated choice data with many choices per person. Its spread may then be distorted in either direction, and for skewed distributions such as the lognormal, whose mean rises with its spread, so may the mean valuation.

These results concern forecasts of single choices at attribute values like those in the estimation data. On the one task outside that range, the fixed COVID-19 task, the panel mixtures recovered market shares far better than both the models without mixing and their own cross-sectional versions (Section \ref{sec:res_shares}), although they did worse there on the mean probability of the chosen alternative. This single finding deserves testing on designs built to shift the attribute distribution, on revealed preference as well as stated choice data.

Several limitations remain. The two likelihood identities hold for any mixing distribution, but the account of why the estimates disagree is local, describing the direction in which the panel likelihood rises away from the cross-sectional estimates, and does not order the two estimates in general. Misspecification is detected rather than resolved, clearly on the DECISIONS data and more mildly on the COVID-19 data, and its source remains open in both. Habit is the obvious candidate on the DECISIONS data, but in a separate attempt, adding the previous choice to the utilities of the mixed logit, on one fold, removes only about a fifth of consistency and widens rather than closes the gap between the two estimators, with $E$ rising from $242$ to $440$ points. The role of panel length is shown within one dataset, for two families and one way of shortening the panels. Both the cross-sectional and the panel maxima may be local, the panel ones in particular for the latent class and logit mixed logit models, so $H^*$ and $E$ are best read as lower bounds and the fit gain and consistency as approximate. Finally, the logit mixed logit models use a single polynomial order and no cross-terms, so a more flexible mixing distribution could raise the heterogeneity terms further, though, by the identities, not make consistency usable.

Mixture models remain the standard tool for representing heterogeneity in preferences and for the behavioural insight that comes with it. For forecasting, what matters is not how well the model fits the sequence of choices it was estimated on but which of the two objects the forecast will be scored on, and whether the choices of the people being forecast will be available when the forecast is made.

\section*{Acknowledgements}

Stephane Hess acknowledges the support of the European Research Council through the advanced grant 101020940-SYNERGY. Sander van Cranenburgh acknowledges the support of the TU Delft AI Labs programme.

{\small
\setlength{\bibsep}{1pt plus 0.3ex}
\bibliographystyle{elsarticle-harv}
\bibliography{ref_library}

@incollection{Chapter26Forecastingchoice,
      author = {Andrew Daly},
      title = {Forecasting choice},
      chapter = 26,
      booktitle = {Handbook of Choice Modelling},
      year = {2024},
      publisher = "Edward Elgar Publishing",
      address = "Cheltenham, UK",
      pages=      "746--765",
      editor = {Stephane Hess and Andrew Daly}
}

@article{HESS2011973,
title = {Recovery of inter- and intra-personal heterogeneity using mixed logit models},
journal = {Transportation Research Part B: Methodological},
volume = {45},
number = {7},
pages = {973-990},
year = {2011},
issn = {0191-2615},
author = {Stephane Hess and Kenneth E. Train},
}

@article{calastri2020we,
  title={We want it all: experiences from a survey seeking to capture social network structures, lifetime events and short-term travel and activity planning},
  author={Calastri, Chiara and dit Sourd, Romain Crastes and Hess, Stephane},
  journal={Transportation},
  volume={47},
  number={1},
  pages={175--201},
  year={2020},
  publisher={Springer}
}

@article{tsoleridis2022deriving,
  title={Deriving transport appraisal values from emerging revealed preference data},
  author={Tsoleridis, Panagiotis and Choudhury, Charisma F and Hess, Stephane},
  journal={Transportation Research Part A: Policy and Practice},
  volume={165},
  pages={225--245},
  year={2022},
  publisher={Elsevier}
}

@article{wrro43607,
         journal = {Transportation Research Record},
            year = {2010},
          volume = {2175},
           pages = {74 -- 83 },
           month = {December},
           title = {Review of evidence for temporal transferability of mode-destination models},
       publisher = {National Academy of Sciences},
             doi = {10.3141/2175-09},
            issn = {0361-1981},
          author = {Fox, J and Hess, S}
          }

@article{FoxDalyHessMiller2014, title={Temporal transferability of models of mode-destination choice for the {Greater Toronto and Hamilton Area}}, volume={7},  number={2}, journal={Journal of Transport and Land Use}, author={Fox, James and Daly, Andrew and Hess, Stephane and Miller, Eric}, year={2014}, month={Jul.}, pages={41-62} }

@article{HESS2022114800,
title = {The path towards herd immunity: Predicting COVID-19 vaccination uptake through results from a stated choice study across six continents},
journal = {Social Science \& Medicine},
volume = {298},
pages = {114800},
year = {2022},
issn = {0277-9536},
author = {Stephane Hess and Emily Lancsar and Petr Mariel and Jürgen Meyerhoff and Fangqing Song and Eline {van den Broek-Altenburg} and Olufunke A. Alaba and Gloria Amaris and Julián Arellana and Leonardo J. Basso and Jamie Benson and Luis Bravo-Moncayo and Olivier Chanel and Syngjoo Choi and Romain {Crastes dit Sourd} and Helena Bettella Cybis and Zack Dorner and Paolo Falco and Luis Garzón-Pérez and Kathryn Glass and Luis A. Guzman and Zhiran Huang and Elisabeth Huynh and Bongseop Kim and Abisai Konstantinus and Iyaloo Konstantinus and Ana Margarita Larranaga and Alberto Longo and Becky P.Y. Loo and Malte Oehlmann and Vikki O'Neill and {Juan de Dios} Ortúzar and María José Sanz and Olga L. Sarmiento and Hazvinei Tamuka Moyo and Steven Tucker and Yacan Wang and Yu Wang and Edward J.D. Webb and Junyi Zhang and Mark H.P. Zuidgeest}
}

@InCollection{70,
  Title                    = {Modeling inter-urban route choice behaviour},
  Author                   = {M. Ben-Akiva and M. E. Bergman and M. J. Daly},
  Booktitle                = {Proceedings from the ninth International Symposium on Transportation and Traffic Theory},
  Publisher                = {VNU Science Press},
  Year                     = {1984},

  Address                  = {Utrecht, Netherlands},
  Editor                   = {Volmuler, J. and Hamerslag, R.},
  Pages                    = {299--330}
}

@Book{50,
  Title                    = {Multinomial probit with a logit kernel and a general parametric specification of the covariance structure},
  Author                   = {M. Ben-Akiva and D. Bolduc},
  Publisher                = {Department of Civil and Environmental Engineering, MIT},
  Year                     = {1996},

  Address                  = {Cambridge, MA},
  Series                   = {Working paper}
}

@Book{9,
  Title                    = {Discrete Choice Analysis: Theory and Application to Travel Demand},
  Author                   = {M. Ben-Akiva and S. R. Lerman},
  Publisher                = {MIT Press},
  Year                     = {1985},

  Address                  = {Cambridge, MA}
}

@Article{Bhat2011,
  Title                    = {The maximum approximate composite marginal likelihood (MACML) estimation of multinomial probit-based unordered response choice models},
  Author                   = {Bhat, Chandra R},
  Journal                  = {Transportation Research Part B: Methodological},
  Year                     = {2011},
  Number                   = {7},
  Pages                    = {923--939},
  Volume                   = {45},
  Publisher                = {Elsevier}
}

@Article{1,
  Title                    = {{Simulation estimation of mixed discrete choice models using randomized and scrambled Halton sequences}},
  Author                   = {C. R. Bhat},
  Journal                  = {Transportation Research Part B},
  Year                     = {2003},
  Number                   = {9},
  Pages                    = {837--855},
  Volume                   = {37}
}

@Article{22,
  Title                    = {{Quasi-random maximum simulated likelihood estimation of the mixed multinomial Logit model}},
  Author                   = {C. R. Bhat},
  Journal                  = {Transportation Research Part B},
  Year                     = {2001},
  Number                   = {7},
  Pages                    = {677--693},
  Volume                   = {35}
}

@Article{59,
  Title                    = {{Covariance Heterogeneity in Nested Logit models: economic structure and application to inter-city travel}},
  Author                   = {C. R. Bhat},
  Journal                  = {Transportation Research Part B},
  Year                     = {1997},
  Number                   = {1},
  Pages                    = {11--21},
  Volume                   = {31}
}

@Article{19,
  Title                    = {{An improved low-discrepancy sequence for multidimensional Quasi-Monte Carlo integration}},
  Author                   = {E. Braaten and G. Weller},
  Journal                  = {Journal of Computational Physics},
  Year                     = {1979},
  Pages                    = {249--258},
  Volume                   = {33}
}

@Article{37,
  Title                    = {Implementation and tests of low-discrepancy sequences},
  Author                   = {P. Bratley and B. L. Fox and H. Niederreiter},
  Journal                  = {ACM Transactions on Modeling and Computer Simulation},
  Year                     = {1992},
  Pages                    = {195--213},
  Volume                   = {2}
}

@Article{5,
  Title                    = {Randomization of number theoretic methods for multiple integrations},
  Author                   = {R. Cranley},
  Journal                  = {Siam Journal on Numerical Analysis},
  Year                     = {1976},
  Number                   = {6},
  Pages                    = {904--914},
  Volume                   = {13}
}

@Article{25,
  Title                    = {{A conditional Probit model for qualitative choice: discrete decisions recognizing interdependence and heterogeneous preferences}},
  Author                   = {J. Hausman and D. Wise},
  Journal                  = {Econometrica},
  Year                     = {1978},
  Number                   = {2},
  Pages                    = {403--426},
  Volume                   = {46}
}

@InCollection{769,
  Title                    = {{Statistical Models for Discrete Panel Data}},
  Author                   = {J. J. Heckman},
  Booktitle                = {{Structural Analysis of Discrete Data with Econometric Applications}},
  Publisher                = {MIT Press},
  Year                     = {1981},

  Address                  = {Cambridge, MA},
  Chapter                  = {3},
  Editor                   = {Manski, C. and McFadden, D.},
  Pages                    = {114--178}
}

@Article{880,
  Title                    = {A systematic comparison of continuous and discrete mixture models},
  Author                   = {S. Hess and M. Bierlaire and J. W. Polak},
  Journal                  = {European Transport \textbackslash{} Trasporti Europei},
  Year                     = {2007},
  Pages                    = {35--61},
  Volume                   = {37}
}

@Article{849,
  Title                    = {Allowing for intra-respondent variations in coefficients estimated on repeated choice data},
  Author                   = {S. Hess and J. M. Rose},
  Journal                  = {Transportation Research Part B},
  Year                     = {2009},
  Number                   = {6},
  Pages                    = {708--719},
  Volume                   = {43}
}

@Article{266,
  Title                    = {{On the use of a Modified Latin Hypercube Sampling (MLHS) method in the estimation of a Mixed Logit model for vehicle choice}},
  Author                   = {S. Hess and K. Train and J. W. Polak},
  Journal                  = {Transportation Research Part B},
  Year                     = {2006},
  Number                   = {2},
  Pages                    = {147--163},
  Volume                   = {40}
}

@InCollection{85,
  Title                    = {Modelling the choice of residential location},
  Author                   = {D. McFadden},
  Booktitle                = {Spatial Interaction Theory and Planning Models},
  Publisher                = {North Holland},
  Year                     = {1978},

  Address                  = {Amsterdam},
  Chapter                  = {25},
  Editor                   = {Karlqvist, A. and Lundqvist, L. and Snickars, F. and Weibull, J. W.},
  Pages                    = {75--96}
}

@InCollection{91,
  Title                    = {Econometric Models of Probabilistic Choice},
  Author                   = {D. McFadden},
  Booktitle                = {Structural Analysis of Discrete Data with Econometric Applications},
  Publisher                = {MIT Press},
  Year                     = {1981},

  Address                  = {Cambridge, MA, USA},
  Chapter                  = {5},
  Editor                   = {Manski, C. and McFadden, D.},
  Pages                    = {198--272}
}

@Article{21,
  Title                    = {{A method of simulated moments for estimation of the multinomial Probit without numerical integration}},
  Author                   = {D. McFadden},
  Journal                  = {Econometrica},
  Year                     = {1989},
  Number                   = {5},
  Pages                    = {995--1026},
  Volume                   = {57}
}

@Book{97,
  Title                    = {Demand Model Estimation and Validation},
  Author                   = {D. McFadden and A. Talvitie and S. Cosslett and I. Hasan and M. Johnson and F. Reid and K. Train},
  Publisher                = {The institute of transportation studies at the University of California},
  Year                     = {1977},

  Address                  = {Berkeley and Irvine, CA},
  Series                   = {Urban Travel Demand Forecasting Project}
}

@Article{31,
  Title                    = {{The Monte Carlo method}},
  Author                   = {N. C. Metropolis and S. M. Ulam},
  Journal                  = {Journal of the American Statistical Association},
  Year                     = {1949},
  Number                   = {247},
  Pages                    = {335--341},
  Volume                   = {44}
}

@InCollection{2,
  Title                    = {New developments in uniform pseudo-random number and vector generation},
  Author                   = {H. Niederreiter},
  Booktitle                = {Monte Carlo and Quasi-Monte Carlo Methods in Scientific Computing},
  Publisher                = {Springer},
  Year                     = {1995},

  Address                  = {New York},
  Editor                   = {Niederreiter, H. and Shiue, J-S.},
  Pages                    = {94--112}
}

@InCollection{27,
  Title                    = {Nets, (t,s)-sequences, and algebraic geometry},
  Author                   = {H. Niederreiter and C. Xing},
  Booktitle                = {Random and Quasi-Random Point Sets, Lecture Notes in Statistics, 138},
  Publisher                = {Springer},
  Year                     = {1998},

  Address                  = {New York},
  Editor                   = {Hellekalek, P. and Larcher, G.},
  Pages                    = {267--302}
}

@Article{7,
  Title                    = {Latin supercube sampling for very high dimensional simulations},
  Author                   = {A. B. Owen},
  Journal                  = {ACM Transactions on Mathematical Software},
  Year                     = {1998},
  Number                   = {1},
  Pages                    = {71--102},
  Volume                   = {8}
}

@Article{6,
  Title                    = {Scrambled net variance for integrals of smooth functions},
  Author                   = {A. B. Owen},
  Journal                  = {The Annals of Statistics},
  Year                     = {1997},
  Number                   = {4},
  Pages                    = {1541--1562},
  Volume                   = {25}
}

@Article{44,
  Title                    = {{Mixed Logit with repeated choices: households' choices of appliance efficiency level}},
  Author                   = {D. Revelt and K. Train},
  Journal                  = {Review of Economics and Statistics},
  Year                     = {1998},
  Number                   = {4},
  Pages                    = {647--657},
  Volume                   = {80}
}

@Article{20,
  Title                    = {{A quasirandom approach to integration in Bayesian statistics}},
  Author                   = {J. E. H. Shaw},
  Journal                  = {The Annals of Statistics},
  Year                     = {1988},
  Number                   = {2},
  Pages                    = {895--914},
  Volume                   = {16}
}

@Article{39,
  Title                    = {{When are Quasi-Monte Carlo algorithms efficient for high dimensional integrals?}},
  Author                   = {J. H. Sloan and H. Wozniakowski},
  Journal                  = {Journal of Complexity},
  Year                     = {1998},
  Number                   = {1},
  Pages                    = {1--33},
  Volume                   = {14}
}

@Article{3,
  Title                    = {Quasi-random methods for estimating integrals using relatively small samples},
  Author                   = {J. Spanier and E. H. Maize},
  Journal                  = {Siam Review},
  Year                     = {1994},
  Pages                    = {18--44},
  Volume                   = {36}
}

@Book{Train2009,
  author    = {K. Train},
  publisher = {Cambridge University Press},
  title     = {Discrete Choice Methods with Simulation},
  year      = {2009},
  address   = {Cambridge, UK},
  edition   = {Second},
}

@Article{45,
  Title                    = {Recreation demand models with taste differences over people},
  Author                   = {K. Train},
  Journal                  = {Land Economics},
  Year                     = {1998},
  Pages                    = {185--194},
  Volume                   = {74}
}

@Article{14,
  Title                    = {Integration and approximation based on scrambled sampling in arbitrary dimensions},
  Author                   = {R-X. Yue and F. J. Hickernell},
  Journal                  = {Journal of Complexity},
  Year                     = {2001},
  Number                   = {4},
  Volume                   = {17}
}

@Article{4,
  Title                    = {{The mathematical basis of Monte Carlo and Quasi-Monte Carlo methods}},
  Author                   = {S. K. Zaremba},
  Journal                  = {Siam Review},
  Year                     = {1968},
  Pages                    = {303--314},
  Volume                   = {10}
}

@article{Hess_et_al_VTT2017,
title = "A framework for capturing heterogeneity, heteroskedasticity, non-linearity, reference dependence and design artefacts in value of time research",
journal = "Transportation Research Part B: Methodological",
volume = "96",
pages = "126 - 149",
year = "2017",
author = "Stephane Hess and Andrew Daly and Thijs Dekker and Manuel Ojeda Cabral and Richard Batley"
}

@Article{hess_palma_apollo,
  author   = {Stephane Hess and David Palma},
  journal  = {Journal of Choice Modelling},
  title    = {Apollo: A flexible, powerful and customisable freeware package for choice model estimation and application},
  year     = {2019},
  issn     = {1755-5345},
  pages    = {100170},
  volume   = {32},
  doi      = {https://doi.org/10.1016/j.jocm.2019.100170},
}

@Article{Krueger2021,
  author  = {Krueger, Rico and Bierlaire, Michel and Daziano, Ricardo A. and Rashidi, Taha H. and Bansal, Prateek},
  journal = {Journal of Choice Modelling},
  title   = {Evaluating the predictive abilities of mixed logit models with unobserved inter- and intra-individual heterogeneity},
  year    = {2021},
  volume  = {41},
  pages   = {100323},
  doi     = {10.1016/j.jocm.2021.100323},
}

@Article{CherchiCirillo2010,
  author  = {Cherchi, Elisabetta and Cirillo, Cinzia},
  journal = {Transportation Research Record},
  title   = {Validation and forecasts in models estimated from multiday travel survey},
  year    = {2010},
  volume  = {2175},
  pages   = {57--64},
  doi     = {10.3141/2175-07},
}

@Article{Zhao2020,
  author  = {Zhao, Xilei and Yan, Xiang and Yu, Alan and Van Hentenryck, Pascal},
  journal = {Travel Behaviour and Society},
  title   = {Prediction and behavioral analysis of travel mode choice: A comparison of machine learning and logit models},
  year    = {2020},
  volume  = {20},
  pages   = {22--35},
  doi     = {10.1016/j.tbs.2020.02.003},
}

@Article{Wang2024,
  author  = {Wang, Shenhao and Mo, Baichuan and Zheng, Yunhan and Hess, Stephane and Zhao, Jinhua},
  journal = {Transportation Research Part B: Methodological},
  title   = {Comparing hundreds of machine learning and discrete choice models for travel demand modeling: An empirical benchmark},
  year    = {2024},
  volume  = {190},
  pages   = {103061},
  doi     = {10.1016/j.trb.2024.103061},
}

@Article{Xie2020,
  author  = {Xie, Yifei and Zhang, Yundi and Akkinepally, Arun Prakash and Ben-Akiva, Moshe},
  journal = {Transportation Research Record},
  title   = {Personalized choice model for managed lane travel behavior},
  year    = {2020},
  volume  = {2674},
  number  = {7},
  pages   = {442--455},
  doi     = {10.1177/0361198120923355},
}

@Article{Danaf2019,
  author  = {Danaf, Mazen and Becker, Felix and Song, Xiang and Atasoy, Bilge and Ben-Akiva, Moshe},
  journal = {Decision Support Systems},
  title   = {Online discrete choice models: Applications in personalized recommendations},
  year    = {2019},
  volume  = {119},
  pages   = {35--45},
  doi     = {10.1016/j.dss.2019.02.003},
}

@Article{Train2016,
  author  = {Train, Kenneth},
  journal = {Journal of Choice Modelling},
  title   = {Mixed logit with a flexible mixing distribution},
  year    = {2016},
  volume  = {19},
  pages   = {40--53},
  doi     = {10.1016/j.jocm.2016.07.004},
}

@Article{Dawid1984,
  author  = {Dawid, A. Philip},
  journal = {Journal of the Royal Statistical Society. Series A (General)},
  title   = {Present position and potential developments: Some personal views. {Statistical} theory: The prequential approach},
  year    = {1984},
  volume  = {147},
  number  = {2},
  pages   = {278--292},
}

@Article{Watanabe1960,
  author  = {Watanabe, Satosi},
  journal = {IBM Journal of Research and Development},
  title   = {Information theoretical analysis of multivariate correlation},
  year    = {1960},
  volume  = {4},
  number  = {1},
  pages   = {66--82},
}

@Article{Gneiting2007,
  author  = {Gneiting, Tilmann and Raftery, Adrian E.},
  journal = {Journal of the American Statistical Association},
  title   = {Strictly proper scoring rules, prediction, and estimation},
  year    = {2007},
  volume  = {102},
  number  = {477},
  pages   = {359--378},
  doi     = {10.1198/016214506000001437},
}

@Book{Ruud2000,
  author    = {Ruud, Paul A.},
  publisher = {Oxford University Press},
  title     = {An Introduction to Classical Econometric Theory},
  year      = {2000},
  address   = {New York},
}

@Article{Zeger1988,
  author  = {Zeger, S. L. and Liang, K.-Y. and Albert, P. S.},
  title   = {Models for longitudinal data: a generalized estimating equation approach},
  journal = {Biometrics},
  year    = {1988},
  volume  = {44},
  number  = {4},
  pages   = {1049--1060},
}

@Article{HeagertyZeger2000,
  author  = {Heagerty, P. J. and Zeger, S. L.},
  title   = {Marginalized multilevel models and likelihood inference},
  journal = {Statistical Science},
  year    = {2000},
  volume  = {15},
  number  = {1},
  pages   = {1--26},
}

@Article{Neuhaus1992,
  author  = {Neuhaus, J. M. and Hauck, W. W. and Kalbfleisch, J. D.},
  title   = {The effects of mixture distribution misspecification when fitting mixed-effects logistic models},
  journal = {Biometrika},
  year    = {1992},
  volume  = {79},
  number  = {4},
  pages   = {755--762},
}

@Article{HeagertyKurland2001,
  author  = {Heagerty, P. J. and Kurland, B. F.},
  title   = {Misspecified maximum likelihood estimates and generalised linear mixed models},
  journal = {Biometrika},
  year    = {2001},
  volume  = {88},
  number  = {4},
  pages   = {973--985},
}

@Article{McCullochNeuhaus2011,
  author  = {McCulloch, C. E. and Neuhaus, J. M.},
  title   = {Misspecifying the shape of a random effects distribution: why getting it wrong may not matter},
  journal = {Statistical Science},
  year    = {2011},
  volume  = {26},
  number  = {3},
  pages   = {388--402},
}

@Article{Lindsay1988,
  author  = {Lindsay, B. G.},
  title   = {Composite likelihood methods},
  journal = {Contemporary Mathematics},
  year    = {1988},
  volume  = {80},
  pages   = {221--239},
}

@Article{VarinReidFirth2011,
  author  = {Varin, C. and Reid, N. and Firth, D.},
  title   = {An overview of composite likelihood methods},
  journal = {Statistica Sinica},
  year    = {2011},
  volume  = {21},
  number  = {1},
  pages   = {5--42},
}

@Article{GaoSong2010,
  author  = {Gao, X. and Song, P. X.-K.},
  title   = {Composite likelihood {B}ayesian information criteria for model selection in high-dimensional data},
  journal = {Journal of the American Statistical Association},
  year    = {2010},
  volume  = {105},
  number  = {492},
  pages   = {1531--1540},
}

@Article{DubeHitschRossi2010,
  author  = {Dub{\'e}, J.-P. and Hitsch, G. J. and Rossi, P. E.},
  title   = {State dependence and alternative explanations for consumer inertia},
  journal = {RAND Journal of Economics},
  year    = {2010},
  volume  = {41},
  number  = {3},
  pages   = {417--445},
}

@article{Shannon1948,
  author = {C. E. Shannon},
  title = {A mathematical theory of communication},
  journal = {Bell System Technical Journal},
  volume = {27},
  number = {3},
  pages = {379--423},
  year = {1948}
}

@book{CoverThomas2006,
  author = {T. M. Cover and J. A. Thomas},
  title = {Elements of Information Theory},
  edition = {2nd},
  publisher = {Wiley},
  address = {Hoboken, NJ},
  year = {2006}
}
}

\clearpage
\section*{Supplementary material: full estimation results for fold 1}
\addcontentsline{toc}{section}{Supplementary material}
\setcounter{table}{0}
\setcounter{subsection}{0}
\renewcommand{\theHtable}{S\arabic{table}}
\renewcommand{\theHsubsection}{S\arabic{subsection}}
\renewcommand{\thetable}{S\arabic{table}}
\renewcommand{\thesubsection}{S\arabic{subsection}}
\setcounter{table}{0}
\renewcommand{\arraystretch}{0.95}
This supplementary material reports the estimates of the panel models used in the paper. Section, table and equation numbers without the prefix S refer to the paper. All models are estimated on five folds (Section \ref{sec:splitting}). For brevity we report the estimates from the first fold. Each table gives the estimate and its $t$-ratio based on robust (sandwich) standard errors computed at these estimates, together with the log-likelihood and the number of estimated parameters. Parameters fixed for normalisation, and the coefficients fixed to zero in the COVID-19 latent class models (Section \ref{sec:randhet}), are omitted. The cross-sectional versions of the mixture models are used in the paper only as a reference point for the decomposition of Section \ref{sec:decomposition} and are not reported, apart from the standard deviations summarised in Table \ref{tab:dispersion}.

The notation is that of Section \ref{sec:utilities}. In the tables for the continuous mixtures, $\mu$ and $\sigma$ are the mean and standard deviation of a normally distributed coefficient (the sign of a standard deviation is not identified), and $a$ and $b$ are the parameters of a log-uniform coefficient (Section \ref{sec:randhet}). The non-random coefficient for unknown duration in the COVID-19 continuous and logit mixed logit models was estimated as $\beta_{\text{unk}}=-e^{a_{\text{unk}}}$ and is reported as $\beta_{\text{unk}}$, with its $t$-ratio obtained by the delta method. In the latent class tables the superscript $(q)$ denotes the class, and $\zeta_q$ and $\omega_{q,w}$ are the constant and the effect of covariate $w$ in the class allocation model, with class 1 as the base. Below the class allocation parameters, each latent class table gives the implied class shares, the mean of the class allocation probabilities $\pi_{nq}$ over the individuals in the estimation sample, which for the models with covariates in the class allocation model depend on each person's covariates.

For the logit mixed logit models we report two tables per dataset. The first gives the 5th percentile, median and 95th percentile of the implied distribution of each coefficient, computed over the $5{,}000$ support points with their estimated probabilities and, for the constants in the specifications with covariates, for the reference profile with all covariates zero. Coefficients without random heterogeneity, or whose support has collapsed, are shown as a single value, and the covariate interactions, which are not random, appear only in the second table. In both datasets the estimated distribution places most of its probability on a small number of support points, and some of the coefficients that enter through an exponential have long, poorly determined tails, which is why we report quantiles rather than means and standard deviations. The coefficient of exemption has a negative median in the logit mixed logit, with a long positive tail, where the parametric models give a positive mean. With a dummy attribute, the shape of its distribution is weakly identified. The second table gives the estimated support and the parameters without random heterogeneity, with robust $t$-ratios. For coefficients that enter through an exponential the table reports the identified quantities $a_k+l_d$ and $u_d-l_d$ (Section \ref{sec:randhet}), and for the other random coefficients the bounds $l$ and $u$ of the support. On the COVID-19 data the support of the constant for the free vaccines in the logit mixed logit without covariates has collapsed to a single point in fold 1, so its two bounds coincide ($-0.712$), and the Legendre coefficients for that dimension are not identified. The Legendre coefficients that give the probabilities of the support points are not reported, since their values have no direct interpretation and those that shape the density where it is close to zero are not separately identified. Because of these coefficients the Hessian is close to singular, and the robust standard errors in the second table are computed with a generalised inverse of the Hessian. This gives the usual sandwich estimator on the directions that the data identify and zero variance along the remaining directions, $12$ to $17$ per model and made up almost entirely of Legendre coefficients, as if they were known. Since $l_d$ is fixed by normalisation (Section \ref{sec:randhet}), $a_k+l_d$ and $u_d-l_d$ differ from $a_k$ and $u_d$ only by a constant and have the same standard errors, which is the delta method with a derivative of one. The $t$-ratios of the support parameters are therefore conditional on the estimated shape of the distribution and are likely to overstate their precision.

\clearpage
{\footnotesize\setlength{\tabcolsep}{4pt}
\begin{longtable}{lrrrrrrrr}
\caption{DECISIONS: models without random heterogeneity and continuous mixtures}\label{tab:A_dec_mnl}\\
\toprule
 & \multicolumn{2}{c}{MNL} & \multicolumn{2}{c}{MNL with socios} & \multicolumn{2}{c}{MMNL} & \multicolumn{2}{c}{MMNL with socios}\\
\cmidrule(lr){2-3} \cmidrule(lr){4-5} \cmidrule(lr){6-7} \cmidrule(lr){8-9}
Parameter & est. & rob.\ $t$ & est. & rob.\ $t$ & est. & rob.\ $t$ & est. & rob.\ $t$\\
\midrule\endfirsthead
\multicolumn{9}{l}{\emph{(continued)}}\\ \toprule  & \multicolumn{2}{c}{MNL} & \multicolumn{2}{c}{MNL with socios} & \multicolumn{2}{c}{MMNL} & \multicolumn{2}{c}{MMNL with socios}\\ Parameter & est. & rob.\ $t$ & est. & rob.\ $t$ & est. & rob.\ $t$ & est. & rob.\ $t$\\ \midrule\endhead
\bottomrule\endfoot
\multicolumn{9}{l}{\emph{Constants}}\\
$\delta_{\text{bus}}$ & $-$2.762 & $-$9.86 & $-$2.374 & $-$7.53 &  &  &  & \\
$\delta_{\text{rail}}$ & $-$1.950 & $-$4.66 & $-$1.449 & $-$2.86 &  &  &  & \\
$\delta_{\text{taxi}}$ & $-$3.873 & $-$10.87 & $-$4.178 & $-$9.03 &  &  &  & \\
$\delta_{\text{cycle}}$ & $-$4.177 & $-$9.71 & $-$3.804 & $-$10.03 &  &  &  & \\
$\delta_{\text{walk}}$ & 0.332 & 1.29 & 0.692 & 2.22 &  &  &  & \\
$\kappa_{\text{bus},\text{fem}}$ &  &  & $-$0.644 & $-$2.37 &  &  & $-$0.768 & $-$1.53\\
$\kappa_{\text{bus},\text{age}}$ &  &  & $-$0.318 & $-$0.68 &  &  & 0.400 & 0.43\\
$\kappa_{\text{bus},\text{inc}}$ &  &  & $-$1.043 & $-$1.21 &  &  & $-$2.153 & $-$4.50\\
$\kappa_{\text{rail},\text{fem}}$ &  &  & $-$0.744 & $-$1.86 &  &  & $-$0.537 & $-$0.70\\
$\kappa_{\text{rail},\text{age}}$ &  &  & $-$0.998 & $-$1.51 &  &  & 0.291 & 0.26\\
$\kappa_{\text{rail},\text{inc}}$ &  &  & 0.025 & 0.04 &  &  & $-$1.418 & $-$0.68\\
$\kappa_{\text{taxi},\text{fem}}$ &  &  & 0.474 & 1.27 &  &  & 0.843 & 0.98\\
$\kappa_{\text{taxi},\text{age}}$ &  &  & $-$0.922 & $-$1.20 &  &  & $-$0.975 & $-$0.90\\
$\kappa_{\text{taxi},\text{inc}}$ &  &  & $-$0.107 & $-$0.23 &  &  & $-$1.700 & $-$0.79\\
$\kappa_{\text{cycle},\text{fem}}$ &  &  & $-$1.419 & $-$2.89 &  &  & $-$2.027 & $-$1.45\\
$\kappa_{\text{cycle},\text{age}}$ &  &  & 0.260 & 0.24 &  &  & $-$0.093 & $-$0.12\\
$\kappa_{\text{cycle},\text{inc}}$ &  &  & 1.541 & 1.79 &  &  & 0.493 & 0.47\\
$\kappa_{\text{walk},\text{fem}}$ &  &  & $-$0.514 & $-$1.66 &  &  & $-$0.362 & $-$0.96\\
$\kappa_{\text{walk},\text{age}}$ &  &  & $-$1.044 & $-$2.86 &  &  & $-$0.728 & $-$1.18\\
$\kappa_{\text{walk},\text{inc}}$ &  &  & $-$0.762 & $-$1.77 &  &  & $-$0.680 & $-$1.21\\
$\mu_{\delta,\text{bus}}$ &  &  &  &  & $-$4.442 & $-$9.08 & $-$4.069 & $-$8.06\\
$\sigma_{\delta,\text{bus}}$ &  &  &  &  & 2.556 & 8.30 & 2.558 & 9.73\\
$\mu_{\delta,\text{rail}}$ &  &  &  &  & $-$3.066 & $-$3.54 & $-$2.467 & $-$1.89\\
$\sigma_{\delta,\text{rail}}$ &  &  &  &  & 2.456 & 4.95 & 2.252 & 3.70\\
$\mu_{\delta,\text{taxi}}$ &  &  &  &  & $-$6.279 & $-$10.50 & $-$6.605 & $-$7.74\\
$\sigma_{\delta,\text{taxi}}$ &  &  &  &  & 2.684 & 5.31 & 2.863 & 6.67\\
$\mu_{\delta,\text{cycle}}$ &  &  &  &  & $-$10.081 & $-$4.26 & $-$8.906 & $-$8.57\\
$\sigma_{\delta,\text{cycle}}$ &  &  &  &  & 4.866 & 3.73 & 4.174 & 6.81\\
$\mu_{\delta,\text{walk}}$ &  &  &  &  & 0.609 & 1.92 & 0.894 & 2.30\\
$\sigma_{\delta,\text{walk}}$ &  &  &  &  & $-$1.612 & $-$10.25 & $-$1.650 & $-$8.43\\
\multicolumn{9}{l}{\emph{Attribute coefficients}}\\
$\beta_{\text{IVT},\text{car}}$ & $-$0.142 & $-$9.77 & $-$0.144 & $-$9.79 &  &  &  & \\
$\beta_{\text{IVT},\text{bus}}$ & $-$0.039 & $-$5.99 & $-$0.039 & $-$5.91 &  &  &  & \\
$\beta_{\text{IVT},\text{rail}}$ & $-$0.037 & $-$2.20 & $-$0.036 & $-$2.16 &  &  &  & \\
$\beta_{\text{IVT},\text{taxi}}$ & $-$0.094 & $-$3.56 & $-$0.095 & $-$3.60 &  &  &  & \\
$\beta_{\text{IVT},\text{cycle}}$ & $-$0.070 & $-$5.11 & $-$0.067 & $-$4.88 &  &  &  & \\
$\beta_{\text{IVT},\text{walk}}$ & $-$0.148 & $-$12.34 & $-$0.148 & $-$12.26 &  &  &  & \\
$\beta_{\text{OVT},\text{bus}}$ & $-$0.119 & $-$4.31 & $-$0.120 & $-$4.35 &  &  &  & \\
$\beta_{\text{OVT},\text{rail}}$ & $-$0.110 & $-$6.78 & $-$0.112 & $-$6.94 &  &  &  & \\
$\beta_{\text{cost}}$ & $-$0.193 & $-$6.16 & $-$0.196 & $-$6.19 &  &  &  & \\
$a_{\text{IVT},\text{car}}$ &  &  &  &  & $-$2.335 & $-$10.36 & $-$2.354 & $-$13.76\\
$a_{\text{IVT},\text{bus}}$ &  &  &  &  & $-$3.723 & $-$14.17 & $-$3.759 & $-$16.02\\
$a_{\text{IVT},\text{rail}}$ &  &  &  &  & $-$3.123 & $-$9.21 & $-$3.226 & $-$7.82\\
$a_{\text{IVT},\text{taxi}}$ &  &  &  &  & $-$2.474 & $-$8.30 & $-$2.528 & $-$7.98\\
$a_{\text{IVT},\text{cycle}}$ &  &  &  &  & $-$3.170 & $-$10.74 & $-$3.163 & $-$12.58\\
$a_{\text{IVT},\text{walk}}$ &  &  &  &  & $-$2.020 & $-$13.90 & $-$2.042 & $-$16.08\\
$a_{\text{OVT},\text{bus}}$ &  &  &  &  & $-$3.208 & $-$13.34 & $-$3.138 & $-$14.51\\
$a_{\text{OVT},\text{rail}}$ &  &  &  &  & $-$3.475 & $-$8.38 & $-$3.308 & $-$6.69\\
$a_{\text{cost}}$ &  &  &  &  & 0.634 & 1.32 & 0.606 & 1.58\\
$b_{\text{IVT}}$ &  &  &  &  & 0.941 & 3.98 & 0.986 & 6.15\\
$b_{\text{OVT}}$ &  &  &  &  & 2.751 & 6.06 & 2.521 & 7.28\\
$b_{\text{cost}}$ &  &  &  &  & $-$5.190 & $-$3.57 & $-$4.705 & $-$3.31\\
\midrule
Log-likelihood & \multicolumn{2}{r}{$-$3,408.68} & \multicolumn{2}{r}{$-$3,308.54} & \multicolumn{2}{r}{$-$2,342.46} & \multicolumn{2}{r}{$-$2,328.83}\\
Parameters & \multicolumn{2}{r}{14} & \multicolumn{2}{r}{29} & \multicolumn{2}{r}{22} & \multicolumn{2}{r}{37}\\
\end{longtable}}

\clearpage
{\footnotesize\setlength{\tabcolsep}{4pt}
\begin{longtable}{lrrrrrr}
\caption{DECISIONS: latent class models}\label{tab:A_dec_lc}\\
\toprule
 & \multicolumn{2}{c}{LC} & \multicolumn{2}{c}{LC, socios in utility} & \multicolumn{2}{c}{LC, socios in class allocation}\\
\cmidrule(lr){2-3} \cmidrule(lr){4-5} \cmidrule(lr){6-7}
Parameter & est. & rob.\ $t$ & est. & rob.\ $t$ & est. & rob.\ $t$\\
\midrule\endfirsthead
\multicolumn{7}{l}{\emph{(continued)}}\\ \toprule  & \multicolumn{2}{c}{LC} & \multicolumn{2}{c}{LC, socios in utility} & \multicolumn{2}{c}{LC, socios in class allocation}\\ Parameter & est. & rob.\ $t$ & est. & rob.\ $t$ & est. & rob.\ $t$\\ \midrule\endhead
\bottomrule\endfoot
\multicolumn{7}{l}{\emph{Constants}}\\
$\delta^{(1)}_{\text{bus}}$ & $-$4.661 & $-$10.18 & $-$4.790 & $-$9.26 & $-$4.630 & $-$9.25\\
$\delta^{(1)}_{\text{rail}}$ & $-$1.895 & $-$3.02 & $-$1.197 & $-$1.61 & $-$1.903 & $-$2.93\\
$\delta^{(1)}_{\text{taxi}}$ & $-$5.320 & $-$9.73 & $-$5.925 & $-$8.08 & $-$5.273 & $-$8.43\\
$\delta^{(1)}_{\text{cycle}}$ & $-$6.969 & $-$11.17 & $-$5.979 & $-$5.25 & $-$6.921 & $-$10.84\\
$\delta^{(1)}_{\text{walk}}$ & 0.622 & 1.14 & 0.829 & 1.42 & 0.510 & 0.93\\
$\delta^{(2)}_{\text{bus}}$ & $-$1.483 & $-$3.22 & $-$1.488 & $-$2.89 & $-$1.446 & $-$3.14\\
$\delta^{(2)}_{\text{rail}}$ & $-$1.690 & $-$1.78 & $-$1.192 & $-$1.02 & $-$1.644 & $-$1.71\\
$\delta^{(2)}_{\text{taxi}}$ & $-$2.381 & $-$2.59 & $-$2.860 & $-$3.94 & $-$2.829 & $-$4.02\\
$\delta^{(2)}_{\text{cycle}}$ & $-$4.886 & $-$5.73 & $-$5.036 & $-$5.83 & $-$4.897 & $-$5.65\\
$\delta^{(2)}_{\text{walk}}$ & 0.128 & 0.29 & 0.208 & 0.44 & 0.226 & 0.54\\
$\delta^{(3)}_{\text{bus}}$ & $-$4.164 & $-$4.76 & $-$4.284 & $-$5.41 & $-$3.727 & $-$5.19\\
$\delta^{(3)}_{\text{rail}}$ & $-$3.554 & $-$3.11 & $-$3.113 & $-$3.54 & $-$2.830 & $-$3.03\\
$\delta^{(3)}_{\text{taxi}}$ & $-$3.115 & $-$5.02 & $-$3.724 & $-$4.85 & $-$3.132 & $-$5.17\\
$\delta^{(3)}_{\text{cycle}}$ & $-$1.996 & $-$3.05 & $-$2.654 & $-$3.78 & $-$2.069 & $-$3.44\\
$\delta^{(3)}_{\text{walk}}$ & 1.001 & 1.78 & 1.007 & 1.78 & 0.967 & 1.75\\
$\kappa_{\text{bus},\text{fem}}$ &  &  & $-$0.013 & $-$0.05 &  & \\
$\kappa_{\text{bus},\text{age}}$ &  &  & $-$0.473 & $-$0.82 &  & \\
$\kappa_{\text{bus},\text{inc}}$ &  &  & $-$0.470 & $-$0.82 &  & \\
$\kappa_{\text{rail},\text{fem}}$ &  &  & $-$0.895 & $-$1.98 &  & \\
$\kappa_{\text{rail},\text{age}}$ &  &  & $-$0.126 & $-$0.13 &  & \\
$\kappa_{\text{rail},\text{inc}}$ &  &  & $-$1.397 & $-$2.68 &  & \\
$\kappa_{\text{taxi},\text{fem}}$ &  &  & 0.822 & 2.17 &  & \\
$\kappa_{\text{taxi},\text{age}}$ &  &  & $-$0.338 & $-$0.39 &  & \\
$\kappa_{\text{taxi},\text{inc}}$ &  &  & $-$0.314 & $-$0.39 &  & \\
$\kappa_{\text{cycle},\text{fem}}$ &  &  & $-$0.277 & $-$0.77 &  & \\
$\kappa_{\text{cycle},\text{age}}$ &  &  & 0.771 & 1.79 &  & \\
$\kappa_{\text{cycle},\text{inc}}$ &  &  & 1.943 & 3.45 &  & \\
$\kappa_{\text{walk},\text{fem}}$ &  &  & $-$0.282 & $-$0.82 &  & \\
$\kappa_{\text{walk},\text{age}}$ &  &  & $-$0.649 & $-$1.46 &  & \\
$\kappa_{\text{walk},\text{inc}}$ &  &  & $-$0.307 & $-$0.54 &  & \\
\multicolumn{7}{l}{\emph{Attribute coefficients}}\\
$\beta^{(1)}_{\text{IVT},\text{car}}$ & $-$0.142 & $-$5.76 & $-$0.153 & $-$5.90 & $-$0.143 & $-$5.72\\
$\beta^{(1)}_{\text{IVT},\text{bus}}$ & $-$0.028 & $-$2.39 & $-$0.032 & $-$2.48 & $-$0.029 & $-$2.32\\
$\beta^{(1)}_{\text{IVT},\text{rail}}$ & $-$0.037 & $-$1.50 & $-$0.039 & $-$1.59 & $-$0.034 & $-$1.34\\
$\beta^{(1)}_{\text{IVT},\text{taxi}}$ & $-$0.100 & $-$1.74 & $-$0.114 & $-$1.67 & $-$0.104 & $-$1.54\\
$\beta^{(1)}_{\text{IVT},\text{cycle}}$ & $-$0.077 & $-$2.85 & $-$0.226 & $-$2.52 & $-$0.080 & $-$2.88\\
$\beta^{(1)}_{\text{IVT},\text{walk}}$ & $-$0.196 & $-$7.47 & $-$0.202 & $-$8.42 & $-$0.188 & $-$7.08\\
$\beta^{(1)}_{\text{OVT},\text{bus}}$ & $-$0.173 & $-$5.64 & $-$0.167 & $-$5.97 & $-$0.194 & $-$4.95\\
$\beta^{(1)}_{\text{OVT},\text{rail}}$ & $-$0.103 & $-$4.78 & $-$0.113 & $-$4.95 & $-$0.103 & $-$4.98\\
$\beta^{(1)}_{\text{cost}}$ & $-$0.142 & $-$3.56 & $-$0.138 & $-$3.24 & $-$0.141 & $-$3.69\\
$\beta^{(2)}_{\text{IVT},\text{car}}$ & $-$0.099 & $-$2.77 & $-$0.102 & $-$3.04 & $-$0.098 & $-$2.73\\
$\beta^{(2)}_{\text{IVT},\text{bus}}$ & $-$0.037 & $-$2.43 & $-$0.037 & $-$2.86 & $-$0.037 & $-$2.55\\
$\beta^{(2)}_{\text{IVT},\text{rail}}$ & $-$0.015 & $-$0.45 & $-$0.017 & $-$0.51 & $-$0.015 & $-$0.47\\
$\beta^{(2)}_{\text{IVT},\text{taxi}}$ & $-$0.074 & $-$1.03 & $-$0.073 & $-$1.41 & $-$0.048 & $-$0.94\\
$\beta^{(2)}_{\text{IVT},\text{cycle}}$ & $-$0.117 & $-$1.96 & $-$0.110 & $-$1.96 & $-$0.118 & $-$1.94\\
$\beta^{(2)}_{\text{IVT},\text{walk}}$ & $-$0.104 & $-$5.41 & $-$0.102 & $-$4.86 & $-$0.105 & $-$5.51\\
$\beta^{(2)}_{\text{OVT},\text{bus}}$ & $-$0.048 & $-$0.92 & $-$0.048 & $-$0.94 & $-$0.048 & $-$0.89\\
$\beta^{(2)}_{\text{OVT},\text{rail}}$ & $-$0.117 & $-$2.63 & $-$0.124 & $-$2.46 & $-$0.117 & $-$2.60\\
$\beta^{(2)}_{\text{cost}}$ & $-$0.401 & $-$6.00 & $-$0.408 & $-$5.76 & $-$0.408 & $-$6.20\\
$\beta^{(3)}_{\text{IVT},\text{car}}$ & $-$0.140 & $-$5.45 & $-$0.147 & $-$5.16 & $-$0.145 & $-$5.44\\
$\beta^{(3)}_{\text{IVT},\text{bus}}$ & $-$0.031 & $-$1.69 & $-$0.033 & $-$2.18 & $-$0.041 & $-$2.72\\
$\beta^{(3)}_{\text{IVT},\text{rail}}$ & $-$0.056 & $-$1.39 & $-$0.056 & $-$1.66 & $-$0.065 & $-$1.98\\
$\beta^{(3)}_{\text{IVT},\text{taxi}}$ & $-$0.072 & $-$1.96 & $-$0.070 & $-$1.90 & $-$0.086 & $-$2.28\\
$\beta^{(3)}_{\text{IVT},\text{cycle}}$ & $-$0.064 & $-$4.32 & $-$0.058 & $-$4.75 & $-$0.064 & $-$5.09\\
$\beta^{(3)}_{\text{IVT},\text{walk}}$ & $-$0.183 & $-$6.79 & $-$0.174 & $-$4.59 & $-$0.192 & $-$7.31\\
$\beta^{(3)}_{\text{OVT},\text{bus}}$ & $-$0.096 & $-$3.74 & $-$0.106 & $-$3.88 & $-$0.098 & $-$4.13\\
$\beta^{(3)}_{\text{OVT},\text{rail}}$ & $-$0.082 & $-$1.50 & $-$0.070 & $-$2.34 & $-$0.100 & $-$2.03\\
$\beta^{(3)}_{\text{cost}}$ & $-$0.166 & $-$3.48 & $-$0.197 & $-$3.70 & $-$0.161 & $-$3.62\\
\multicolumn{7}{l}{\emph{Class allocation}}\\
$\zeta_{2}$ & $-$0.548 & $-$3.08 & $-$0.474 & $-$2.67 & $-$0.179 & $-$0.73\\
$\zeta_{3}$ & $-$1.114 & $-$4.06 & $-$0.901 & $-$4.00 & $-$0.947 & $-$2.63\\
$\omega_{2,\text{fem}}$ &  &  &  &  & $-$0.507 & $-$1.59\\
$\omega_{2,\text{age}}$ &  &  &  &  & 0.271 & 0.36\\
$\omega_{2,\text{inc}}$ &  &  &  &  & $-$2.141 & $-$1.99\\
$\omega_{3,\text{fem}}$ &  &  &  &  & 0.090 & 0.22\\
$\omega_{3,\text{age}}$ &  &  &  &  & $-$1.015 & $-$0.95\\
$\omega_{3,\text{inc}}$ &  &  &  &  & $-$0.604 & $-$0.73\\
\multicolumn{7}{l}{\emph{Implied class shares (sample mean of $\pi_{nq}$)}}\\
class 1 & \multicolumn{2}{r}{0.524} & \multicolumn{2}{r}{0.493} & \multicolumn{2}{r}{0.507}\\
class 2 & \multicolumn{2}{r}{0.303} & \multicolumn{2}{r}{0.307} & \multicolumn{2}{r}{0.301}\\
class 3 & \multicolumn{2}{r}{0.172} & \multicolumn{2}{r}{0.200} & \multicolumn{2}{r}{0.191}\\
\midrule
Log-likelihood & \multicolumn{2}{r}{$-$2,687.72} & \multicolumn{2}{r}{$-$2,642.59} & \multicolumn{2}{r}{$-$2,681.81}\\
Parameters & \multicolumn{2}{r}{44} & \multicolumn{2}{r}{59} & \multicolumn{2}{r}{50}\\
\end{longtable}}

\clearpage
{\footnotesize\setlength{\tabcolsep}{4pt}
\begin{longtable}{lrrrrrr}
\caption{DECISIONS: logit mixed logit, implied distribution of each coefficient}\label{tab:A_dec_lml}\\
\toprule
 & \multicolumn{3}{c}{LML} & \multicolumn{3}{c}{LML with socios}\\
\cmidrule(lr){2-4} \cmidrule(lr){5-7}
Coefficient & 5\% & median & 95\% & 5\% & median & 95\%\\
\midrule\endfirsthead
\multicolumn{7}{l}{\emph{(continued)}}\\ \toprule Coefficient & 5\% & median & 95\% & 5\% & median & 95\%\\ \midrule\endhead
\bottomrule\endfoot
\multicolumn{7}{l}{\emph{Constants}}\\
$\delta_{\text{bus}}$ & $-$6.400 & $-$2.964 & $-$0.925 & $-$6.136 & $-$4.940 & $-$0.459\\
$\delta_{\text{rail}}$ & $-$5.066 & $-$1.788 & 1.196 & $-$3.899 & $-$2.113 & 0.834\\
$\delta_{\text{taxi}}$ & $-$8.478 & $-$5.750 & $-$2.043 & $-$8.654 & $-$6.265 & $-$2.590\\
$\delta_{\text{cycle}}$ & $-$13.473 & $-$9.622 & $-$2.271 & $-$15.745 & $-$12.430 & $-$2.444\\
$\delta_{\text{walk}}$ & $-$0.612 & 1.383 & 2.223 & $-$1.797 & 1.857 & 4.155\\
\multicolumn{7}{l}{\emph{Attribute coefficients}}\\
$\beta_{\text{IVT},\text{car}}$ & $-$0.229 & $-$0.143 & $-$0.100 & $-$0.243 & $-$0.159 & $-$0.042\\
$\beta_{\text{IVT},\text{bus}}$ & $-$0.066 & $-$0.041 & $-$0.029 & $-$0.053 & $-$0.035 & $-$0.009\\
$\beta_{\text{IVT},\text{rail}}$ & $-$0.097 & $-$0.061 & $-$0.042 & $-$0.053 & $-$0.035 & $-$0.009\\
$\beta_{\text{IVT},\text{taxi}}$ & $-$0.215 & $-$0.134 & $-$0.093 & $-$0.201 & $-$0.132 & $-$0.034\\
$\beta_{\text{IVT},\text{cycle}}$ & $-$0.131 & $-$0.082 & $-$0.057 & $-$0.119 & $-$0.078 & $-$0.020\\
$\beta_{\text{IVT},\text{walk}}$ & $-$0.364 & $-$0.227 & $-$0.159 & $-$0.350 & $-$0.229 & $-$0.060\\
$\beta_{\text{OVT},\text{bus}}$ & $-$0.564 & $-$0.188 & $-$0.005 & $-$0.356 & $-$0.217 & $-$0.082\\
$\beta_{\text{OVT},\text{rail}}$ & $-$0.415 & $-$0.138 & $-$0.004 & $-$0.299 & $-$0.182 & $-$0.069\\
$\beta_{\text{cost}}$ & $-$1.599 & $-$0.265 & $-$0.003 & $-$1.438 & $-$0.294 & $-$0.005\\
\end{longtable}}

\clearpage
{\footnotesize\setlength{\tabcolsep}{4pt}
\begin{longtable}{lrrrr}
\caption{DECISIONS: logit mixed logit, support and parameters without random heterogeneity}\label{tab:A_dec_lmlpar}\\
\toprule
 & \multicolumn{2}{c}{LML} & \multicolumn{2}{c}{LML with socios}\\
\cmidrule(lr){2-3} \cmidrule(lr){4-5}
Parameter & est. & rob.\ $t$ & est. & rob.\ $t$\\
\midrule\endfirsthead
\multicolumn{5}{l}{\emph{(continued)}}\\ \toprule  & \multicolumn{2}{c}{LML} & \multicolumn{2}{c}{LML with socios}\\ Parameter & est. & rob.\ $t$ & est. & rob.\ $t$\\ \midrule\endhead
\bottomrule\endfoot
\multicolumn{5}{l}{\emph{Constants}}\\
$\kappa_{\text{bus},\text{fem}}$ &  &  & $-$0.544 & $-$1.71\\
$\kappa_{\text{bus},\text{age}}$ &  &  & 0.612 & 0.82\\
$\kappa_{\text{bus},\text{inc}}$ &  &  & $-$2.109 & $-$5.99\\
$\kappa_{\text{rail},\text{fem}}$ &  &  & $-$1.144 & $-$2.50\\
$\kappa_{\text{rail},\text{age}}$ &  &  & $-$0.148 & $-$0.17\\
$\kappa_{\text{rail},\text{inc}}$ &  &  & $-$3.553 & $-$5.96\\
$\kappa_{\text{taxi},\text{fem}}$ &  &  & 0.855 & 2.28\\
$\kappa_{\text{taxi},\text{age}}$ &  &  & $-$0.068 & $-$0.06\\
$\kappa_{\text{taxi},\text{inc}}$ &  &  & $-$0.946 & $-$2.04\\
$\kappa_{\text{cycle},\text{fem}}$ &  &  & $-$1.686 & $-$4.20\\
$\kappa_{\text{cycle},\text{age}}$ &  &  & 0.928 & 1.69\\
$\kappa_{\text{cycle},\text{inc}}$ &  &  & 1.534 & 6.72\\
$\kappa_{\text{walk},\text{fem}}$ &  &  & $-$0.562 & $-$1.51\\
$\kappa_{\text{walk},\text{age}}$ &  &  & $-$0.435 & $-$0.55\\
$\kappa_{\text{walk},\text{inc}}$ &  &  & $-$1.259 & $-$2.45\\
$l_{\delta,\text{bus}}$ & $-$6.829 & $-$10.96 & $-$6.921 & $-$15.81\\
$l_{\delta,\text{rail}}$ & $-$7.051 & $-$3.36 & $-$4.483 & $-$5.24\\
$l_{\delta,\text{taxi}}$ & $-$9.781 & $-$7.97 & $-$9.308 & $-$17.73\\
$l_{\delta,\text{cycle}}$ & $-$14.004 & $-$10.33 & $-$16.114 & $-$9.28\\
$l_{\delta,\text{walk}}$ & $-$0.675 & $-$1.14 & $-$1.909 & $-$3.96\\
$u_{\delta,\text{bus}}$ & $-$0.639 & $-$2.09 & $-$0.145 & $-$0.47\\
$u_{\delta,\text{rail}}$ & 1.321 & 2.11 & 1.284 & 2.19\\
$u_{\delta,\text{taxi}}$ & $-$1.819 & $-$5.02 & $-$2.435 & $-$6.71\\
$u_{\delta,\text{cycle}}$ & $-$2.208 & $-$2.99 & $-$2.371 & $-$8.57\\
$u_{\delta,\text{walk}}$ & 3.041 & 6.38 & 4.217 & 10.81\\
\multicolumn{5}{l}{\emph{Attribute coefficients}}\\
$a_{\text{IVT},\text{car}}+l_{\text{IVT}}$ & $-$4.069 & $-$12.81 & $-$3.723 & $-$17.10\\
$a_{\text{IVT},\text{bus}}+l_{\text{IVT}}$ & $-$5.312 & $-$15.00 & $-$5.242 & $-$17.35\\
$a_{\text{IVT},\text{rail}}+l_{\text{IVT}}$ & $-$4.929 & $-$12.36 & $-$5.246 & $-$11.59\\
$a_{\text{IVT},\text{taxi}}+l_{\text{IVT}}$ & $-$4.135 & $-$12.21 & $-$3.914 & $-$14.80\\
$a_{\text{IVT},\text{cycle}}+l_{\text{IVT}}$ & $-$4.628 & $-$15.79 & $-$4.435 & $-$17.84\\
$a_{\text{IVT},\text{walk}}+l_{\text{IVT}}$ & $-$3.607 & $-$10.45 & $-$3.359 & $-$16.04\\
$a_{\text{OVT},\text{bus}}+l_{\text{OVT}}$ & $-$6.661 & $-$7.90 & $-$5.474 & $-$13.39\\
$a_{\text{OVT},\text{rail}}+l_{\text{OVT}}$ & $-$6.968 & $-$7.84 & $-$5.646 & $-$13.69\\
$a_{\text{cost}}+l_{\text{cost}}$ & $-$5.718 & $-$15.32 & $-$5.331 & $-$17.97\\
$u_{\text{IVT}}-l_{\text{IVT}}$ & 2.789 & 6.66 & 2.575 & 10.65\\
$u_{\text{OVT}}-l_{\text{OVT}}$ & 6.867 & 6.33 & 5.405 & 10.19\\
$u_{\text{cost}}-l_{\text{cost}}$ & 12.999 & 18.32 & 11.155 & 19.25\\
\midrule
Log-likelihood & \multicolumn{2}{r}{$-$2,263.17} & \multicolumn{2}{r}{$-$2,227.35}\\
Parameters & \multicolumn{2}{r}{70} & \multicolumn{2}{r}{85}\\
\end{longtable}}

\clearpage
{\footnotesize\setlength{\tabcolsep}{4pt}
\begin{longtable}{lrrrrrrrr}
\caption{COVID-19: models without random heterogeneity and continuous mixtures}\label{tab:A_cov_nl}\\
\toprule
 & \multicolumn{2}{c}{NL} & \multicolumn{2}{c}{NL with socios} & \multicolumn{2}{c}{Mixed NL} & \multicolumn{2}{c}{Mixed NL with socios}\\
\cmidrule(lr){2-3} \cmidrule(lr){4-5} \cmidrule(lr){6-7} \cmidrule(lr){8-9}
Parameter & est. & rob.\ $t$ & est. & rob.\ $t$ & est. & rob.\ $t$ & est. & rob.\ $t$\\
\midrule\endfirsthead
\multicolumn{9}{l}{\emph{(continued)}}\\ \toprule  & \multicolumn{2}{c}{NL} & \multicolumn{2}{c}{NL with socios} & \multicolumn{2}{c}{Mixed NL} & \multicolumn{2}{c}{Mixed NL with socios}\\ Parameter & est. & rob.\ $t$ & est. & rob.\ $t$ & est. & rob.\ $t$ & est. & rob.\ $t$\\ \midrule\endhead
\bottomrule\endfoot
\multicolumn{9}{l}{\emph{Constants}}\\
$\delta_{\text{pos}}$ & 0.017 & 3.06 & 0.017 & 3.05 & 0.030 & 2.91 & 0.030 & 2.91\\
$\delta_{\text{free}}$ & 0.983 & 10.25 & 1.111 & 7.23 &  &  &  & \\
$\delta_{\text{paid}}$ & 0.798 & 7.81 & 0.817 & 5.03 &  &  &  & \\
$\kappa_{\text{free},\text{fem}}$ &  &  & $-$0.270 & $-$1.86 &  &  & $-$0.814 & $-$3.56\\
$\kappa_{\text{free},\text{age}}$ &  &  & 0.141 & 0.92 &  &  & $-$0.167 & $-$0.66\\
$\kappa_{\text{free},\text{inc}}$ &  &  & $-$0.005 & $-$0.03 &  &  & $-$0.131 & $-$0.55\\
$\kappa_{\text{paid},\text{fem}}$ &  &  & $-$0.243 & $-$1.64 &  &  & $-$0.792 & $-$3.34\\
$\kappa_{\text{paid},\text{age}}$ &  &  & 0.247 & 1.58 &  &  & $-$0.095 & $-$0.36\\
$\kappa_{\text{paid},\text{inc}}$ &  &  & 0.188 & 1.22 &  &  & 0.143 & 0.59\\
$\mu_{\delta,\text{free}}$ &  &  &  &  & $-$0.560 & $-$1.83 & 0.032 & 0.09\\
$\mu_{\delta,\text{paid}}$ &  &  &  &  & $-$0.696 & $-$2.28 & $-$0.239 & $-$0.70\\
$\sigma_{\delta,\text{free}}$ &  &  &  &  & 0.019 & 0.18 & $-$0.026 & $-$0.29\\
$\sigma_{\delta,\text{paid}}$ &  &  &  &  & 0.702 & 5.37 & 0.708 & 5.39\\
\multicolumn{9}{l}{\emph{Attribute coefficients}}\\
$\beta_{\text{inf}}$ & $-$0.085 & $-$10.43 & $-$0.085 & $-$10.46 &  &  &  & \\
$\beta_{\text{ill}}$ & $-$0.060 & $-$11.91 & $-$0.060 & $-$11.93 &  &  &  & \\
$\beta_{\text{unk}}$ & $-$0.169 & $-$6.23 & $-$0.169 & $-$6.24 & $-$0.229 & $-$4.38 & $-$0.228 & $-$4.36\\
$\beta_{\text{prot}}$ & 0.009 & 10.72 & 0.009 & 10.73 &  &  &  & \\
$\beta_{\text{mild}}$ & $-$0.025 & $-$9.15 & $-$0.025 & $-$9.14 &  &  &  & \\
$\beta_{\text{sev}}$ & $-$12.925 & $-$8.04 & $-$12.948 & $-$8.04 &  &  &  & \\
$\beta_{\text{wait}}$ & $-$0.019 & $-$10.60 & $-$0.019 & $-$10.60 &  &  &  & \\
$\beta_{\text{fee}}$ & $-$0.002 & $-$10.15 & $-$0.002 & $-$10.17 &  &  &  & \\
$\beta_{\text{cov}}$ & 0.009 & 3.95 & 0.009 & 3.94 &  &  &  & \\
$\beta_{\text{ex}}$ & 0.205 & 1.38 & 0.210 & 1.41 &  &  &  & \\
$\mu_{\text{cov}}$ &  &  &  &  & 0.245 & 6.50 & 0.244 & 5.95\\
$\mu_{\text{ex}}$ &  &  &  &  & 2.412 & 2.31 & 2.227 & 2.16\\
$\sigma_{\text{cov}}$ &  &  &  &  & $-$0.168 & $-$6.86 & $-$0.167 & $-$6.08\\
$\sigma_{\text{ex}}$ &  &  &  &  & $-$3.540 & $-$3.36 & $-$3.324 & $-$3.15\\
$a_{\text{inf}}$ &  &  &  &  & $-$0.531 & $-$1.73 & $-$0.516 & $-$1.76\\
$a_{\text{ill}}$ &  &  &  &  & $-$0.248 & $-$1.22 & $-$0.229 & $-$0.95\\
$a_{\text{prot}}$ &  &  &  &  & $-$1.773 & $-$6.03 & $-$1.773 & $-$6.31\\
$a_{\text{mild}}$ &  &  &  &  & $-$5.136 & $-$6.77 & $-$5.212 & $-$6.65\\
$a_{\text{sev}}$ &  &  &  &  & 5.438 & 30.65 & 5.420 & 30.49\\
$a_{\text{wait}}$ &  &  &  &  & $-$1.121 & $-$6.42 & $-$1.157 & $-$6.46\\
$a_{\text{fee}}$ &  &  &  &  & $-$2.023 & $-$7.65 & $-$2.143 & $-$9.05\\
$b_{\text{inf}}$ &  &  &  &  & $-$4.184 & $-$4.01 & $-$4.257 & $-$4.31\\
$b_{\text{ill}}$ &  &  &  &  & $-$5.521 & $-$10.29 & $-$5.576 & $-$8.74\\
$b_{\text{prot}}$ &  &  &  &  & $-$6.591 & $-$6.24 & $-$6.578 & $-$6.61\\
$b_{\text{mild}}$ &  &  &  &  & 3.255 & 3.17 & 3.348 & 3.16\\
$b_{\text{sev}}$ &  &  &  &  & $-$6.603 & $-$8.20 & $-$6.549 & $-$8.29\\
$b_{\text{wait}}$ &  &  &  &  & $-$5.106 & $-$14.38 & $-$5.055 & $-$14.16\\
$b_{\text{fee}}$ &  &  &  &  & $-$5.596 & $-$15.53 & $-$5.449 & $-$16.89\\
\multicolumn{9}{l}{\emph{Nesting parameter}}\\
$\theta$ & 0.376 & 11.59 & 0.377 & 11.59 & 0.398 & 6.37 & 0.397 & 6.25\\
\midrule
Log-likelihood & \multicolumn{2}{r}{$-$11,067.03} & \multicolumn{2}{r}{$-$11,009.06} & \multicolumn{2}{r}{$-$9,598.88} & \multicolumn{2}{r}{$-$9,582.14}\\
Parameters & \multicolumn{2}{r}{14} & \multicolumn{2}{r}{20} & \multicolumn{2}{r}{25} & \multicolumn{2}{r}{31}\\
\end{longtable}}

\clearpage
{\footnotesize\setlength{\tabcolsep}{4pt}
\begin{longtable}{lrrrrrr}
\caption{COVID-19: latent class models}\label{tab:A_cov_lc}\\
\toprule
 & \multicolumn{2}{c}{LC NL} & \multicolumn{2}{c}{LC NL, socios in utility} & \multicolumn{2}{c}{LC NL, socios in class allocation}\\
\cmidrule(lr){2-3} \cmidrule(lr){4-5} \cmidrule(lr){6-7}
Parameter & est. & rob.\ $t$ & est. & rob.\ $t$ & est. & rob.\ $t$\\
\midrule\endfirsthead
\multicolumn{7}{l}{\emph{(continued)}}\\ \toprule  & \multicolumn{2}{c}{LC NL} & \multicolumn{2}{c}{LC NL, socios in utility} & \multicolumn{2}{c}{LC NL, socios in class allocation}\\ Parameter & est. & rob.\ $t$ & est. & rob.\ $t$ & est. & rob.\ $t$\\ \midrule\endhead
\bottomrule\endfoot
\multicolumn{7}{l}{\emph{Constants}}\\
$\delta_{\text{pos}}$ & 0.039 & 3.53 & 0.040 & 3.15 & 0.039 & 3.52\\
$\delta^{(1)}_{\text{free}}$ & 1.380 & 5.09 & 1.604 & 4.99 & 1.354 & 4.94\\
$\delta^{(1)}_{\text{paid}}$ & 1.739 & 6.15 & 1.891 & 5.64 & 1.711 & 5.96\\
$\delta^{(2)}_{\text{free}}$ & 1.090 & 4.01 & 1.314 & 4.26 & 1.099 & 4.03\\
$\delta^{(2)}_{\text{paid}}$ & 0.568 & 1.95 & 0.654 & 1.92 & 0.577 & 1.98\\
$\delta^{(3)}_{\text{free}}$ & $-$3.663 & $-$7.49 & $-$3.372 & $-$6.35 & $-$3.641 & $-$7.48\\
$\delta^{(3)}_{\text{paid}}$ & $-$4.190 & $-$7.39 & $-$4.002 & $-$6.62 & $-$4.150 & $-$7.33\\
$\kappa_{\text{free},\text{fem}}$ &  &  & $-$0.197 & $-$0.97 &  & \\
$\kappa_{\text{free},\text{age}}$ &  &  & 0.057 & 0.25 &  & \\
$\kappa_{\text{free},\text{inc}}$ &  &  & $-$0.230 & $-$1.09 &  & \\
$\kappa_{\text{paid},\text{fem}}$ &  &  & $-$0.109 & $-$0.49 &  & \\
$\kappa_{\text{paid},\text{age}}$ &  &  & 0.124 & 0.49 &  & \\
$\kappa_{\text{paid},\text{inc}}$ &  &  & 0.108 & 0.45 &  & \\
\multicolumn{7}{l}{\emph{Attribute coefficients}}\\
$\beta^{(1)}_{\text{inf}}$ & $-$0.166 & $-$5.08 & $-$0.212 & $-$4.62 & $-$0.168 & $-$5.01\\
$\beta^{(1)}_{\text{ill}}$ & $-$0.083 & $-$5.40 & $-$0.105 & $-$4.82 & $-$0.084 & $-$5.32\\
$\beta^{(1)}_{\text{unk}}$ & $-$0.395 & $-$4.41 & $-$0.489 & $-$4.06 & $-$0.401 & $-$4.41\\
$\beta^{(1)}_{\text{prot}}$ & 0.013 & 5.17 & 0.017 & 4.66 & 0.014 & 5.11\\
$\beta^{(1)}_{\text{mild}}$ & $-$0.058 & $-$5.32 & $-$0.072 & $-$4.83 & $-$0.058 & $-$5.27\\
$\beta^{(1)}_{\text{sev}}$ & $-$22.772 & $-$4.25 & $-$29.259 & $-$3.97 & $-$22.902 & $-$4.22\\
$\beta^{(1)}_{\text{wait}}$ & $-$0.054 & $-$5.11 & $-$0.070 & $-$4.61 & $-$0.055 & $-$5.06\\
$\beta^{(1)}_{\text{fee}}$ & $-$0.003 & $-$5.11 & $-$0.004 & $-$4.66 & $-$0.003 & $-$5.07\\
$\beta^{(1)}_{\text{cov}}$ & 0.027 & 1.96 & 0.013 & 0.82 & 0.029 & 2.01\\
$\beta^{(2)}_{\text{inf}}$ & $-$0.152 & $-$5.29 & $-$0.143 & $-$4.82 & $-$0.151 & $-$5.31\\
$\beta^{(2)}_{\text{ill}}$ & $-$0.148 & $-$6.72 & $-$0.139 & $-$5.88 & $-$0.147 & $-$6.75\\
$\beta^{(2)}_{\text{unk}}$ & $-$0.352 & $-$3.74 & $-$0.346 & $-$3.76 & $-$0.347 & $-$3.70\\
$\beta^{(2)}_{\text{prot}}$ & 0.021 & 5.68 & 0.020 & 5.06 & 0.021 & 5.69\\
$\beta^{(2)}_{\text{mild}}$ & $-$0.042 & $-$4.42 & $-$0.041 & $-$4.29 & $-$0.041 & $-$4.42\\
$\beta^{(2)}_{\text{sev}}$ & $-$20.895 & $-$3.94 & $-$20.170 & $-$3.87 & $-$20.743 & $-$3.94\\
$\beta^{(2)}_{\text{wait}}$ & $-$0.037 & $-$5.52 & $-$0.035 & $-$5.06 & $-$0.037 & $-$5.54\\
$\beta^{(2)}_{\text{fee}}$ & $-$0.028 & $-$5.50 & $-$0.026 & $-$5.00 & $-$0.028 & $-$5.47\\
$\beta^{(2)}_{\text{cov}}$ & 0.008 & 0.81 & 0.011 & 0.98 & 0.007 & 0.75\\
$\beta^{(3)}_{\text{inf}}$ & $-$0.141 & $-$3.55 & $-$0.138 & $-$3.58 & $-$0.142 & $-$3.55\\
$\beta^{(3)}_{\text{ill}}$ & $-$0.106 & $-$4.76 & $-$0.103 & $-$4.68 & $-$0.104 & $-$4.60\\
$\beta^{(3)}_{\text{prot}}$ & 0.020 & 4.49 & 0.018 & 3.82 & 0.020 & 4.60\\
$\beta^{(3)}_{\text{mild}}$ & $-$0.027 & $-$1.87 & $-$0.029 & $-$2.16 & $-$0.026 & $-$1.84\\
$\beta^{(3)}_{\text{sev}}$ & $-$24.582 & $-$2.29 & $-$22.413 & $-$2.21 & $-$25.204 & $-$2.34\\
$\beta^{(3)}_{\text{wait}}$ & $-$0.009 & $-$0.99 & $-$0.010 & $-$1.20 & $-$0.009 & $-$0.92\\
$\beta^{(3)}_{\text{fee}}$ & $-$0.004 & $-$2.48 & $-$0.004 & $-$2.42 & $-$0.004 & $-$2.47\\
$\beta^{(3)}_{\text{cov}}$ & 0.017 & 4.47 & 0.017 & 4.42 & 0.017 & 4.47\\
$\beta^{(3)}_{\text{ex}}$ & 0.446 & 1.67 & 0.509 & 1.89 & 0.436 & 1.64\\
\multicolumn{7}{l}{\emph{Nesting parameter}}\\
$\theta^{(1)}$ & 0.588 & 5.45 & 0.774 & 4.75 & 0.591 & 5.38\\
$\theta^{(2)}$ & 0.887 & 6.09 & 0.826 & 5.39 & 0.883 & 6.11\\
$\theta^{(3)}$ & 0.523 & 4.18 & 0.467 & 3.57 & 0.527 & 4.26\\
\multicolumn{7}{l}{\emph{Class allocation}}\\
$\zeta_{2}$ & 0.437 & 5.24 & 0.432 & 5.13 & 0.872 & 6.68\\
$\zeta_{3}$ & $-$1.373 & $-$9.66 & $-$1.351 & $-$9.66 & $-$1.204 & $-$5.10\\
$\omega_{2,\text{fem}}$ &  &  &  &  & $-$0.102 & $-$0.84\\
$\omega_{2,\text{age}}$ &  &  &  &  & $-$0.476 & $-$3.70\\
$\omega_{2,\text{inc}}$ &  &  &  &  & $-$0.716 & $-$5.52\\
$\omega_{3,\text{fem}}$ &  &  &  &  & 0.329 & 1.49\\
$\omega_{3,\text{age}}$ &  &  &  &  & $-$0.590 & $-$2.48\\
$\omega_{3,\text{inc}}$ &  &  &  &  & $-$0.638 & $-$2.56\\
\multicolumn{7}{l}{\emph{Implied class shares (sample mean of $\pi_{nq}$)}}\\
class 1 & \multicolumn{2}{r}{0.357} & \multicolumn{2}{r}{0.357} & \multicolumn{2}{r}{0.357}\\
class 2 & \multicolumn{2}{r}{0.553} & \multicolumn{2}{r}{0.550} & \multicolumn{2}{r}{0.552}\\
class 3 & \multicolumn{2}{r}{0.090} & \multicolumn{2}{r}{0.092} & \multicolumn{2}{r}{0.091}\\
\midrule
Log-likelihood & \multicolumn{2}{r}{$-$9,798.16} & \multicolumn{2}{r}{$-$9,784.85} & \multicolumn{2}{r}{$-$9,775.51}\\
Parameters & \multicolumn{2}{r}{39} & \multicolumn{2}{r}{45} & \multicolumn{2}{r}{45}\\
\end{longtable}}

\clearpage
{\footnotesize\setlength{\tabcolsep}{4pt}
\begin{longtable}{lrrrrrr}
\caption{COVID-19: logit mixed logit, implied distribution of each coefficient}\label{tab:A_cov_lml}\\
\toprule
 & \multicolumn{3}{c}{LML NL} & \multicolumn{3}{c}{LML NL with socios}\\
\cmidrule(lr){2-4} \cmidrule(lr){5-7}
Coefficient & 5\% & median & 95\% & 5\% & median & 95\%\\
\midrule\endfirsthead
\multicolumn{7}{l}{\emph{(continued)}}\\ \toprule Coefficient & 5\% & median & 95\% & 5\% & median & 95\%\\ \midrule\endhead
\bottomrule\endfoot
\multicolumn{7}{l}{\emph{Constants}}\\
$\delta_{\text{pos}}$ &  & 0.039 &  &  & 0.049 & \\
$\delta_{\text{free}}$ &  & $-$0.712 &  & $-$0.812 & $-$0.584 & $-$0.241\\
$\delta_{\text{paid}}$ & $-$1.367 & $-$0.543 & $-$0.088 & $-$1.510 & $-$0.839 & 0.202\\
\multicolumn{7}{l}{\emph{Attribute coefficients}}\\
$\beta_{\text{cov}}$ & $-$0.000059 & 0.283 & 0.388 & $-$0.023 & 0.249 & 0.444\\
$\beta_{\text{ex}}$ & $-$0.606 & $-$0.216 & 1.665 & $-$1.335 & $-$0.554 & 3.423\\
$\beta_{\text{inf}}$ & $-$0.399 & $-$0.084 & $-$0.022 & $-$0.578 & $-$0.160 & $-$0.039\\
$\beta_{\text{ill}}$ & $-$21.841 & $-$0.058 & $-$0.005 & $-$1.592 & $-$0.079 & $-$0.010\\
$\beta_{\text{prot}}$ & 0.00043 & 0.007 & 0.416 & 0.001 & 0.014 & 3.762\\
$\beta_{\text{mild}}$ & $-$0.077 & $-$0.039 & $-$0.012 & $-$0.327 & $-$0.028 & $-$0.00040\\
$\beta_{\text{sev}}$ & $-$409.376 & $-$6.612 & $-$0.432 & $-$213.554 & $-$9.276 & $-$1.839\\
$\beta_{\text{wait}}$ & $-$0.301 & $-$0.039 & $-$0.001 & $-$0.408 & $-$0.044 & $-$0.002\\
$\beta_{\text{fee}}$ & $-$3.327 & $-$0.018 & $-$0.00032 & $-$5.473 & $-$0.016 & $-$0.001\\
$\beta_{\text{unk}}$ &  & $-$0.241 &  &  & $-$0.279 & \\
\multicolumn{7}{l}{\emph{Nesting parameter}}\\
$\theta$ &  & 0.417 &  &  & 0.552 & \\
\end{longtable}}

\clearpage
{\footnotesize\setlength{\tabcolsep}{4pt}
\begin{longtable}{lrrrr}
\caption{COVID-19: logit mixed logit, support and parameters without random heterogeneity}\label{tab:A_cov_lmlpar}\\
\toprule
 & \multicolumn{2}{c}{LML NL} & \multicolumn{2}{c}{LML NL with socios}\\
\cmidrule(lr){2-3} \cmidrule(lr){4-5}
Parameter & est. & rob.\ $t$ & est. & rob.\ $t$\\
\midrule\endfirsthead
\multicolumn{5}{l}{\emph{(continued)}}\\ \toprule  & \multicolumn{2}{c}{LML NL} & \multicolumn{2}{c}{LML NL with socios}\\ Parameter & est. & rob.\ $t$ & est. & rob.\ $t$\\ \midrule\endhead
\bottomrule\endfoot
\multicolumn{5}{l}{\emph{Constants}}\\
$\delta_{\text{pos}}$ & 0.039 & 3.91 & 0.049 & 3.73\\
$\kappa_{\text{free},\text{fem}}$ &  &  & $-$0.648 & $-$2.65\\
$\kappa_{\text{free},\text{age}}$ &  &  & $-$0.017 & $-$0.07\\
$\kappa_{\text{free},\text{inc}}$ &  &  & 0.086 & 0.34\\
$\kappa_{\text{paid},\text{fem}}$ &  &  & $-$0.555 & $-$2.20\\
$\kappa_{\text{paid},\text{age}}$ &  &  & 0.166 & 0.67\\
$\kappa_{\text{paid},\text{inc}}$ &  &  & 0.333 & 1.23\\
$l_{\delta,\text{free}}$ & $-$0.712 & $-$2.55 & $-$0.976 & $-$2.11\\
$l_{\delta,\text{paid}}$ & $-$1.408 & $-$4.85 & $-$1.569 & $-$3.64\\
$u_{\delta,\text{free}}$ & $-$0.712 & $-$2.55 & $-$0.230 & $-$0.62\\
$u_{\delta,\text{paid}}$ & 0.012 & 0.04 & 0.437 & 1.23\\
\multicolumn{5}{l}{\emph{Attribute coefficients}}\\
$\beta_{\text{unk}}$ & $-$0.241 & $-$4.82 & $-$0.279 & $-$4.91\\
$a_{\text{inf}}+l_{\text{inf}}$ & $-$4.772 & $-$15.65 & $-$4.001 & $-$15.97\\
$a_{\text{ill}}+l_{\text{ill}}$ & $-$5.888 & $-$24.76 & $-$5.240 & $-$15.90\\
$a_{\text{prot}}+l_{\text{prot}}$ & $-$8.036 & $-$14.36 & $-$7.196 & $-$29.12\\
$a_{\text{mild}}+l_{\text{mild}}$ & $-$4.665 & $-$4.56 & $-$8.546 & $-$9.26\\
$a_{\text{sev}}+l_{\text{sev}}$ & $-$1.344 & $-$3.00 & $-$0.521 & $-$1.37\\
$a_{\text{wait}}+l_{\text{wait}}$ & $-$7.207 & $-$20.80 & $-$7.837 & $-$15.04\\
$a_{\text{fee}}+l_{\text{fee}}$ & $-$8.263 & $-$17.88 & $-$7.642 & $-$32.77\\
$u_{\text{inf}}-l_{\text{inf}}$ & 8.403 & 17.11 & 7.551 & 13.93\\
$u_{\text{ill}}-l_{\text{ill}}$ & 9.985 & 21.15 & 9.197 & 12.29\\
$u_{\text{prot}}-l_{\text{prot}}$ & 10.642 & 8.62 & 10.099 & 30.82\\
$u_{\text{mild}}-l_{\text{mild}}$ & 2.337 & 1.54 & 7.652 & 7.36\\
$u_{\text{sev}}-l_{\text{sev}}$ & 12.368 & 18.64 & 11.347 & 27.05\\
$u_{\text{wait}}-l_{\text{wait}}$ & 10.589 & 18.68 & 12.236 & 11.80\\
$u_{\text{fee}}-l_{\text{fee}}$ & 11.045 & 10.17 & 10.271 & 28.28\\
$l_{\text{cov}}$ & $-$0.003 & $-$0.50 & $-$0.028 & $-$5.21\\
$l_{\text{ex}}$ & $-$0.792 & $-$0.79 & $-$2.090 & $-$4.42\\
$u_{\text{cov}}$ & 0.419 & 6.19 & 0.470 & 8.20\\
$u_{\text{ex}}$ & 1.870 & 1.91 & 3.890 & 3.21\\
\multicolumn{5}{l}{\emph{Nesting parameter}}\\
$\theta$ & 0.417 & 7.84 & 0.552 & 8.61\\
\midrule
Log-likelihood & \multicolumn{2}{r}{$-$9,438.81} & \multicolumn{2}{r}{$-$9,399.38}\\
Parameters & \multicolumn{2}{r}{91} & \multicolumn{2}{r}{97}\\
\end{longtable}}

\end{document}